\documentclass[a4paper,12pt,twoside]{ThesisStyle}

\pdfoutput=1
\usepackage{vmargin}
\setpapersize{A4}
\setmarginsrb{37mm}{15mm}{25mm}{20mm}{12pt}{11mm}{0pt}{11mm}%

\usepackage{amsmath,amssymb}             
\usepackage{natbib}
\usepackage{aastex_hack}

\usepackage[latin1]{inputenc}
\usepackage[T1]{fontenc}

\usepackage[nottoc, notlof, notlot]{tocbibind}
\usepackage{minitoc}
\usepackage{aecompl}

\usepackage[intoc]{nomencl}

\makenomenclature

\usepackage{ifpdf}

\ifpdf
  \usepackage[pdftex]{graphicx}
  \DeclareGraphicsExtensions{.jpg}
  \usepackage[a4paper,pagebackref,hyperindex=true]{hyperref}
\else
  \usepackage{graphicx}
  \DeclareGraphicsExtensions{.ps,.eps}
  \usepackage[a4paper,dvipdfm,pagebackref,hyperindex=true]{hyperref}
\fi

\graphicspath{{.}{images/}}

\usepackage{color}
\definecolor{linkcol}{rgb}{0,0,0.4} 
\definecolor{citecol}{rgb}{0.5,0,0} 

\hypersetup
{
bookmarksopen=true,
pdftitle="The Evolution of the Hubble Sequence: morpho-kinematics of distant galaxies",
pdfauthor="Rodney DELGADO SERRANO", 
pdfsubject="The Evolution of the Hubble Sequence", 
pdfmenubar=true, 
pdfhighlight=/O, 
colorlinks=true, 
pdfpagemode=None, 
pdfpagelayout=SinglePage, 
pdffitwindow=true, 
linkcolor=linkcol, 
citecolor=citecol, 
urlcolor=linkcol 
}

\usepackage{rotating}                    
\usepackage{fancyhdr}                    

\let\footruleORIG\footrule
\renewcommand{\footrule}{\color{black} \footruleORIG}

\let\headruleORIG\headrule
\renewcommand{\headrule}{\color{black} \headruleORIG}

\usepackage{colortbl}
\arrayrulecolor{black}

\fancypagestyle{plain}{
  \fancyhead{}
  \fancyfoot{}

}

\usepackage{algorithm}
\usepackage[noend]{algorithmic}

\makeatletter

\def\cleardoublepage{\clearpage\if@twoside \ifodd\c@page\else%
  \hbox{}%
  \thispagestyle{empty}
  \newpage%
  \if@twocolumn\hbox{}\newpage\fi\fi\fi}

\makeatother
 
{%

\hrulefill
\vspace*{0.5cm}%
\end{minipage}
}

\let\minitocORIG\minitoc
\renewcommand{\minitoc}{\minitocORIG \vspace{1.5em}}

\usepackage{multirow}
\usepackage{slashbox}

{ \begin{list}%
	{$\bullet$}%
	{\setlength{\labelwidth}{25pt}%
	 \setlength{\leftmargin}{30pt}%
	 \setlength{\itemsep}{\parsep}}}%
{ \end{list} }

\renewcommand{\epsilon}{\varepsilon}

\newenvironment{vcenterpage}
{\newpage\vspace*{\fill}\thispagestyle{empty}}
{\vspace*{\fill}}

\usepackage{color}
\usepackage{enumerate}
\usepackage[latin1]{inputenc} 
\usepackage[T1]{fontenc}      
\usepackage{multicol}

\newenvironment{changemargin}{\begin{list}{}{%
\setlength{\headheight}{0cm}%
\setlength{\topmargin}{0cm}
\setlength{\headsep}{0cm}
\setlength{\oddsidemargin}{0cm}
\setlength{\evensidemargin}{0cm}
\setlength{\leftmargin}{0cm}%
\setlength{\rightmargin}{0cm}%
\setlength{\footskip}{0cm}%
}\item }{\end{list}}

\begin{document}

\begin{titlepage}
\begin{center}
\noindent {\large \textbf{OBSERVATOIRE DE PARIS}} \\
\vspace*{0.4cm}
\noindent {\large \textbf{DOCTORAL SCHOOL}} \\
\noindent \textbf{ASTRONOMY and ASTROPHYSICS OF ILE DE FRANCE} \\
\vspace*{0.3cm}
\noindent  Observatoire de Paris-Meudon \\
\noindent  Laboratoire Galaxies, Etoiles, Physique et Instrumentation - UMR 8111\\
\vspace*{0.7cm}
\noindent \Huge \textbf{T H E S I S} \\
\vspace*{0.9cm}
\vspace*{0.2cm}
\noindent \large {presented to obtain the degree of} \\
\vspace*{0.2cm}
\noindent \Large PhD of the Observatoire de Paris \\
\noindent \Large \textbf{Specialty : \textsc{Astronomy and Astrophysics}}\\
\vspace*{0.6cm}
\noindent \large {by\\}
\vspace*{0.1cm}
\noindent \LARGE Rodney \textsc{Delgado Serrano} \\
\vspace*{1.0cm}
\noindent {\Huge \textbf{The Evolution of the Hubble Sequence: morpho-kinematics of distant galaxies}} \\
\vspace*{0.8cm}
\vspace*{0.5cm}
\end{center}
\begin{center}
\begin{tabular}{lll}
\noindent \textbf{Examination board:} & &\\ \\
      \textit{President :}	& Ana \textsc{Gomez}     	     & Observatoire de Paris-Meudon\\
      \textit{Reviewers :}	& Sidney \textsc{van den Bergh}	     & Dominion Astrophysical Observatory\\
				& Denis \textsc{Burgarella}	     & Observatoire de Marseille Provence\\
      \textit{Examiners :}      & Lia \textsc{Athanassoula}          & Observatoire de Marseille Provence\\
      				& Thierry \textsc{Contini}	     & Observatoire Midi-Pyr{\'e}n{\'e}es\\
      \textit{Supervisors :}	& Francois \textsc{Hammer}	     & Observatoire de Paris-Meudon\\
                             	& Hector \textsc{Flores}	     & Observatoire de Paris-Meudon\\
\end{tabular}
\end{center}

\vspace*{0.35cm}
\begin{center}
\noindent October 06, 2010 \\
\end{center}

\end{titlepage}

\begin{changemargin}
\thispagestyle{empty}
\newpage
\thispagestyle{empty}
\includegraphics{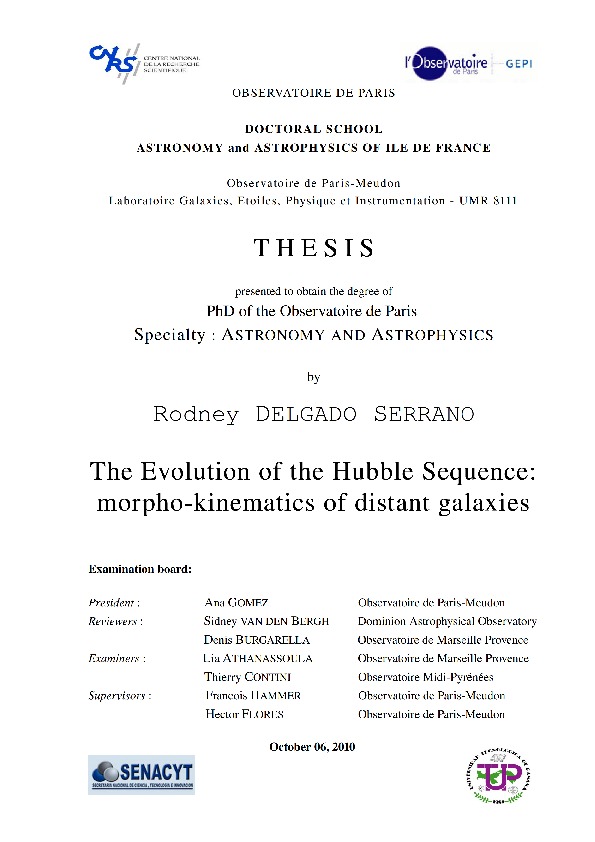}
\end{changemargin}

\sloppy

\titlepage

\dominitoc

\pagenumbering{roman}

 \cleardoublepage


\chapter*{Abstract}
\addstarredchapter{Abstract (English)}

{\small
The main objective of my thesis was to provide us, for the first time, with a reliable view of the distant Hubble sequence, and its evolution over the past 6 Gyr. To achieve this goal, we have created a new morphological classification method which (1) includes all the available observational data, (2) can be easily reproduced, and (3) presents a negligible subjectivity. This method allows us to study homogeneously the morphology of local and distant galaxies, and has the main advantage of presenting a good correlation between the morphological type and dynamical state of each galaxy. 
\\
The first step has been to study the evolution of galaxies using the IMAGES survey. This survey allowed us to establish the kinematic state of distant galaxies, to study the chemical evolution of galaxies over the past 8 Gyr, and to test important dynamical relations such as the Tully-Fisher relation. The information gained from kinematics is, indeed, crucial to guarantee a robust understanding of the different physical processes leading to the present day Hubble sequence. Using Integral Field Spectroscopy, which provides a complete kinematic diagnosis, we have been able to test our new morphological classification against the kinematic state of each galaxy. We found that the morpho-kinematic correlation is much better using our classification than other morphological classifications. Applying our classification to a representative sample of galaxies at $z \sim 0.6$, we found that 4/5 of spiral galaxies are rotating disks, while more than 4/5 peculiar galaxies are not in a dynamical equilibrium.  
\\
Applying our morphological classification to a representative sample of both local and distant galaxies, having equivalent observational data, we obtained a Hubble sequence both in the local and distant Universe. We found that spiral galaxies were 5/2 times less abundant in the past, which is compensated exactly by the strong decrease by a factor 5 of peculiar galaxies, while the fractional number of elliptical and lenticular galaxies remains constant. It strongly suggests that more than half of the present-day spirals had peculiar morphologies, 6 Gyr ago. 
\\
Finally, I present further studies concerning the history of individual galaxies at $z < 1$, combining kinematic and morphological observations. I also present the first ever-estimated distant baryonic Tully-Fisher relation, which does not appear to evolve over the past 6 Gyr. In the coming years, our morphological classification and these studies will be extended to galaxies at $z >> 1$, thanks to the future ELTs.
\\
{\large\textbf{Keywords:}}
Galaxy Formation and Evolution - Galaxy Morphology and kinematics.
\\

}

\chapter*{Abstract}
\addstarredchapter{Abstract (French)}

{\small
L'objectif principal de ma th{\`e}se etait de fournir, pour la premi{\`e}re fois, une vision fiable de la s{\'e}quence de Hubble distante, et de son {\'e}volution au cours des derni{\`e}res 6 milliards d'ann{\'e}es. Pour atteindre cet objectif, nous avons construit une nouvelle classification morphologique qui (1) prend en compte toutes les donn{\'e}es observationnelles disponibles, (2) peut {\^e}tre facilement reproduite, et (3) pr{\'e}sente une subjectivit{\'e} n{\'e}gligeable. Cette m{\'e}thode peut {\^e}tre appliqu{\'e}e {\`a} l'{\'e}tude morphologique des galaxies locales et distantes, et son principal avantage est de pr{\'e}senter une bonne corr{\'e}lation entre le type morphologique et l'{\'e}tat dynamique de chaque galaxie.
\\
La premi{\`e}re {\'e}tape a {\'e}t{\'e} d'{\'e}tudier l'{\'e}volution des galaxies gr{\^a}ce au relev{\'e} IMAGES. Ce relev{\'e} nous a permis d'{\'e}tablir l'{\'e}tat cin{\'e}matique des galaxies lointaines, d'{\'e}tudier leur {\'e}volution chimique depuis 8 milliards d'ann{\'e}es, et de tester des relations dynamiques importantes telles que celle de Tully-Fisher. La cin{\'e}matique est, en effet, une information cruciale n{\'e}cessaire pour garantir une compr{\'e}hension solide des diff{\'e}rents processus physiques conduisant {\`a} la s{\'e}quence de Hubble actuelle. En utilisant la spectroscopie int{\'e}grale de champ, qui fournit un diagnostic cin{\'e}matique complet, nous avons {\'e}t{\'e} en mesure de tester notre m{\'e}thode de classification morphologique en fonction de l'{\'e}tat cin{\'e}matique de chaque galaxie. En utilisant notre classification, nous avons trouv{\'e} une corr{\'e}lation morpho-cin{\'e}matique bien meilleur qu'avec d'autres classifications. En appliquant notre classification {\`a} un {\'e}chantillon repr{\'e}sentatif de galaxies {\`a} z $\sim$ 0.6, nous avons constat{\'e} que 4/5 des galaxies spirales sont des disques en rotation, tandis que plus de 4/5 des galaxies particuli{\`e}res ne sont pas en {\'e}quilibre dynamique.
\\
En appliquant notre classification morphologique {\`a} deux {\'e}chantillon repr{\'e}sentatif de galaxies locales et distants, disposant des donn{\'e}es observationnelles {\'e}quivalentes, nous avons obtenu une s{\'e}quence de Hubble dans l'Univers local et dans l'Univers distant. Nous avons trouv{\'e} que les galaxies spirales {\'e}taient 5/2 fois moins abondantes dans le pass{\'e}, ce qui est compens{\'e} exactement par la forte diminution, par un facteur 5, des galaxies particuli{\`e}res, alors que la fraction de galaxies elliptiques et lenticulaires reste constante. Ce resultat sugg{\`e}re fortement que plus de la moiti{\'e} des spirales d'aujourd'hui avait des morphologies particuli{\`e}res, 6 milliards d'ann{\'e}es auparavant.
\\
Enfin, je pr{\'e}sente de nouvelles {\'e}tudes concernant l'histoire de galaxies individuelles {\`a} $z < 1$, en combinant des observations morphologiques et cin{\'e}matiques. Je pr{\'e}sente aussi la premi{\`e}re {\'e}tude concernant la relation Tully-Fisher baryonique lointaine, qui montre une absence d'{\'e}volution depuis 6 milliards d'ann{\'e}es. Dans les ann{\'e}es {\`a} venir, notre classification morphologique et ces {\'e}tudes seront {\'e}tendues aux galaxies {\`a} z $>> 1 $, gr{\^a}ce aux futurs ELTs.
\\
{\large\textbf{Keywords:}}
Formation et {\'E}volution des Galaxies - Morphologie et Cin{\'e}matique des Galaxies.
\\

}

\chapter*{Acknowledgment}
\addstarredchapter{Acknowledgment}

{\small
First of all, I want to thank my parents, Wilfredo Delgado and Ana Matilde Serrano, my brother, Will, and my best friend (and now my wife), Jossela Calder{\'o}n, for being unconditionally there with me.
\\
I would like to also thank very much Francois Hammer and Hector Flores. Thank you for everything, for giving me the opportunity to be in this team, for trusting me the job I have done during these past few years, which  finally succeeded to derive the first Distant Hubble Sequence with a highlighted A$\&$A article and an ESA/NASA press release. Evidently, it was not so easy, but finally we could probe what we were looking for. Thank you very much Masters for all the knowledge you have shared with me and for teaching me all the meticulous details of the astronomical observations, which are not only "observations" but astronomical analytic examinations. Furthermore, thank you for giving me the opportunity to do observations at ESO in Chili. Anyway, you know this is not a goodbye.
\\
Thanks to the DU at the Paris Observatory for letting me participate and help in such a formation, offering to students of different ages the opportunity to learn more about the world of astronomical analysis. 
\\
Thanks to my PhD friends: Raphael Galicher, Myriam Rodrigues, Lo\"{\i}c Le Tiran, Manuel Gonz{\'a}lez, Noah Schwartz, Pierre Inizan, Sandrine Guerlet, Jean Coupon, Carlos Correia, Pierre Guillard,  with whom the idea of organizing conferences "by the PhD students, for the PhD students" of {\^I}le-de-France was born, creating thus the first Elbereth conferences (Elbereth 2008) with the help of the Paris Observatory (Gepi), IAP, ONERA, and Phase. Moreover, thanks to Jacqueline Plancy, Myriam Rodrigues, Raphael Galicher, Lo\"{\i}c Le Tiran, Daniel Rouan, Manuel Gonz{\'a}lez and Ivan Debono for letting me also participate and organize with them the "parrainage" of foreign PhD students at the doctoral school of {\^I}le-de-France. I hope such efforts will continue producing great results.
\\
Special thanks to the "Uranoscope de France", the French Embassy in Panama, and the Technological University of Panama for their efforts in giving to Panama the first Panamanian Astronomical Observatory. It has been 10 long years, but we have finally succeeded. As the first director of the new Panama's Observatory, I will put all of my best to make it as good as everyone is expecting.
\\
Thanks to the post-docs Mathieu Puech, Yanbin Yang, Paola di Matteo, Benoit Neichel, Isaura Fuentes and S{\'e}bastien Peirani, for the numerous rich discussions, and with whom I have learned a lot. In a similar way, thanks to Chantal Balkowsky, David Valls-Gabaud, Marc Huertas and Matthew Lehnert. Mathieu and Yanbin, who are not post-docs any more, thank you very much for your patience and for always having time for me.  
\\
Thanks to the PhD students Myriam Rodrigues, Lo\"{\i}c Le Tiran, Irene Balmes, Anand Raichoor, Rami Gasmi, Sylvain Fouquet and Yan Qu, the post-docs Jianling Wang and Susanna Vergani, the master student Karen Disseau, for the great moments we shared together, including the "tea time". Thanks to all the PhD and post-doc students of the "Observatoire de Paris" I had the great opportunity to know and share quite important recreation times. Thanks to my thesis sister Myriam Rodrigues, with whom I have shared the office during these three years of PhD. Furthermore, thanks to the summer football team with whom I had great moments in our Observatory football field: Juan Cabrera, Xuhui Han, Rui Pinto, Joao Marques, Juan Gutierrez, Nicolas Vasset, Matteo Cerruti, Lorenzo Matteini, Julio Ramirez, Alberto Escalante, Alvaro Alvarez, and all the others.
\\
I also want to thank Vivienne Wild and Jaime Fields for having the kindness of reading my thesis and correcting the grammar and orthographic mistakes.
\\
Of course I cannot forget to thank the incredible help of the Gepi and PhD school secretaries. I am thinking especially about Sabine Kimmel, Laurence Gareaux, Pascale Hamm{\`e}s, Pascale Lain{\'e}, Daniel Michoud, Jacqueline Plancy. Thank you for your friendship and your patience each time I went to ask for your help. Thanks also to Nadine Denis for all the support in the administration procedures during these last months of PhD.
\\
I have also a very special thought for Ana Gomez, Christophe Sauty, Jacqueline Plancy and S{\'e}bastien Fontaine, whose great help and support was essential since I came to Paris for the first time. I cannot hide the sadness that leaving Paris, and the Paris Observatory, causes me. There are, a lot, a lot of things I am leaving here. It has been five years since I met Ana Gomez, Christophe Sauty and Jacqueline Plancy for the first time, during the reception for the new M1 students of the Paris Observatory, and six years since I met S{\'e}bastien. Do you remember that? Nevertheless, within this sadness, I am also very happy because I am going back to my country, with a lot of projects and responsibilities, to try to make a better Panama.
\\
Thanks to SENACYT-IFHARU and the Technological University of Panama for the financial support during these five years. Of course, without it, I could not have come to the Paris Observatory, nor gained my PhD. A special thanks to Bernardo Fernandez, who has supported and helped me since I was a student at the University of Panama.
\\
Finally, but no less importantly, I want to express my most sincere gratefulness to the members of my thesis jury for their comments and great interest. Please, let me give a special thanks to my reviewers, Sidney van den Bergh and Denis Burgarella, whose interest in my thesis was evidently enormous and with whom I had the great pleasure to share a lot of rich discussions through the numerous emails we interchanged.
\\
I hope I did not forget anyone of those who were important during my five years in Paris. If I did, sorry for that, and be sure I will make it known.
\\
{\footnotesize Acknowledgment to http://olivier.commowick.org/thesis$\_$template.php for providing the thesis template used here.}
}

\tableofcontents

\listoffigures  
\listoftables   

\setlength{\nomitemsep}{-\parsep}
\printnomenclature[3.5cm]
\mtcaddchapter

\nomenclature{$\Lambda$CDM}{Lambda Cold Dark Matter}
\nomenclature{ACBAR}{    Arcminute Cosmology Bolometer Array Receiver}
\nomenclature{ACS}{      Advanced Camera for Surveys}
\nomenclature{ADU}{      Analog-to-Digital Unit}
\nomenclature{BOOMERANG}{Balloon Observations of Millimetric Extragalactic Radiation and Geophysics}
\nomenclature{AGN}{      Active Galactic Nuclei}
\nomenclature{CCD}{      Charge-Coupled Device}
\nomenclature{CDF-S}{    Chandra Deep Field - South}
\nomenclature{CDM}{      Cold Dark Matter}
\nomenclature{CERN}{     European Organization for Nuclear Research}
\nomenclature{CFRS}{     Canada-France Redshift Survey}
\nomenclature{CK}{       Complex Kinematics}
\nomenclature{CMB}{      Cosmic Microwave Background}
\nomenclature{CNRS}{     Centre National de la Recherche Scientifique}
\nomenclature{COBE}{     Cosmic Background Explorer}
\nomenclature{CoI}{      Co-Investigator}
\nomenclature{CSS}{      Collisional Starburst Scenario}
\nomenclature{DDO}{      David Dunlap Observatory}
\nomenclature{DM}{       Dark Matter}
\nomenclature{dss}{       Digitized Sky Survey}
\nomenclature{ELT}{      Extremely Large Telescope}
\nomenclature{ERO}{      Extremely Red Object}
\nomenclature{ESA}{      European Space Agency}
\nomenclature{ESO}{      European Southern Observatory}
\nomenclature{ETG}{      Early-Type Galaxy}
\nomenclature{EW}{       Equivalent Width}
\nomenclature{FWHM}{Full Width at Half Maximum}
\nomenclature{GIM2D}{Galaxy IMage 2D}
\nomenclature{GOODS}{Great Observatories Origins Deep Survey}
\nomenclature{GTO}{Guaranteed Time Observations} 
\nomenclature{HDFS}{Hubble Deep Field Survey}
\nomenclature{HDM}{Hot Dark Matter}
\nomenclature{HST}{Hubble Space Telescope}
\nomenclature{HUDF}{HST Ultra Deep Field}
\nomenclature{IDL}{Interactive Data Language}
\nomenclature{IFU}{Integral Field Unit}
\nomenclature{IMAGES}{Intermediate MAss Galaxies Evolution Sequence}
\nomenclature{INSU}{Institut National des Sciences de l'Univers}
\nomenclature{IRAF}{Image Reduction and Analysis Facility}
\nomenclature{LBG}{Lyman Break Galaxy}
\nomenclature{LCBG}{Luminous Compact Blue Galaxy}
\nomenclature{LCG}{Luminous Compact Galaxy}
\nomenclature{LF}{Luminosity Function}
\nomenclature{LIR}{InfraRed Luminosity}
\nomenclature{LIRG}{Luminous InfraRed Galaxy}
\nomenclature{MCG}{Morphological Catalogue of Galaxies}
\nomenclature{MDLF}{Morphological-Dependent Luminosity Function}
\nomenclature{MIPS}{Multiband Imaging Photometer for Spitzer} 
\nomenclature{MOS}{Multi-Object Spectroscopy}
\nomenclature{MW}{Milky Way}
\nomenclature{NASA}{National Aeronautics and Space Administration}
\nomenclature{NTT}{New Technology Telescope}
\nomenclature{P.A.}{Position Angle}
\nomenclature{PACS}{Photodetector Array Camera and Spectrometer}
\nomenclature{PCM}{Primordial Collapse Model}
\nomenclature{PR}{Perturbed Rotation}
\nomenclature{PSF}{Point Spread Function}
\nomenclature{RD}{Rotating Disk}
\nomenclature{SDSS}{Sloan Digital Sky Survey}
\nomenclature{SED}{Spectral Energy Distribution}
\nomenclature{SFR}{Star Formation Rate}
\nomenclature{TFR}{Tully-Fisher Relation}
\nomenclature{TF}{Tully-Fisher}
\nomenclature{UDF}{Ultra Deep Field}
\nomenclature{ULIRG}{Ultra-Luminous InfraRed Galaxy}
\nomenclature{VLT}{Very Large Telescope}
\nomenclature{VVDS}{VIMOS VLT Deep Survey}
\nomenclature{WFPC2}{Wide Field Planetary Camera 2}
\nomenclature{WFI}{Wide-Field Imager}
\nomenclature{WMAP}{Wilkinson Microwave Anisotropy Probe}
\nomenclature{ZP}{Zero-Point}

\mainmatter


\part{Galaxy evolution in a cosmological context}


\chapter{Cosmology and galaxy classification}
\minitoc

\section{From the Big-Bang to the formation of galaxies}

\begin{figure}[!h]
\centering
\includegraphics[width=0.8\textwidth]{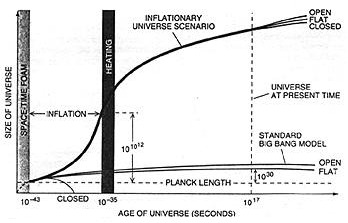}
\includegraphics[width=0.8\textwidth]{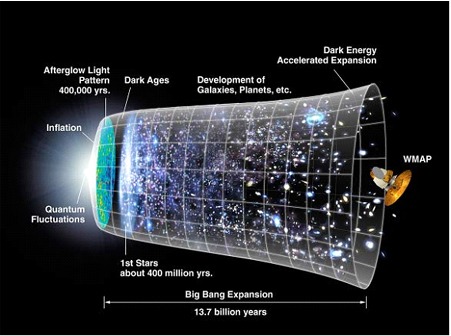}
\caption[Universe history: from the Big-Bang to the present galaxies]{{\it Top}: plot showing the Universe evolution scenario (see details in appendix \ref{CosmologyTheHistoryOfTheUniverse}). {\it Bottom}: an artistic view of the Universe history, from the Big-Bang, through the inflation and the emission of the present Cosmic Microwave Background (CMB), until the present galaxies (Credit: NASA/WMAP Science Team).}
\label{Evol_UniverseScenarii}
\end{figure}

In appendix \ref{CosmologyTheHistoryOfTheUniverse}, I make a more detailed explanation of the different approaches attempting to explain the evolution of the Universe from its "beginning" until the present (see figure \ref{Evol_UniverseScenarii}), as well as the principal models explaining the formation of the different structures of the Universe. In this section (and in appendix \ref{CosmologyTheHistoryOfTheUniverse}), I refer especially to a framework based on the Big-Bang theory, which generally explain the evolution of the Universe beginning by a big "explosion", and a subsequent expansion (which continues infinitely, or later becomes a contraction with a Big-Crunch end). This theory is the more accepted within the scientific community until the present days. \\

There are two principal model developed in order to explain the formation of large and small structures in the Universe: the "primordial collapse model" (PCM), and the "hierarchical model" (HM). Both of them confront each other. The PCM proposes a {\it top-down} configuration, with the larger structures forming first and evolving to form the smaller ones.  The HM has been developed with a {\it bottom-up} configuration: the smaller structures form first, and evolve to form the largest ones. However, it is the observational cosmology which has the last word. It seems that a large part of observational results support the hierarchical model (see appendix \ref{CosmologyTheHistoryOfTheUniverse}). Nonetheless, such a theoretical scenario has evolved significantly since its creation (in the 60s) until the present days, in order to establish the best possible combination of parameters in agreement with the existing observational data. \\

\begin{figure}[!h]
\centering
\includegraphics[width=1.0\textwidth]{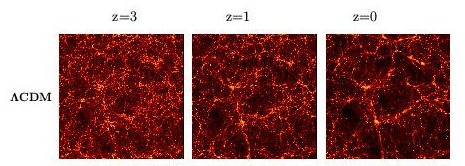}
\caption[$\Lambda$CDM model simulation]{$\Lambda$CDM model simulation on the Virgo cluster \citep[][for the Virgo Supercomputing Consortium]{1998ApJ...499...20J,1998MNRAS.296.1061T}.}
\label{LCDM_Model_VirgoCluster}
\end{figure}

Nowadays the most accepted model, adopted by a large part of the scientific community, is the so called $\Lambda$CDM (Lambda-Cold Dark Matter) model (see figure \ref{LCDM_Model_VirgoCluster}). It is an extension of the original CDM model proposed in the 80s \citep[e.g.,][]{1984fegl.proc..163P}, even though it includes new parameter values to be in agreement with the most recent observations. The $\Lambda$CDM model supposes an expanding Universe composed of baryonic and cold dark matter (see appendix \ref{CosmologyTheHistoryOfTheUniverse}). Such an expansion is accelerating thanks to a certain dark energy, which is represented by the cosmological constant $\Lambda$. This scenario predicts that the first baryonic structures to be formed have a typical size of the present globular clusters (10$^{5}$-10$^{6}$ M$_{\odot}$). Interestingly, these are also known to be amongst the oldest objects in the Universe. Inside these halos of baryonic matter, the first stars are formed. Then, such structures evolve to form the largest ones (see more details in appendix \ref{CosmologyTheHistoryOfTheUniverse}). \\

      \subsection{The cosmological principle}

This principle states that the Universe is homogeneous and isotropic. The first property refers to the fact that, at a given time, all points in the Universe are equivalent to each other. The second one means that the Universe looks the same in all directions. These two properties drive us to the original statement of the cosmological principle: "we do not live at a privileged location in the Universe". Even if, at first glance, the Universe is not strictly homogeneous and isotropic (for example, if we only consider the Solar System, or the Galaxy), it appears to be so, with a good approximation, on sufficiently large scales.\\

\begin{figure}[!h]
\centering
\includegraphics[width=1.0\textwidth]{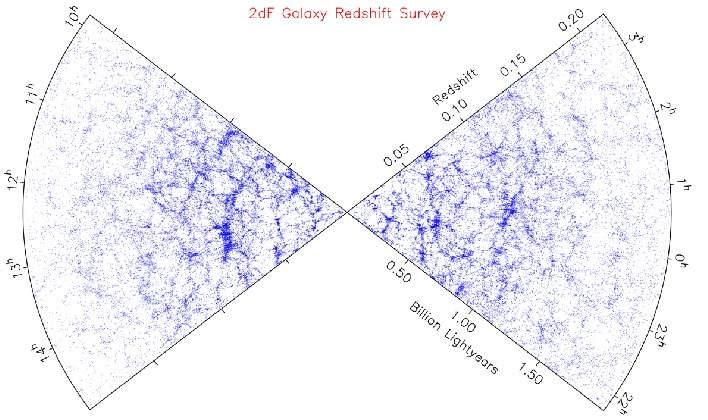}
\caption[Redshift galaxy distribution from the 2dFGRS]{Image showing the distribution of galaxies in a redshift cone from the 2dFGRS (2 degree Field Galaxy Redshift Survey)\citep[][]{2001MNRAS.328.1039C}.}
\label{Galaxy_Survey}
\end{figure}

However, the last affirmation in the precedent paragraph does not seem to be the case either. During decades, astronomers have been compiling catalogs giving the positions and magnitudes of galaxies. The result is showed in figure \ref{Galaxy_Survey}. As we can see galaxies appear to lie along filaments or within flattened structures (sheets) separated by voids. Moreover, concerning the isotropy of the Universe, one can well imagine that an observer traveling very fast with respect to the local matter will see galaxies (until a certain scale) moving toward him in one direction and away from him in another direction. The isotropic property of the Universe is thus not true for all observers. In addition to the above, one must consider that the Universe evolves, while we are observing it at different epochs.\\

Is therefore the cosmological principle wrong? No, because of the following reasons. On the one hand, one must replace the classic idea of homogeneity by the notion of "statistical homogeneity" (whereby the distribution is smooth only in an average sense), as well as understand how the Universe deviates from perfect smoothness. In this context, if we have a homogeneous distribution and we add deviations on it, which we could call inhomogeneities, then this will cause differences between locations to appear by chance, and any statistical measure of the inhomogeneities is independent of position \citep[][]{Liddle...Loveday...TOCC}. As an example, we can consider two people rolling two dice (each one) a large number of times. Even if their individual sequences will differ, they will still share statistical properties such as the mean value and its standard deviation \citep[][]{Liddle...Loveday...TOCC}. On the other hand, one also must consider that random velocities of galaxies, as well as the velocities of the stars inside them, are small if we compare these velocities to the relative velocities of galaxies separated by larger distances ($\sim$several tens of Mpc). Thus, any observer looking at the Universe at sufficiently large scales is, in a good approximation, a fundamental observer.\\

\begin{figure}[!h]
\centering
\includegraphics[width=0.9\textwidth]{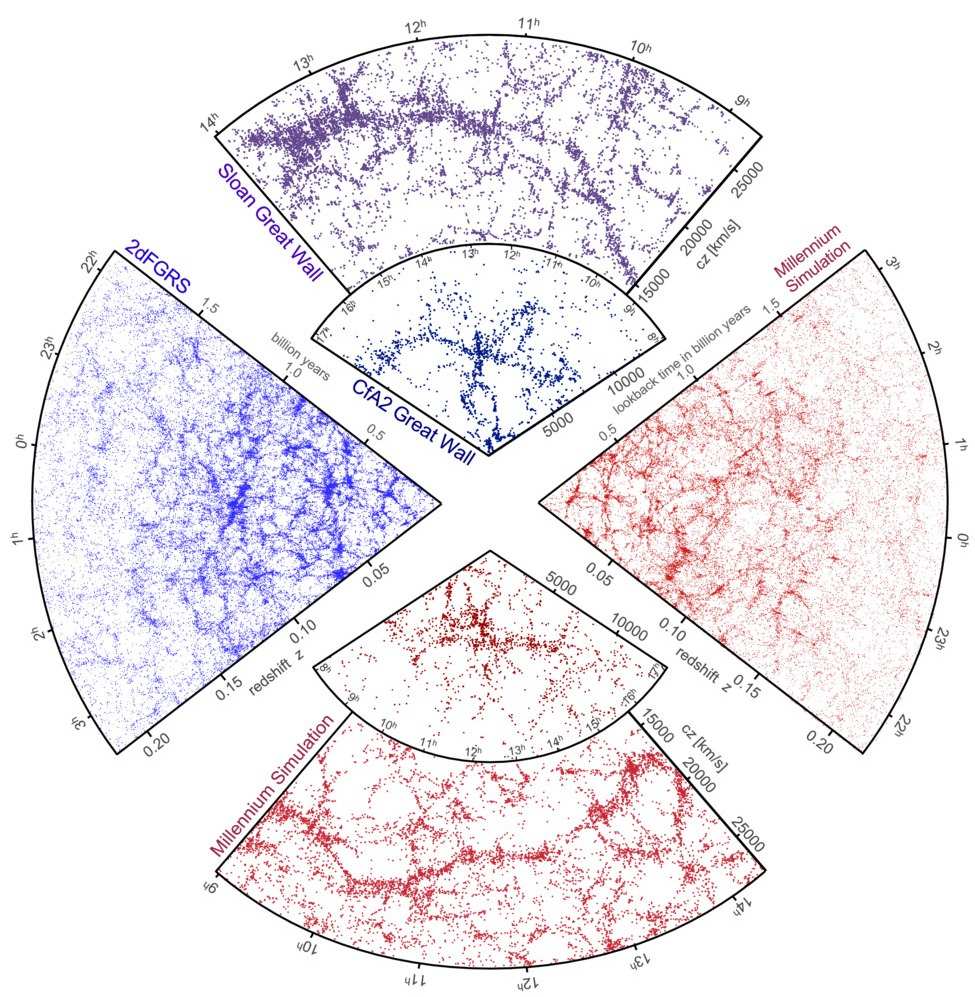}
\caption[Comparison of galaxy redshift distribution between simulations and observations]{Comparison of galaxy redshift distribution between simulations (in red) and observations (in bleu) \citep[][]{2006Natur.440.1137S}.}
\label{Simu_redshift_survey}
\end{figure}

\begin{figure}[!h]
\centering
\includegraphics[width=0.5\textwidth]{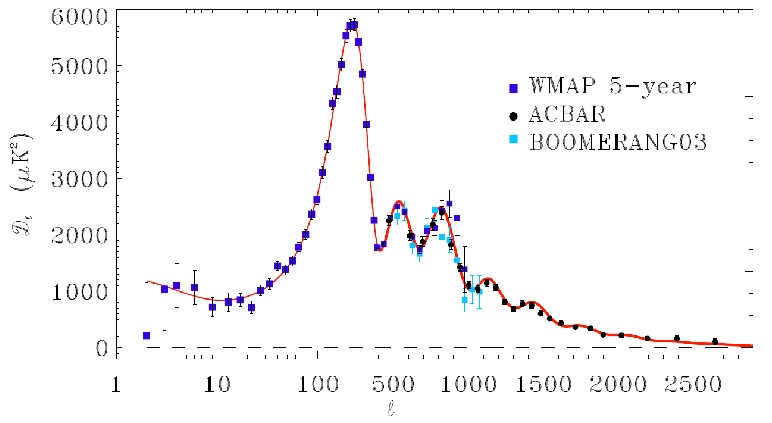}
\includegraphics[width=0.5\textwidth]{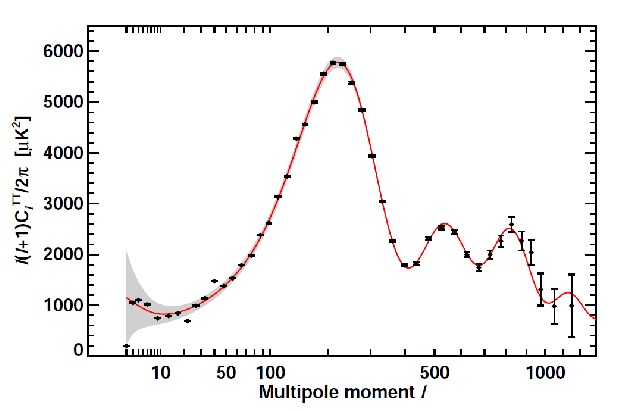}
\caption[Comparison between the power spectrum derived from a $\Lambda$CDM model and the values obtained by observations]{{\it Left}: Comparison between the temperature anisotropy power spectrum derived from the $\Lambda$CDM model and values from WMAP5, BOOMERANG and ACBAR observations \citep[][]{2009ApJ...694.1200R}\citep[see also][for a general explanation]{2009AIPC.1132...86C}.{\it Right}:Comparison between the temperature anisotropy power spectrum derived from the $\Lambda$CDM model and values determined from seven years of WMAP data \citep[][]{2010arXiv1001.4635L}.}
\label{PowerSpect1}
\end{figure}

Therefore, based on \citet{1978obco.meet....1G}, we can rewrite the cosmological principle as follows: the Universe appears statistically homogeneous and isotropic to all fundamental observers at a given cosmic time. Two important observations in agreement with this are the CMB (see figure \ref{CMB_WMAP5}) and the fact that the Hubble law seems to be independent of the direction of observation. Furthermore, the well known $\Lambda$CDM model simulations (see section \ref{The_hierarchical_model}) are in good approximation with observations of large scale structures (see figure \ref{Simu_redshift_survey} and \ref{PowerSpect1}).  \\

This fundamental principle thus allows us to connect distant objects to local ones. This is a key condition for the goal I wanted to reach with my thesis, which assumes that galaxies seen at a look-back time of $\sim$6.5 Gyr are the progenitors of the present-day Hubble sequence (see chapter 6). Therefore, the cosmological principle allows us to link these two epochs independently of the line of sight\footnote{However, the problem of the cosmic variance still remains to be solved.}, establishing that one can compare distant galaxies and local ones.\\

\section{Morphological classification of galaxies}
\label{MorphologicalClassificationSubsect}

Galaxies are complex objects containing several tens of billions stars, as well as gas and dust. Present-day galaxies are regular and relaxed systems, and are made of a dispersion-supported bulge surrounded by a rotationally-supported disk. They all fit into the initial morphological scheme proposed by \citet{1926ApJ....64..321H}, which is called the Hubble sequence and will be described below. \\

A morphological classification of galaxies lies in classifying galaxy shapes on the basis of their images. Looking for some similar features, we can separate galaxies into differents groups. However, if we want to catch a physical meaning, this task is much more complicated than it could be expected at first glance.\\

Before starting to talk about the morphological classification of galaxies some questions need to be answered: how we define the sample?, is this sample representative of one or all kinds of galaxy populations?, which methodology should be used to classify the galaxy morphologies?, which parameters will be included in such methodology?, can it be reproduced in other samples?, and can it be independent of human judgment (opinion)? \\

   \subsection{Considerations on the classification of a galaxy sample}
   \label{ConsiderationsforClassificationSubsect}

Since the beginning of the past century, astronomers have been gathering images of galaxies from the Local Universe (and very recently from the Distant Universe). The number of Local galaxies being photographed have thus increased enormously from a few hundreds in the 1920s \citep[][]{1926ApJ....64..321H}\footnote{I refer here to galaxy images with enough resolution and surface brightness to be morphologically classified. Galaxies have been photographed since the 18th century, as they were among the objects known as "nebulae" in the Messier catalog (1784). Furthermore, Sir William Herschel (1738-1822) and his son, Sir John (1792-1871), constructed a larger catalog (the first of the entire -north and south- sky) of about 4 630 nebulae (1864), which was replaced later by the Dreyer's New General Catalog (NGC) in 1890.} to almost a million in the 2000s \citep[e.g.,][\footnote{The Third Reference Catalog of Bright Galaxies.}; and the  SDSS catalog]{1991trcb.book.....D}. We can then infer that compared to the "sample" of local galaxies existing in the 1920s, at present we now have available a truly exhaustive "library" of Local galaxies. \\

The methodology followed by Hubble to classify the morphology of galaxies was the simplest one. He first separated objects with regular forms from those with irregular ones, these last representing 2 or 3 per cent of his sample. He then separated the regular objects in different groups, each one following a common feature. This gave two main groups: elliptical and spiral galaxies. Finally, he subdivided each group in subgroups (see subsection \ref{TheLocalHubbleSequenceClassif}). At this stage, one can ask oneself if there could be some kinds of galaxy morphology that could be missed in the relative small Hubble's sample, as he did not have access to the whole library of the local galaxies. Indeed, if we consider that lenticular galaxies were added after Hubble papers, we could consider so. However, it was just a problem of definition \citep[][]{1975gaun.book.....S,2009ApJ...694L.120V}. Then, we can conclude that, thanks to the cosmological principle, the probability of missing some morphological types is very small (even if the size of the sample is relatively small and the sky zone being observed is limited)\footnote{For example, if we consider only the elliptical galaxies, which represent the smallest fraction ($\sim 3\%$) of the local galaxy population (see chapter 6), the probability of \citet{1926ApJ....64..321H}, using a sample of 400 galaxies, to miss the elliptical population is roughly 0.000005. Note that if we consider the elliptical/lenticular population with a fraction of $\sim 17\%$, following the Shapley-Ames Catalog, the probability of missing such a population by Hubble goes down to 4.28{\it x}10$^{-33}$. I assumed here that galaxies are randomly distributed over the whole sky.}. Nonetheless, \citet{1996ApJS..105..209I}, by studying a sample of faint galaxies, concluded that when samples are limited by a selection bias (the flux in this case), they can exclude large number of objects having structures which may be important for the understanding of the galaxy evolution (rings, bars, etc.). Taking into account that the main selection bias during the beginning of the past century was the surface brightness limit imposed by the small telescope diameters at such epochs, it explains the evolution of the local Hubble sequence diagram from 1936 ({\it The Realm of the Nebulae} by Hubble\footnote{\citet{1936rene.book.....H}.}) to the revised one by \citet{1959HDP....53..275D} (see subsections \ref{TheLocalHubbleSequenceClassif} and \ref{DeVaucouleursRevisedClassif}).  \\

A fundamental consideration when studying a sample of galaxies is therefore: "{\bf the selection criteria}". Rather than being only concerned by the number of galaxies in the sample, could a selection criterion lead to a representative sample\footnote{see subsection \ref{NeedOfRepresentativenessSubsect} for a discussion about the representativeness of a sample, and its possible effects on the galaxy evolution studies.} that misses a galaxy type population (e.g., E/S0 galaxies)? Yes, it does. One example I experienced myself during my thesis is the comparison of the work we made in \citet{2008A&A...484..159N} and that in \citet{2010A&A...509A..78D} (see chapters 5 and 6). By taking a selection criterion (equivalent width of the O[II] emission line larger than 15$\mathring{A}$), which is not related to the morphology of galaxies, we found in Neichel et al. only spiral and peculiar galaxies. This further influenced the methodology used to classify the galaxies defined in both papers, which was limited in the first one. Therefore, the definition of a specific selection criterion can lead to a specific sample of galaxy images, which could miss some morphological types populations, and even could affect the methodology defined for the galaxy classification. \\

The classification methodology could also be affected by the observational data available. One example is the comparison between a sample of galaxies where each galaxy has images with different filters, and another sample where each galaxy has only one filter observation. In the first case, the color of each galaxy could play an important role in the classification process, while in the second case, the color is not taken into account. \\

Additionally to the above reflections, I want to draw the reader's attention to the characteristics of the images used to classify galaxies. Images play the main role for the morphological classification process, and should be considered seriously:\\

\begin{figure}[!h]
\centering
\includegraphics[width=0.4\textwidth]{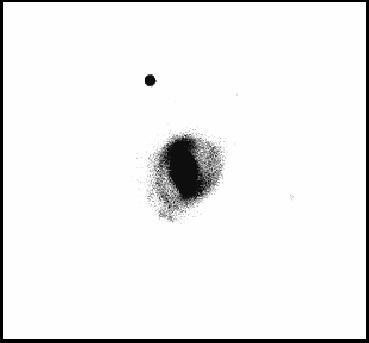}
\includegraphics[width=0.4\textwidth]{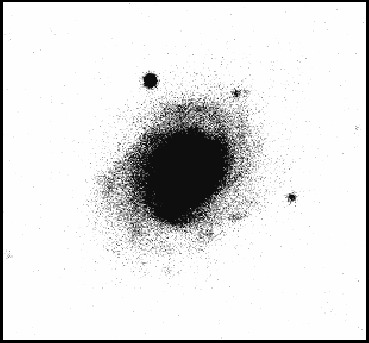}
\caption[Images of the same object with difference depth]{The difference in depth of images of the same object could imply a different morphological classification. As an example, I show here the same galaxy image from the SDSS. Only by playing with the contrast, we could think they are two different galaxies.}
\label{depthIm}
\end{figure}

\begin{figure}[!h]
\centering
\includegraphics[width=0.6\textwidth]{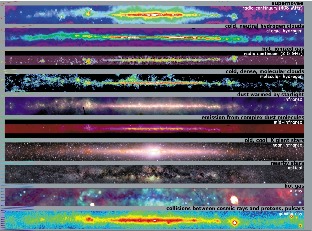}
\caption[Images of the Milky Way in different wavelength bands]{Images of the Milky Way in different wavelength bands. As we can see, each band gives us different information about a galaxy. (Credit: http://www.astro.wisc.edu/).}
\label{wavebandIm}
\end{figure}

\begin{figure}[!h]
\centering
\includegraphics[width=0.9\textwidth]{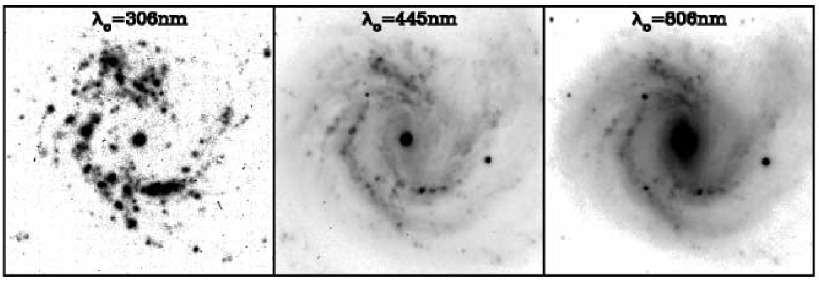}
\caption[Different appearances a galaxy could have if observed at different wavelengths]{Different appearances a galaxy could have if observed at different wavelengths from \citet{2003ApJ...592L..13S}. From left to right: NGC 4303 in UV, B-band, and I-band, respectively.}
\label{wavebandIm2}
\end{figure}

\begin{itemize}
\item Depth: Figure \ref{depthIm} shows two images of the same galaxy with identical scale and orientation. Nevertheless, as we can see, one could think that these two images illustrate two different objects. Thus, morphological classification strongly depends on depth. 
 
\item Wavelength range: Observations of the same galaxy in different wavebands reveals different information, as it is shown in figures \ref{wavebandIm} and \ref{wavebandIm2}. Therefore, even if each band image is very useful to study and to better understand all the components of a particular galaxy, there is no meaning in comparing two galaxies using different bands. If such was the case, one could classify nearly identical objects in different classes just because we are not comparing the same information. This problem should be treated carefully when comparing galaxies at different redshifts, since the Doppler effect\footnote{Note: "Doppler effect" isn't really correct, as the "cosmological redshift" has the same effect as the Doppler effect but the cause (an expanding Universe) is completely different.} shifts the observable wavelength compared to the rest-frame wavelength. For example, an R-band (SDSS) image for a local galaxy and a z-band (ACS/HST) for a galaxy at {\it z}$=$0.7 allows us to sample the same rest-frame wavelengths. 
  
\item Resolution: The spatial resolution of the images can also change the way we can see a galaxy. It depends on the PSF (FWHM), which gives the smallest resolution element of the observation. Evidently, for a given galaxy, the better the spatial resolution of its images, the better the details we can distinguish on it. Therefore, if we compare two different galaxies with different spatial resolution images, we are not looking at the same degree of details\footnote{such details will also depend on the number of photons by element of resolution, which depend on the exposition time, and is defined by the "Depth" of the image.}. As a consequence, we may not be "measuring" the same information. Furthermore, if the galaxies are at different redshift (distances), one must pay attention to make the difference between the angular spatial resolution (arsec/pixel) of the instrument, and the linear proper spatial resolution (kpc/resolution element) of the galaxy being observed.

\end{itemize}

The last considerations to be evoked here is the reproducibility and subjectivity of the morphological classification method. For this, I just want to draw the reader's attention on the study carried out by \citet{1995MNRAS.274.1107N}. They took 831 galaxies to be morphologically classified (by eye) by six astronomers, investigating the question whether different astronomers place a given object in the same class when they work from identical material. Interestingly, in many cases, two astronomers placed the same galaxy in different classes with a non-negligible scatter. I could call this the {\it Human Feeling (HF) bias}. Such a HF bias makes advisable to look for a quantitative analysis which could verify the qualitative one. \\

Finally, according to the above considerations for a morphological classification of galaxies, I can cite the following conclusions: \\
{\bf (1)} any galaxy sample must be homogeneous, in the sense that all galaxies in it must have the same kind of information, \\
{\bf (2)} the criteria used to separate the galaxy morphologies into classes must lead to a unique classification, making it a reproducible process. When different parameters are used, the order of their application must be carefully established to reduce its inherent subjectivity (to avoid that a same galaxy could be placed into more than one class). With this objective, the parameters should also be quantifiable.  \\
{\bf (3)} Such parameters should be as close as possible to physical properties (e.g., D/T, which represents the ratio of the disk flux to the total galaxy flux, is correlated with the amount of rotational support). \\

Since the last century, different propositions for the morphological classification of galaxies have been built up. Even if they all share Hubble's original notion\footnote{the morphological sequence should attest to an underlaying sequence of numerous physical processes}, each of them has its special characteristics. The most relevant ones are presented below.

      \subsection{The local Hubble sequence}
      \label{TheLocalHubbleSequenceClassif}

\begin{figure}[!h]
\centering
\includegraphics[width=0.9\textwidth]{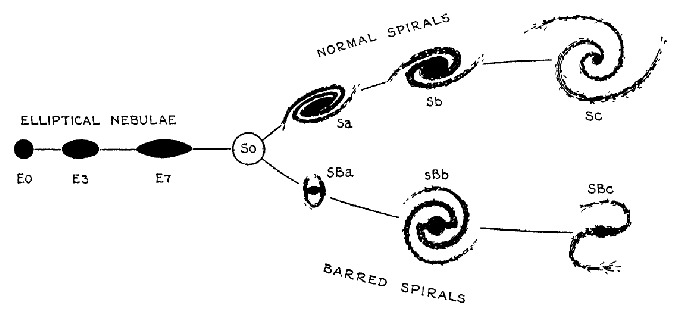}
\caption[Hubble tuning-fork diagram (1936)]{This is "The Sequence of Nebular Types" diagram published by \citet{1936rene.book.....H}.}
\label{firstHubblediag}
\end{figure}

The most used morphological classification scheme is the "Hubble sequence", which was initially introduced by \citet{1926ApJ....64..321H}. In his first version, based on the optical appearance of galaxy images on photographic plates, Hubble proposed three general classes: ellipticals (En\footnote{n=1,2,...,7 indicates the ellipticity of the galaxy}), spirals (S or SB\footnote{B identifies a spiral galaxy with a bar in the center}), and irregulars (Irr). In 1936\footnote{in his book {\it The Realm of the Nebulae}}, he presented the first "Hubble tuning-fork diagram", as it is shown in figure \ref{firstHubblediag}. Here, Hubble added the "more or less hypothetical type S0", presented as a necessary intermediate state between early type galaxies (ellipticals) and late types galaxies (spirals). The final Hubble sequence thus consisted of four galaxy types: elliptical (E), lenticular (S0), spiral (S or SB), and irregular (Irr). However, this last type of galaxies (irregulars) was considered at that time as not finding a "place in the sequence of classification" because they "show no evidence of rotational symmetry". For this reason it does not appear in the original diagram (figure \ref{firstHubblediag}). It is important to note that in 1936 scientists believed that the Hubble diagram revealed a time evolution of galaxies, and it was thus called "a sequence". Indeed, it was believed that "elliptical galaxies formed first in the universe history and becomes spirals with time, but after being lenticulars". We keep calling elliptical galaxies as early type galaxies and spiral galaxies as late type galaxies for this reason.\\

Later studies developed a more detailed Hubble diagram \citep[e.g.,][]{1975gaun.book.....S,1981rsac.book.....S,1994cagv.book.....S} using larger catalogs of galaxies. However, they all follow the same principle: galaxies are organised from pure bulges (ellipticals) to increasing disk contribution to their light or mass (S0 to Sc). Contrary to figure \ref{firstHubblediag}, in these later diagrams real examples of S0 galaxies are found.\\

The main types in this classification are the following:
\begin{itemize}
\item Ellipticals: these galaxies are characterized by their ellipticity e=(a-b/a), where a and b are respectively the major and minor axes of the ellipse formed by the projection of the galaxy on the sky's plane. They are also noted as En, where n=10e (n value goes from 0 to 7).
\item Lenticulars: S0 galaxies guarantee the transition between elliptical and spiral galaxies. They have a bulge and a disk. Nevertheless, they don't present spiral arms.
\item Spirals: they are composed by a bulge and a disk, and this latter (the disk) could present spiral arms which are attached to the bulge. For instant, this type of galaxies can be divided in subclasses (Sa to Sd) according to the increase resolution of spiral arms and the decrease importance of the bulge compared to the disk. 
\item Irregulars: Originally, these galaxies were separated from the others mainly because of the absence of a rotational symmetry\citep{1936rene.book.....H}. Morphologically, these objects are characterized by a lack of a conspicuous nuclei and/or by a non symmetric shape.
\end{itemize}

      \subsection{De Vaucouleurs's Revised Classification Scheme}
      \label{DeVaucouleursRevisedClassif}

\begin{figure}[!h]
\centering
\includegraphics[width=0.43\textwidth]{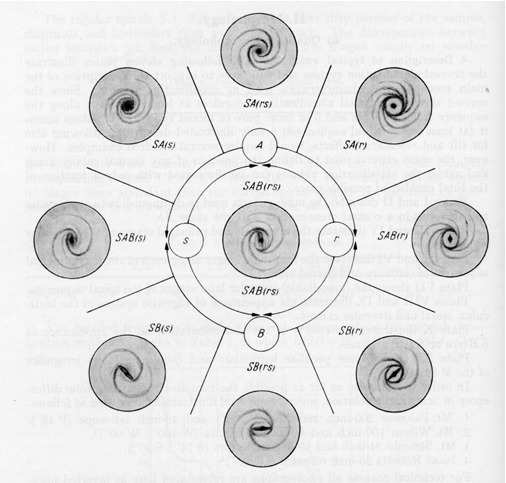}
\includegraphics[width=0.52\textwidth]{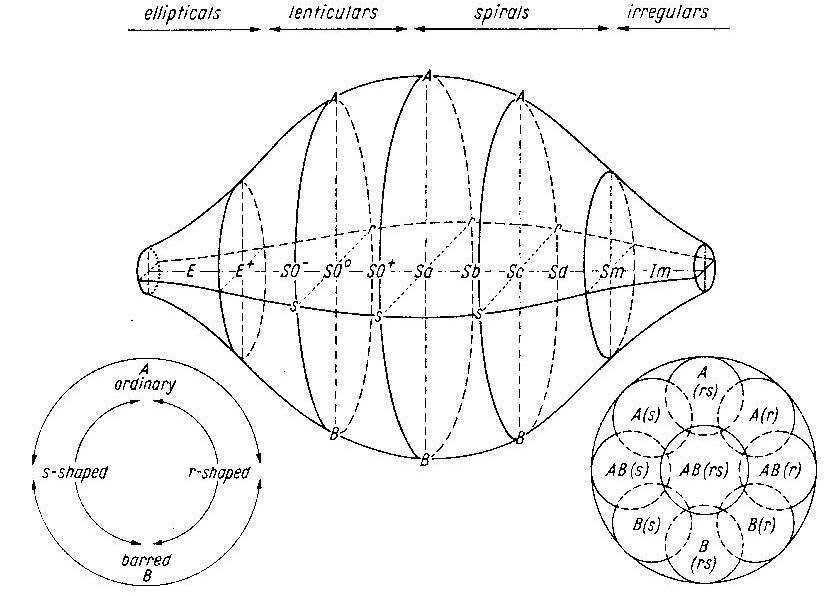}
\caption[de Vaucouleurs morphological scheme (1959)]{This is "The Sequence of Nebular Types" diagram published by de Vaucouleurs in 1959 as part of his article {\it Classification and Morphology of External Galaxies}.}
\label{deVaucouleursdiag}
\end{figure}

\citet{1959HDP....53..275D}\footnote{"Classification and Morphology of External Galaxies", where he also makes a remarkable summary of almost all the galaxy classification history until 1959.} classification keeps the basic division of Hubble one. However, he introduced a more detailed classification for spiral galaxies, since there are at least three more characteristics which deserve more attention: bars, rings, and the spiral arms. It is a consistent scheme which includes all or most of the recent revisions and additions to the standard classification. While the Hubble's diagram is two dimensional, de Vaucouleurs developed a classification that can be seen as three dimensional (see figure \ref{deVaucouleursdiag}). This new "classification volume" is explained in the following.\\

The presence of a bar is noted by a "B", while its absence by an "A". Then, a spiral galaxy is represented by "SB" or "SA" when it has or not a bar in its center, respectively. An "SBA" representation is adopted as intermediate class, when the galaxy has mixed characteristics (weakly barred). Similarly, lenticular galaxies could be "SA0" (unbarred), "SB0" (barred) or SAB0 (intermediate class). However, when it is impossible to tell if the galaxy has a bar or not, it is noted only by an "S" in the case of a spiral galaxy or by "S0" if the galaxy is lenticular. \\

The presence of a ring is distinguished with an "(r)", while its absence with an (s). An intermediate type is noted (rs). Nevertheless, de Vaucouleurs expressed himself that the "(s)" symbol warns, more exactly, about the presence of spirals (an "S-shaped" type). Then, the "distinction between the two families A and B and between the two varieties (r) and (s) is most clearly marked at the transition stage S0/a between the S0 and S classes. It vanishes at the transition stage between E and S0 on the one hand, and at the transition stage between S and I on the other" (see figure \ref{deVaucouleursdiag}).\\

Taking into account the tightness of the spiral arms, de Vaucouleurs extends the Hubble's diagram. After de Vaucouleurs, we can distinguish four stages along each of the four spiral principal sequences SA(r), SA(s), SB(r), SB(s), noted a, b, c, d for "early", "intermediate", "late" and "very late". Intermediate stages are Sab, Sbc, Scd. We can thus  have, for example, a SB(s)c or a SA(r)ab types. For the transition to the magellanic irregulars the notation Sm is used (e.g., the Large Magellanic Cloud is a SB(s)m type). In the case of elliptical or lenticular galaxies, the signs "+" and "-" are used to denote "early" and "late" subdivisions. "In both the SA0 and SB0 sub-classes three stages, noted S0 -, S0 $\deg$, S0 + are thus distinguished; the transition stage between S0 and Sa, noted S0/a by Hubble, may also be noted Sa-". \\

\begin{figure}[!h]
\centering
\includegraphics[width=1.0\textwidth]{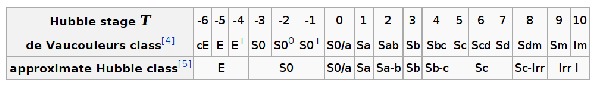}
\caption[Table comparing Hubble's morphological types and de Vaucouleurs's ones]{Table comparing Hubble's morphological types and de Vaucouleurs's ones.}
\label{HubbledeVaucouleurscom}
\end{figure}

\begin{figure}[!h]
\centering
\includegraphics[width=0.323\textwidth]{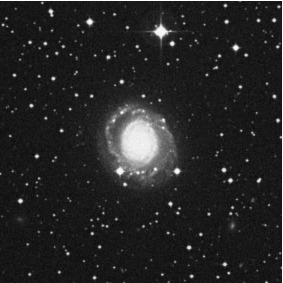}
\includegraphics[width=0.32\textwidth]{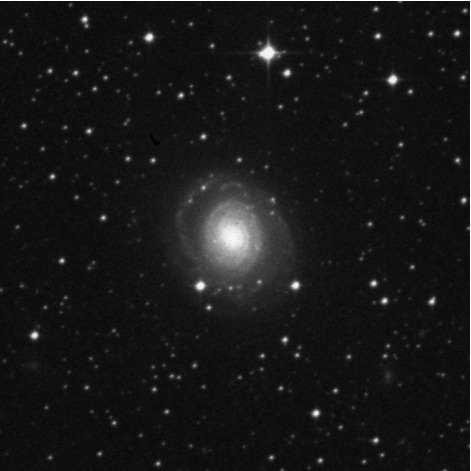}
\includegraphics[width=0.329\textwidth]{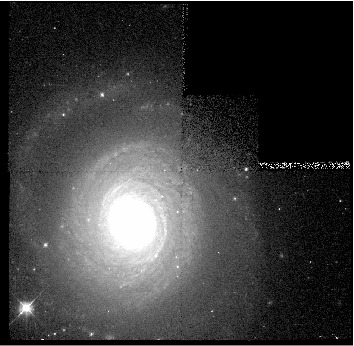}
\caption[Images of NGC 6753 in different bands and with different instruments]{Images of NGC 6753. {\it Left}: band blue ($\lambda$ = 440 nm) from dss. {\it Middle}: band red ($\lambda$ = 640 nm) from dss. {\it Right}: band i ($\lambda$ = 801.2 nm) from HST/WFPC2.}
\label{NGC6753image}
\end{figure}

Finally, irregular galaxies, as the Small Magellanic Cloud, are noted Im. Figure \ref{HubbledeVaucouleurscom} shows a table comparing Hubble's and de Vaucouleurs's classes. In here, as well as in figure \ref{deVaucouleursdiag}, there is no mention of the (R) "type" galaxies (e.g., an (R)SA galaxy). This (R), preceding the symbol of the class, was used by de Vaucouleurs to distinguish those galaxies with an outer ring-like structure which appears in all four sequences near the transition stage S0/a. However, this particularity was consider as not so important along any definite line of evolution. It was more characteristic of a certain stage of evolution. One example of galaxy having an outer and an inner ring could be NGC 6753, which is noted an (R)SA(r)ab galaxy by de Vaucouleurs (see figure \ref{NGC6753image}).

      \subsection{Yerkes's Classification Scheme}

This is a classification of galaxies introduced by W.W. Morgan \citep[][]{1958PASP...70..364M,1959PASP...71..394M}, who worked in the Yerkes Observatory in Wisconsin (University of Chicago). It is based on the central luminosity of galaxies and its spectral features. These allow to derive the stellar populations locked in the inner region of the galaxy, using the MKK system \citep[][]{1943QB881.M6.......,1973ARA&A..11...29M}. At that time, the main idea was to distinguish each galaxy by their stellar composition, which was not taken into account by the Hubble classification alone. Was this new classification really necessary? Let Morgan answer that question himself: "{\it a valid and precise definition of an empirical system of classification can only be in terms of the observed properties of the specimens classified in each category. As the amount of observational evidence increases, the ideas on which the new classification is based are subjected to increasingly critical tests.}"\citep[W.W.][]{1959PASP...71..394M}. In the same article, he pointed out that the final answer will depend on the usefulness it would prove in the next decades.\\

The Morgan's classification was based on the work of \citet{1957PASP...69..291M}. Here, the authors made a classification of a group of galaxies by their spectroscopic categories from A to K (systems A-F-G-K). Galaxies with a strong contribution of A star in their integrated light (galaxies with an earliest spectral type) were called A-systems, galaxies where the principal contributors are F stars are called F-systems, and so on. We can also find mixed systems as AF or FG, for example. They find in this study that most of the galaxies follow an interesting correlation between the spectral classification and the degree of central concentration of light in the system. In other words, irregulars and spirals galaxies with little or no central concentration of light appears to be A or F systems, while spirals with large nucleus light concentration, lenticular and elliptical galaxies have G and K spectral types. Galaxies with a relative intermediate stage of central light concentration are in the F-G categories.\\

Because one of the principal criteria used by Hubble for his morphological classification is the degree of central concentration of light, we can therefore consider the new Morgan classification as a modification of that made by Hubble. However, the new classification makes a remarkably distinction between the different central stellar populations in the different morphological types. "{\it The value of a system of classification depends on its usefulness; the justification of the present attempt must lie in the direction of furnishing information additional to that of the classical Hubble system. It should be emphasized that the present system is put forward for the purpose of giving information additional to that included in Hubble's system}" \citep[W.W.][]{1958PASP...70..364M}.\\

Finally, the Yerkes classification can be detailed as follows. It is described by three parameters. The fundamental one is noted by seven categories which are "a", "af", "f", "fg", "g", "gk", and "k". An "a" galaxy has a contribution to its central luminosity principally due to B-, A-, and F-types stars, while a "k" galaxy has a principal contribution coming from giant K-type stars\footnote{Morgan (1958) clarifies that such contributions are studied in the violet spectral region.}. The second parameter called "form family" is denoted by the initials S, B, E, and I, for Spiral, Barred spiral, Elliptical, and Irregular galaxy, respectively.  Moreover, four additional morphological type were added: Ep systems which are elliptical galaxies with well-marked dust absorption; D galaxies which have a characteristic rotational symmetry but no spiral or elliptical structures, or non-spiral with symmetrically distributed absorption from dust clouds; L systems have low surface brightness; and N galaxies contain a small and bright nucleus superimposed on a considerably fainter background. Intermediate stages for the "form family" are also possible. Finally, we have the purely geometrical inclination parameter or "inclination class", which goes from 1 to 7. A face-on galaxy is noted by 1, and an edge-on galaxy by 7. Zero is not used for this parameter to avoid any ambiguity with class S0 in the Hubble classification. A "p" letter is used in the case of peculiar systems (I, L), where an inclination is difficult to determine.\\

\begin{figure}[!h]
\centering
\includegraphics[width=0.321\textwidth]{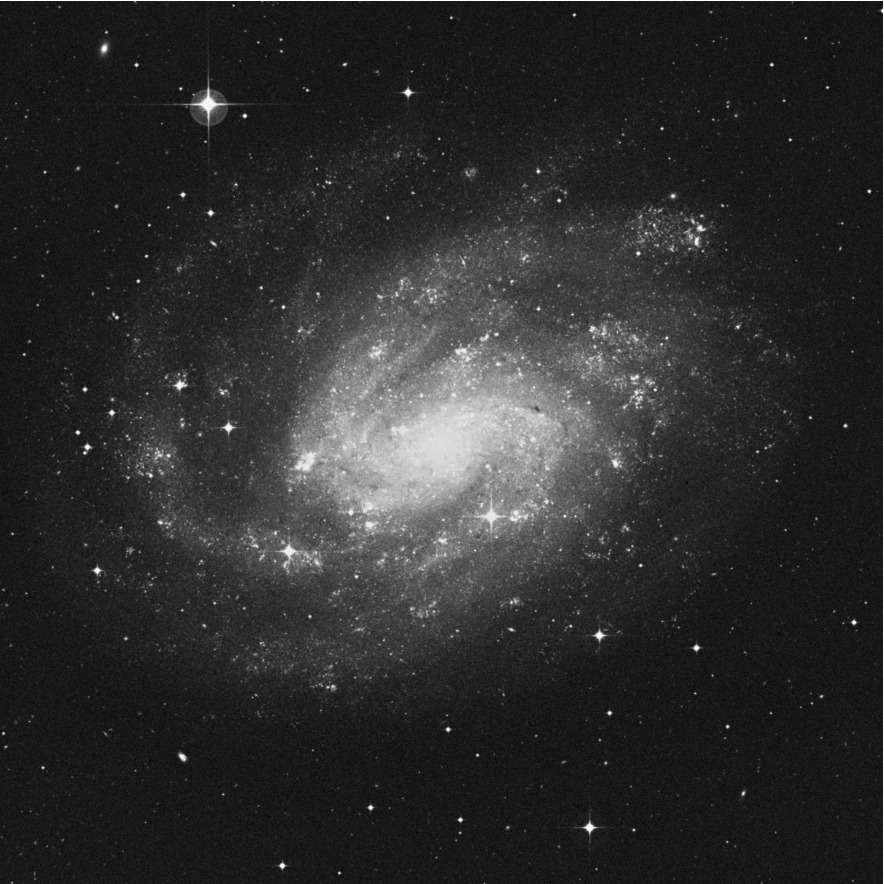}
\includegraphics[width=0.321\textwidth]{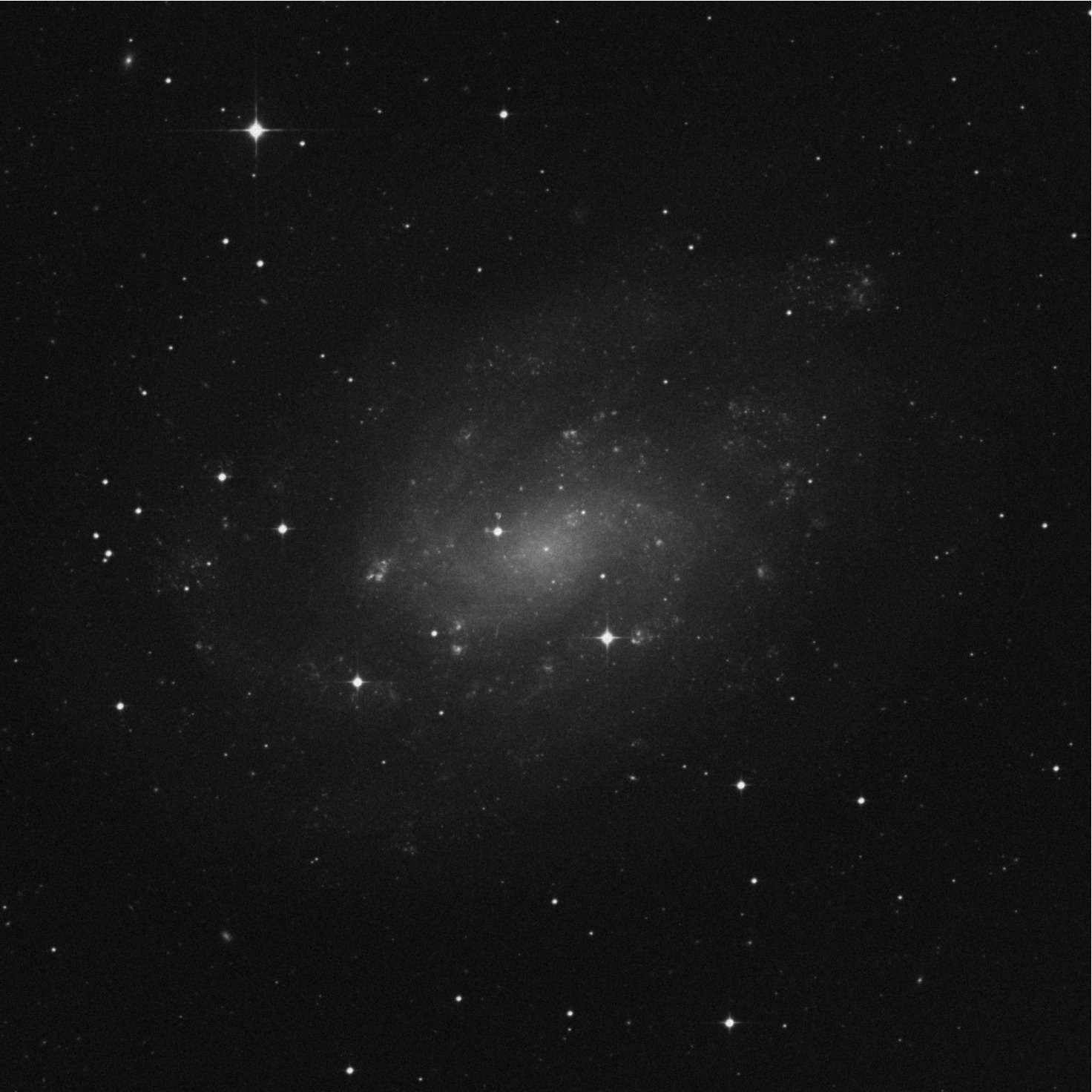}
\includegraphics[width=0.34\textwidth]{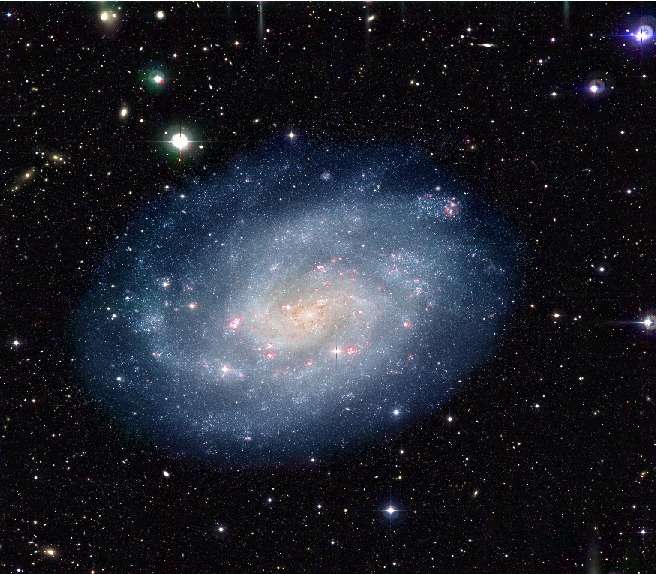}
\caption[Images of NGC 300 in different bands and with different instruments]{Images of NGC 300. {\it Left}: band blue ($\lambda$ = 440 nm) from dss. {\it Middle}: band red ($\lambda$ = 640 nm) from dss. {\it Right}: color image from the Wide-Field Imager (WFI) on the MPG/ESO 2.2-m telescope at the La Silla Observatory (Image Credit: ESO).}
\label{NGC300image}
\end{figure}

\begin{figure}[!h]
\centering
\includegraphics[width=0.32\textwidth]{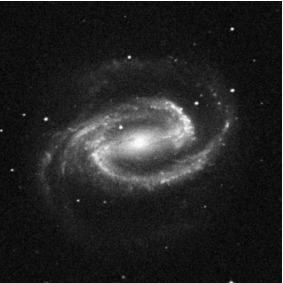}
\includegraphics[width=0.319\textwidth]{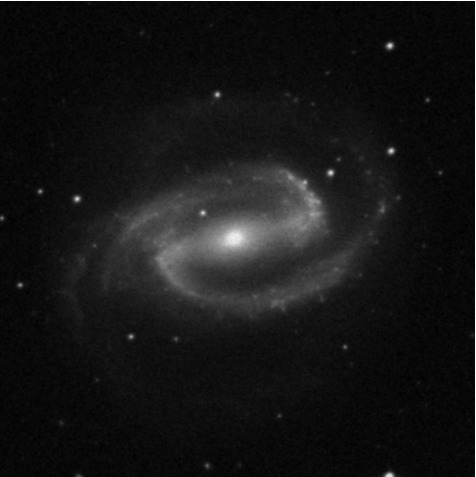}
\includegraphics[width=0.343\textwidth]{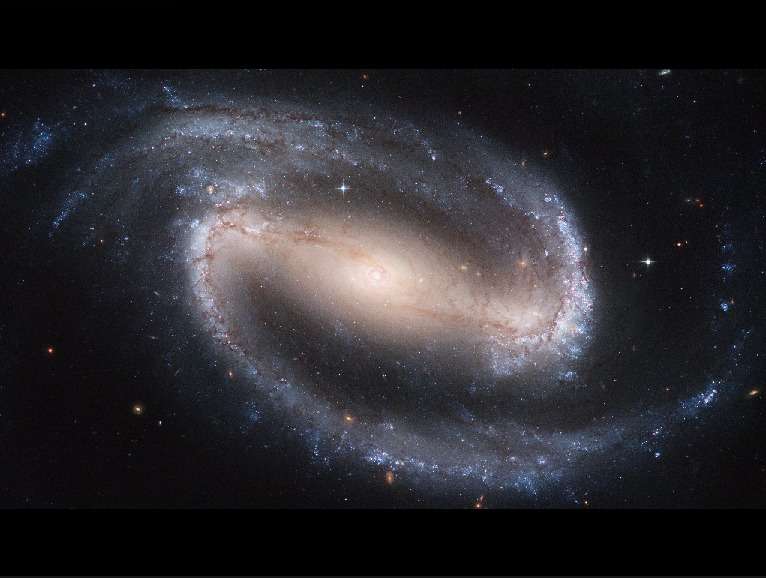}
\caption[Images of NGC 1300 in different bands and with different instruments]{Images of NGC 1300. {\it Left}: band blue ($\lambda$ = 440 nm) from dss. {\it Middle}: band red ($\lambda$ = 640 nm) from dss. {\it Right}: color image from HST/WFPC2 (Image Credit: NASA, ESA, The Hubble Heritage Team - STScI/AURA. Acknowledgment: P. Knezek - WIYN).}
\label{NGC1300image}
\end{figure}

As an example, I can mention galaxies NGC300 and NGC1300. The first was classed as Sc by Hubble, while in the new system is an aS4 (see figure \ref{NGC300image}). The second is a SBb in Hubble's classification, and a fB2 in Morgan's classification (see figure \ref{NGC1300image}). As we can notice, no diagram has been created to show Morgan's classification as it was the case for Hubble and de Vaucouleurs (see figures \ref{firstHubblediag} and \ref{deVaucouleursdiag}). Maybe because the Morgan's classification is a bit more difficult to organize in a sequence?

      \subsection{van den Bergh's Classification Scheme (DDO)}

This is another extension of the Hubble morphological classification developed by Sidney van den Bergh at the David Dunlap Observatory (DD0). In this case, the classification is based on the correlation between absolute magnitude and the degree of development of spiral arms. Thus, it was, since the beginning, employed on spiral and irregular galaxies. \citet{1960ApJ...131..215V} found that galaxies with the highest luminosity are also those with the most strongly developed spiral structures. He then concluded that one limitation of the Hubble system is the existence of luminosity effects on the contrast and development of spiral arms.\\

This finding allowed \citet{1960ApJ...131..215V,1960ApJ...131..558V} to classified galaxies following two criteria: (1) the galactic type, and (2) the luminosity class. The first one follows the Hubble morphological classification, recognizing Sa, Sb, Sc, and Irr types. The second one divides galaxies from I to V by decreasing luminosity. Luminosity classes were chosen to agree with those used to classify stellar luminosities in the Yerkes system. Type I refers to super-giant galaxies, type II to bright giant galaxies, type III to normal giant, type IV to sub-giant galaxies, and type V to dwarf galaxies. Super-giant galaxies could have, approximately, an absolute blue magnitude of -20.5 and dwarf galaxies a blue magnitude of -14.\\

The DDO scheme adds further notations. Sub-giant spirals with low and high arms resolution\footnote{It refers to the degree the arms are resolved.} are noted "S$^{-}$" and "S$^{+}$", respectively. Objects between barred and not barred spirals, corresponding approximately to the SAB type in de Vaucouleurs diagram, are noted "S(B)". When a galaxy has fuzzy or nebulous arms an "n" is added to the galactic type, while an "*" is added if the spiral arms have a patchy structure. The letter "t" at the end of the galactic type indicates distorted spiral arms, which are maybe coming from past or present tidal interactions. In extreme cases, "nn", "**", "tt" could be also used. Finally, the symbol "SD" represents the disk-shaped galaxies. In the DDO classification, the irregulars galaxy type exclude colliding and interacting galaxies, deformed elliptical galaxies, and ellipticals exhibiting dust patches. \\

\begin{figure}[!h]
\centering
\includegraphics[width=0.8\textwidth]{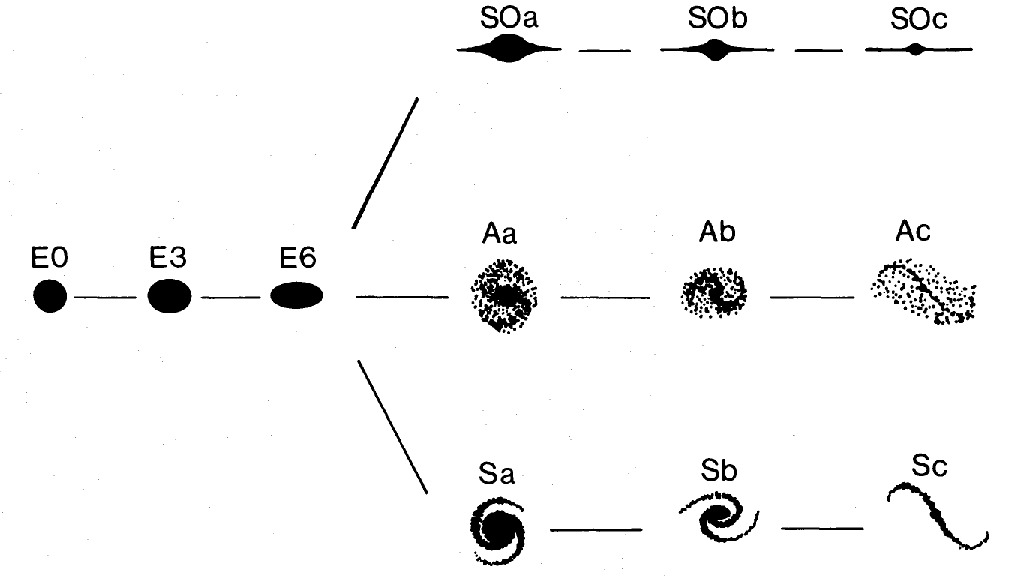}
\caption[van den Bergh scheme of galaxy morphology]{van den Bergh scheme of galaxy morphology from \citet{1976ApJ...206..883V}.}
\label{vandenBerghGalaxyScheme}
\end{figure}

A later revision of the classification led van den Bergh to identify transition cases between S0 and spiral galaxies. These transitions appeared to be spiral galaxies with little star formation in their arms. Such detections and other reasons (e.g., flattening and B/D ratio) made \citet{1976ApJ...206..883V} disagree with the placement of S0 galaxies in the Hubble sequence (see figure \ref{firstHubblediag}). For van den Bergh, S0 galaxies must be placed in a sequence parallel to spirals, instead of being in the "transition region" between ellipticals and spirals (see figure \ref{vandenBerghGalaxyScheme}). It was called the RDDO (Revised DDO) system\footnote{The separation of S0 and Sp galaxies by van den Bergh is based on the B/D ratio as a classification criterion. Further studies by \citet{1986ApJ...302..564S} on the bulge/disk decomposition and correlation between B/T ratio and Hubble type found that B/T of S0 galaxies is generally intermediate between pure spheroidal systems and spirals, supporting the placement of S0 galaxies between ellipticals and spirals.}.

      \subsection{Discussion about the physical Hubble sequence}
      \label{PhyHubbleSeq}

In the previous section, I have cited the most known morphological classification systems. Noteworthy, other morpho-classifications have been also developed. Amongst them we have: a.) the Vorontsov-Velyaminov classification, which is a purely descriptive (or MCG) system \citep{1975gaun.book.....S}; b.) the arm classes system of \citet{1982MNRAS.201.1021E,1987ApJ...314....3E}; c.) the \citet{1996ApJ...464L.119K} revised system of Hubble classification for elliptical galaxies, based on their inclination and structures rather than any intrinsic property; and d.) \citet{2006MNRAS.373.1389C} system, taking into account the star formation, galaxy interaction and mass. \\

All these systems make reference to the different "species" of galaxies we can find using a morphological classification. However, one might wonder whether this is even important for understanding galaxy evolution. First, only with the intention of starting the discussion, I will dare to make a light analogy (if there could be any) with paleontology. Even if living beings could have many intrinsic individual characteristics, their shapes and structures have been an essential, and maybe the most important feature to trace the evolution of organisms on earth (e.g., see "On the Origin of Species" by Charles Darwin\footnote{\citet{Darwin...Charles...OTOOS}}).\\

As a result, we are essentially in agreement with \citet{1994.2729...deVau} who says that "taxonomy ... is nevertheless an essential first step in all fields of science". Now, we could also say that galaxy taxonomy through the morphological sequence attests to an underlaying sequence of numerous physical processes (see table \ref{GalProperVSmorpho}). As we have seen, this notion has been widely accepted for the past decades, making morphological classification of large numbers of galaxies important for better modeling and understanding the galaxy structure and evolution. \\

\begin{table}[!h]
\begin{center}
\begin{tabular}{|c|c|}
\hline
Parameter/property & from early to late type  \\
\hline
V/$\sigma$ & $\nearrow$ \\ 
\hline
R$_{disk}$ & $\searrow$ \\ 
\hline
B/T & $\searrow$ \\
\hline
(B-V) color & $\searrow$  \\
\hline
HI abundance & $\nearrow$  \\
\hline
HII abundance & $\nearrow$ \\
\hline
H$_{2}$ abundance & $\searrow$ \\
\hline
Metal abundance & $\searrow$ \\
\hline
K/A$^{*}$ & $\searrow$ \\
\hline
Luminosity & $\searrow$ \\
\hline
Dynamical mass & $\searrow$ \\
\hline
Baryonic mass & $\searrow$ \\ 
\hline

\end{tabular}
\caption[Galaxy properties vs morphology]{Galaxy properties vs morphology. Relation between different physical parameters/properties of galaxies and their morphology (from elliptical/lenticular to spiral/irregular galaxies, see figures \ref{firstHubblediag} and \ref{deVaucouleursdiag}). These relations have been studied amongst the local galaxies by \citet{1994ESOC...49..197R,1994ARA&A..32..115R}, \citet{2004AJ....127.2511N}, \citet{2003MNRAS.343..978S}, \citet{2005ApJ...621..215S}, \citet{2003MNRAS.341...33K}, \citet{2001AJ....122.1861S}, \citet{2001AJ....122.1238S}, \citet{1992ApJ...388..310K}, \citet{1985ApJS...59..115K}, \citet{1981MNRAS.194...63T}, \citet{1969AJ.....74..859R}, \citet{2004A&A...424..447P}, \citet{2001Ap&SS.276..509V}, \citet{2006MNRAS.373.1389C}. Nevertheless, due to the considerable scatter in some of the relations, these trends may not be evidence of causal connection, until new results provide better understanding of both trends and scatter. One example is the integral galaxy color, which scatter is still under discussion \citep[see section 6 in][]{2007MNRAS.380..585C}. Another consideration to be taken into account is the existence of exceptions to generalities, as it is the case for the R$_{Disk}$ or the total mass. The general trend of the relation between these two properties with the morphology is affected by the existence of elliptical dwarf galaxies, for example. (* K/A represents the ratio between the {\it old stellar population (K) fraction and the young stellar population (A) fraction)}.}
\label{GalProperVSmorpho}
\end{center}
\end{table}

We can say that galaxy morphology is useful because it does not only succeed, to some extent, in distinguishing galaxies which are physically different. The basic classification scheme (Hubble scheme) succeeds, although with some caveats\footnote{even if properties are generally well correlated with the morphological types, in some cases the considerable scatter must be carefully considered (see legend of table \ref{GalProperVSmorpho}). Somehow such scatter should be pondered when using them to morphologically classify galaxies.}, in separating galaxies according to their physical and kinematic properties. Furthermore, galaxy morphology is a direct consequence of the underlying physics governing their formation and evolution. \\

After establishing the main importance of galaxy morphology, we are now confronted with a new paradigm: how many morphological details do we need to include in a morphological classification? As we saw in the above sections, each morphological classification took the basis of the Hubble sequence. However, each one takes, more or less, different physical details to achieve the classification of galaxies because "there are several possible approaches to the problem of the classification of galaxies: morphologic, photometric, colorimetric, spectroscopic" \citep[][]{1963ApJS....8...31D}\footnote{even if \citet{1994.2729...deVau}, referring to the "low" number of morphological features taking into account in the original Hubble sequence, argued that "this would be erudition, but not science which aims at the general, not the particular", sometimes science can be lost in so many details.}. \\

Details are important in the description of the evolution of individual systems. However, how much importance do they deserve when trying to unravel the evolution of galaxies in the Universe? It depends on the scale at which we are looking at, as well as the interrelation that could have the different scales. On the one hand, we have the morphological/physical details to be taken into account for the morphological classification. On the other hand, we have the physical/dynamical properties that correlate with such morphologies. Where is the limit between the two? A key point will then be the definition of the methodology to be followed during the morphological classification (see chapters 3 and 6). This methodology, in the ideal case, must be easily reproducible and not subjective, must take into account a reasonable number of principal morphological/physical details, and must result in the best possible correlation with the main galaxy properties (for example, the kinematics, the important role of which to understand the evolution of galaxies is well established). "{\it As Long as only a few criteria define a system, and if image material of a similar quality to that which formed the basis of the system is used, then there will be a greater ease of applicability and reproducibility of that system by independent observers. If one later finds correlations between fundamental observables and classifications, then the system could lead to physical insight ...}" \citet{1992ASSL..178....1B}. Finally, a good classification must drive morphology to the physical/dynamical properties (e.g., table \ref{GalProperVSmorpho}), but such properties must not be used to drive (make) the classification\footnote{this is a modification to words by \citet{2005ARA&A..43..581S}.} (see subsections \ref{ColordeterminationMyPaper} and \ref{ColorDistributionFromMyPaper} for an example).  \\

Since the past century, morphological classification has had great success in helping us to understand local galaxies. One of the objectives of my thesis is to extend this successful analysis technique into the distant Universe, and look for a possible time evolution of the Hubble sequence. \\


\chapter{The galaxy formation and evolution}
\minitoc

{\textbf{\itshape Scenario of galaxy formation and evolution}}\\

How can we obtain the different shapes of galaxies we see in the local Universe? How do galaxies get different components with different dynamics? Why some galaxies appear flattened like disks while others are spheroidal? These questions remains a major unsolved problem in galaxy formation. \\

Some attempts have been made in the past and continues in the present. The principal idea presents this problem as a 'nature' versus 'nurture' dichotomy: is the morphology of a galaxy imprinted in the initial conditions, or does it result from environmental processes such as mergers, tidal interactions, or gas infall? On the one hand, it has been proposed that the determining factor of the galaxy formation process is the protogalactic angular momentum \citep[][]{1970ApJ...160..831S}. On the other hand, different mechanisms that could transform one morphology type into another have been studied. Amongst them we can mention ram-pressure stripping by the hot intracluster medium \citep[][]{1972ApJ...176....1G}, or galaxy 'harassment' by impulsive encounters in clusters \citep[][]{1996Natur.379..613M}. Furthermore, other studies \citep[][and references therein]{2009A&A...507.1313H} support the idea that galaxy mergers are a possible explanation for the origin of the Hubble types.\\

In the following, I describe the principal scenarios of galaxy formation.\\

  \section{Downsizing: a scenario of galaxy formation or a natural consequence of the primordial collapse model?}
  \label{DownsizingarevivaloftheprimorColl}

Downsizing was introduced by \citet{1996AJ....112..839C}. Studying the redshift, mass (through M$_{K}$) and star formation of 280 galaxies since redshift 1.6, they concluded that the maximal mass of galaxies having a burst of star formation increases with redshift, and that most of the galaxy formation process took place between redshift 0.8 and 1.6. This led them to suggest that the galaxy mass assembly occurred in a 'downsizing' way (referring to a top-down universe evolution) with the most massive galaxies forming at higher redshift, while less massive ones formed at later cosmic times. \\

According to this downsizing effect, massive galaxies are thus expected to exist at high redshifts ($z >> 1$). Observations show that it is the case \citep{1988ApJ...331L..77E,1999ApJ...519..610D,2000ApJ...531..624D,2002ApJ...578L..19I,2004Natur.430..184C,2005ApJ...626..680D,2005MNRAS.357L..40S,2006ApJ...649L..71K,2008A&A...482...21C}. Those large systems have being detected by their red colors or their bright emission at sub-millimeter wavelengths. Some of them already have old stellar populations, some with significant recent star formation.\\ 

The downsizing picture has been widely debated because of the apparent contradiction between downsizing and the hierarchical evolution expected from the $\Lambda$CDM cosmological model (see chapter 1). Nevertheless, this contradiction is only apparent. While the downsizing picture describes only the evolution of baryonic structures, the $\Lambda$CDM model focuses on the dark matter halos evolution. Moreover, the interactions between baryons and dark matter remains largely uncertain. The hierarchical evolution of dark matter halos does not imply, "a priori", a hierarchical evolution of baryonic structures.\\

Actually, the first fluctuations of baryonic mass to be condensed are those inside the most massive dark matter halos, because these last are the catalysts of the formation of the first baryonic structures \citep{2006MNRAS.372..933N,2006A&A...459..371M}. It is in these halos where the stellar formation activity would first begin. Consequently, the apparent paradox between downsizing and $\Lambda$CDM model is mostly due to a confusion in the literature between the evolution of dark matter halos (hierarchical) and the evolution of baryonic structures (anti-hierarchical).\\

\begin{figure}[!h]
\centering
\includegraphics[width=0.8\textwidth]{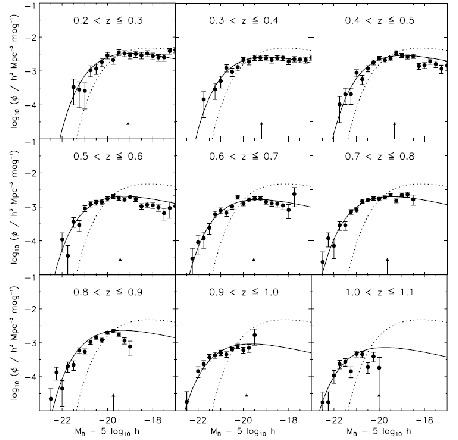}
\caption[Evolution of the ETGs luminosity function with redshift]{Redshift evolution of the red-sequence luminosity function, in straight line, from \citet{2004ApJ...608..752B}. Here, we notice the increasing faint end from z$\sim$1 to z$\sim$0.25. For comparison they plot, in dotted line, the LF of a local sample they select from the SDSS EDR \citep[Early Data Release,][]{2002AJ....123..485S}.}
\label{Belletal2004aLFetg}
\end{figure}

After the finding of the downsizing effect by Cowie et al., other works seemed to point in the same direction. One example is the finding of lack of density evolution of large elliptical and spiral galaxies between redshift 1.3 and 0 \citep{1998ApJ...500...75L,1999ApJ...525...31S}. Moreover, \citet{2000ApJ...536L..77B}, using a sample of 321 galaxies morphologically classified (visually and with automated classifiers), showed how the stellar mass density changes by morphological type as a function of redshift. They found that spiral galaxies do not present any change in their stellar mass density between redshift 1 and 0. However, that of irregular galaxies has a remarkable decline, while that of elliptical galaxies present a modest increase. Later, \citet{2005ApJ...625..621B} found that spiral and irregular morphologies dominate among galaxies of low mass in all redshifts, while the E/S0 dominate among galaxies of larger mass range. They then suggest a morphological extension of the downsizing principle: galaxies that form at high redshift are the most massive ones and have elliptical morphology. They also found that the fraction of massive galaxies at lower redshifts are dominated by elliptical types, which could suggest an evolution to this morphological type. Another important fact that supports the downsizing effect is the extrapolated evolution of the faint end of the red sequence luminosity function (LF) since z$\sim$1 \citep[see figures 6 and figure 3 in][respectively; see also figure \ref{Belletal2004aLFetg}] {2007MNRAS.380..585C,2006ApJ...651..120B,2004ApJ...608..752B}, which supposes that the population of dwarf ellipticals (dE) galaxies increases from the past to the present. \\

\begin{figure}[!h]
\centering
\includegraphics[width=1.0\textwidth]{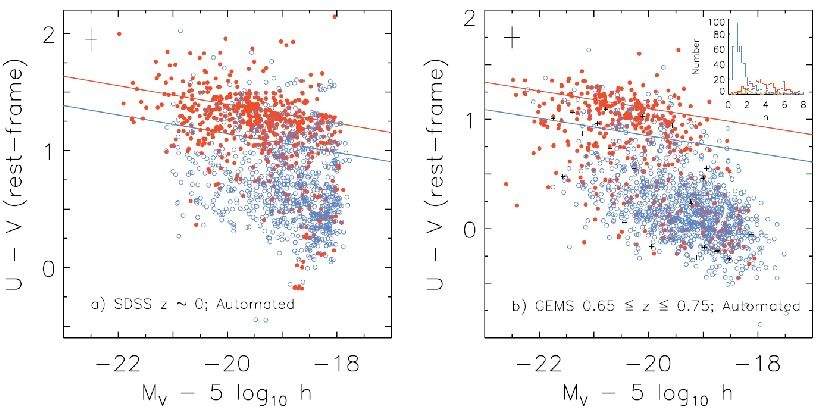}
\caption[Morphological types of galaxies as a function of their rest-frame V-band absolute magnitude and U-V color]{Color contamination from \citet{2004ApJ...600L..11B}. This figure shows the morphological types of galaxies as a function of their rest-frame V-band absolute magnitude and U-V color over the last half of the cosmic history. Blue open circles denote morphologically classified late-type galaxies, whereas red solid symbols show early-types. The red line shows a fit to the red sequence and the blue line the adopted cut between red and blue galaxies. As we can easily notice just by looking to the fainter galaxies in both plots, the contamination of late-type galaxies into the red-sequence for local fainter galaxies (left plot) is higher than those at intermediate redshift (right plot). Could it be the cause of the evolution inferred from figure \ref{Belletal2004aLFetg}?}
\label{Belletal2004bColorCont}
\end{figure}

These observational results have led to the downsizing effect being considered as a real scenario of galaxy formation\footnote{by a dissipative collapse of baryons in dark matter halos.}, predicting that the galaxy population which would be formed in the present is mostly composed of dE types. However, such a scenario would present some problems in a cosmological frame. First, the formation of small galaxies (principally, dE galaxies) at present epochs, and a passive evolution of intermediate mass spiral and elliptical galaxies since z$\sim$1, can not explain the high stellar mass fraction ($\sim 50\%$) formed since z$\sim$1 \citep{2003ApJ...587...25D}, and could be in contradiction with the fact that most of the present-day stars are in structures more massive than dE galaxies \citep[elliptical and bulge of spiral galaxies;][]{1998ApJ...503..518F,2000ApJ...536L..77B}. It is also in contradiction with the fact that the bulk of the star formation density at intermediate redshift is mainly in intermediate mass galaxies \citep{2004Natur.428..625H,2005A&A...430..115H,2005ApJ...625...23B}, and with the metal evolution found since z$\sim$1 \citep{2006A&A...447..113L}. In addition the application of the downsizing scenario seems to be in disagreement with the increasing frequency of mergers with redshift \citep{2000MNRAS.311..565L,2003AJ....126.1183C,2004ApJ...601L.123B,2006ApJ...636..592L,2008ApJ...681.1089R}, as mergers form bigger structures rather than smaller ones. Finally, we could go even further, and argue that the assumption of the evolution of the faint end of the red sequence LF are based on the hypothesis that such a LF corresponds to the LF for early-type galaxies (ETGs). Making the hypothesis that all red color galaxies are ETGs could be too rough, as it is well known, even for local galaxies, that there is a non-negligible "contamination" of late-type galaxies amongst red color galaxies: \citet{2001AJ....122.1861S} found that $\sim$20 per cent of late-type galaxies have red colors compatible with the red sequence in the local Universe \citep[see also][and figure \ref{Belletal2004bColorCont}]{2004ApJ...600L..11B}. This contamination becomes even stronger with redshift \citep[$\sim$35 per cent at z$\sim$1;][]{2007A&A...465..711F}\citep[see also][and figure \ref{Belletal2004bColorCont} for z < 1]{2004ApJ...600L..11B}. This could be related to the discussion, about the correlation between morphology and different physical galaxy properties, started in subsection \ref{PhyHubbleSeq}, and will be further detailed in subsection \ref{ColordeterminationMyPaper}.\\

Now, we may distinguish between the {\it downsizing effect}, which is a real tendency, and the {\it downsizing scenario}, which disagrees with a number of observations. \\

  \section{Secular evolution}
  \label{Secular_evolutionScenaSubsec}

In the 1960s-70s, observations of the Milky Way (MW) suggested a scenario in which spiral disks acquired their angular momentum at early epochs \citep[][and references therein]{1969ApJ...155..393P}, and then evolved in a slow, "secular" manner by gradual accretion of material from their environment. In this secular evolution process, galaxies are hence considered as open systems which can get, by accretion, the gas coming from the intergalactic medium. This gas is then transformed into stars through time, while the various galaxy components interact with each other. For example, the bulge is formed slowly from the disk (and external gas) through the bar action \citep{2005MNRAS.358.1477A,2004ARA&A..42..603K,1992MNRAS.259..345A,1990A&A...233...82C}. The total accretion process happens in a nonviolent way. The redistribution of energy and angular momentum inside the system could be also driven by nonaxisymetries other than a bar. These are the spiral arms, which can also cause an internal, slow secular evolution rearranging the disk structure.\\

\begin{figure}[!h]
\centering
\includegraphics[width=0.8\textwidth]{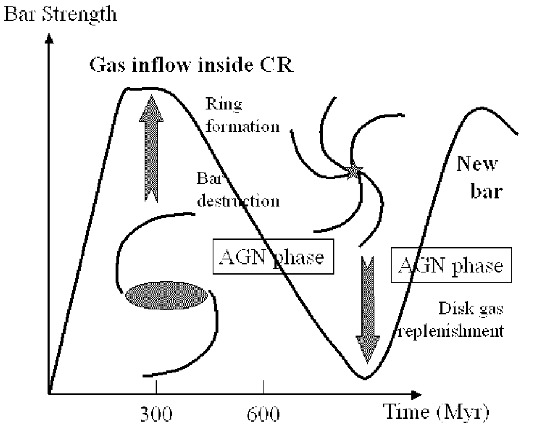}
\caption[Schematic scenario of two bar episodes, and the corresponding phases for the AGN fueling]{Schematic scenario of two bar episodes, and the corresponding phases for the AGN fueling \citep{2005astro.ph..6265C}. It shows the relatively short lifetime of the bar.}
\label{Short_bar_lifetimeCombes2005}
\end{figure}

This is useful to distinguish between internal secular evolution and environmental secular evolution. The first is responsible for the formation of pseudobulges in disk galaxies \citep{2004ARA&A..42..603K} and could explain the abundance of pure disk galaxies with no evidence for merger-built bulges in the local Universe \citep{2008ASPC..396..297K}. The second is responsible for the formation of spheroidal galaxies whose observable parameters are similar to those of late-type galaxies, even if they are morphologically similar to ellipticals. These authors suggest that spheroidals are defunct late-type galaxies transformed by internal processes such as supernova-driven gas ejection and environmental processes such as secular harassment and ram-pressure stripping.\\

\begin{figure}[!h]
\centering
\includegraphics[width=0.8\textwidth]{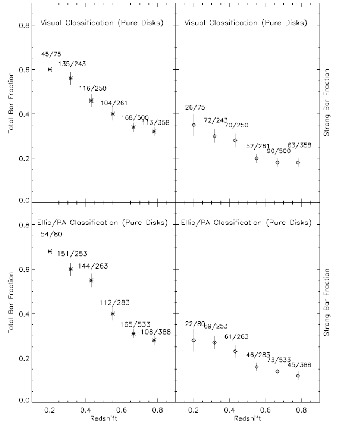}
\caption[Bar fraction evolution with redshift]{Bar fraction evolution with redshift from \citet{2008ApJ...675.1141S}.}
\label{Bar_fraction_evolutionSheth2008}
\end{figure}

One piece of evidence in support of secular evolution is showed by \citet{2001AJ....122.1298G}. They studied the color gradients, and the correlation between bulge and disk colors, in 257 late-type barred galaxies, reaching the conclusion that their results could be only explained if the bulge of those barred objects are formed from a secular evolution. \citet{2005MNRAS.364L..18B} \citep[and][]{2002A&A...392...83B} also showed that if, during the formation of the bulge, the central concentration of matter exceeds a few percents of the total mass, it could even destroy the bar, which would have a lifetime of a few (1-2) Gyr (see figure \ref{Short_bar_lifetimeCombes2005}). Given the roughly constant fraction of barred galaxies from the local Universe to the distant one \citep{2005A&A...435..507Z,2004ApJ...615L.105J,2003ApJ...592L..13S}, it would imply that bars are continuously reconstructed. Such periodic reconstructions could only be possible through gas infall \citep{2005astro.ph..6265C,2002A&A...394L..35B} coming from filaments \citep[][see also figures \ref{CDM_models}, \ref{CDM_models_scales0}, \ref{Millenium_simulation}]{2005A&A...441...55S}. Nevertheless, \citet{2005MNRAS.363..496A} propose that bars are not sporadic events, but instead their lifetime could be larger than 5 Gyr, even with an increase of the galaxy central mass of a few percents of the total mass. This is supported by recent studies that show, indeed, a quite important evolution of the fraction of barred galaxies since z$\sim$0.8 \citep[][see figure \ref{Bar_fraction_evolutionSheth2008}]{2008ApJ...675.1141S,2008ASPC..396..333A}.\\

The secular evolution scenario is challenged by observations of distant galaxies, which emitted their light 6 Gyr ago. They reveal that nearly half of them had peculiar morphologies and anomalous kinematics \citep{2002PASP..114..797V,2006A&A...455..107F}. In other words, they are not yet dynamically relaxed. This is at odds with the hypothesis of secular evolution of most of spiral galaxies during the past 6 Gyr (for further drawbacks of the secular scenario see also subsection \ref{DiscussionScenariosGalForm}). Additionally, contrary to the early predictions that a secular evolution leads to a transformation of late-types to early ones, \citet{2005ApJ...632..217S} studied the gas inflow induced by the bar in spiral galaxies, and concluded that "{\it despite the evidence for bar-driven inflows in both early and late Hubble-type spirals, the data indicates that it is highly unlikely for a late-type galaxy to evolve into an early type via bar-induced gas inflow. Nonetheless, secular evolutionary processes are undoubtedly present, and pseudobulges are inevitable; evidence for pseudobulges is likely to be clearest in early-type galaxies because of their high gas inflow rates and higher star formation activity.}"\\

Finally, an important issue is the origin of the gas accreted during the secular evolution process. \citet{1992ApJ...401L..79B} showed that in some S0 galaxies the gas rotates in the opposite direction of stars, which led them to predict an external origin of the gas. Possible origins are filaments (see figure \ref{CDM_models}) or a high occurrence of minor mergers during Hubble time. However, could it be just the infall of gas due to major merger systems? In any case, identifying from where the gas was lying before being transformed into present day stars, as well as determining its momentum, could solve the issue of the high angular momentum observed in present-day disks which can not be reproduced by a large number of simulations. It has been called "the angular momentum catastrophe".\\

{\textbf{\itshape Minor mergers}}\\

The infall of low mass satellite galaxies has been consider as an extension of the secular evolution scenario in order to solve its disadvantages. Indeed, the first numerical simulations showed that minor mergers (generally having progenitors with a mass ratio smaller than 1:5) do not destroy the disk of the largest progenitor, but this last undergoes a morphological evolution to an earliest type by the increase of the velocity dispersion and height of the disk \citep{1996ApJ...460..121W,1999MNRAS.304..254V}. Later, \citet{2003ApJ...597...21A} showed that the thick disk obtained in their simulations was mainly composed of remains of satellite galaxies, while the thin disk was composed of stars formed "in situ" after the last merger. In addition, \citet{2003MNRAS.341..343B} showed that the interactions between a satellite galaxy with a barred galaxy could drive to the formation of structural features such as rings, as well as a faster displacement and destruction of the bar. Depending on the impact radius, this kind of interactions can produce a thickening of the disk. \\

One of the first galaxy evolution scenarios involving mergers is the "collisional starburst scenario" (CSS), proposed by \citet{2001MNRAS.320..504S}, where the galaxies evolve principally through minor merger interactions. These are supposed to happen more frequently than major mergers. This scenario explains very well the observed properties of LBGs (Lyman Break Galaxies): color, velocity dispersion, luminosity functions between redshift 3 and 4, for example. However, when looking at lower redshifts, the CSS gets in trouble. It underestimates the number of EROs (Extremely Red Objects) at z=1, as well as the number of objects in the near-IR (K<22) by a factor of 3 at z>1.7, and a factor 10 at z>2. Minor mergers can not explain the strong morphological and kinematic evolution of galaxies at z<1 either, as mentioned above. This is because minor mergers should have a much lower efficiency to distort galaxy morphologies and kinematics \citep{2009ApJ...691.1168H}. Additionally, minor mergers present moderate (and even negligible) star formation rates \citep{2008MNRAS.384..386C}, which are not consistent with the observed high values of the cosmic star formation rate density \citep{2005ApJ...632..169L,1999ApJ...517..148F}. Another major difficulty that can be found as well is the prediction of disks with too low angular momentum compared to present-day spirals. \\

\begin{figure}[!h]
\centering
\includegraphics[width=1.0\textwidth]{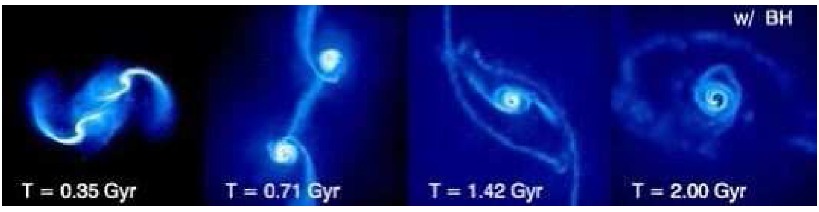}
\caption[Major merger simulation]{Major merger simulation from \citet{2006ApJ...645..986R}.}
\label{MajorMergerSimulationRob2005}
\end{figure}

\section{Violent encounters: major mergers}

Contrary to minor mergers, major mergers can destroy the progenitor disks and may induce high SFRs \citep{2008MNRAS.384..386C,2006MNRAS.373.1013C,2005A&A...433L..17H}. Furthermore, they could potentially solve the angular momentum catastrophe\footnote{if one accounts for the additional effect of the orbital angular momentum provided by major mergers.} \citep{2002MNRAS.329..423M}. Up to now, the majority of the major merger simulations that have been carried out lead to the formation of elliptical galaxies. While mergers of stellar disks with mass ratios between 3:1 and 4:1 produce elliptical-discy galaxies with rapid rotation, mergers between 1:1 and 1:2 lead to the formation of elliptical-boxy galaxies with a very low rotation \citep{2003ApJ...597..893N}. Later studies have also shown that the final elliptical galaxy type (discy or boxy) depends on the progenitors morphology, which allows the reproduction of the different populations of elliptical galaxies that have been observed in the local Universe \citep{2005MNRAS.359.1379K}. What about the large percentage of late-type galaxies in the local Universe? It could be just a matter of  gas fraction \citep{2009ApJ...691.1168H,2009MNRAS.397..802H,2005ApJ...622L...9S,2006ApJ...645..986R,2005MNRAS.361..776S,2005Natur.433..604D,2006ApJ...637..255J}. These authors are challenging the conventional origin of disk galaxies and show that many spirals could be indeed formed after major merger events. Due to the feedback by supernova and AGN, a significant fraction of gas could be expelled towards the galaxy outskirts, allowing a gas reservoir for the creation of a new disk (see figure \ref{MajorMergerSimulationRob2005}). These simulations also suggest a solution to the angular momentum problem. \citet{2006ApJ...645..986R} demonstrates that if large feedback was occurring, this may lead to rejuvenated disks with high angular momenta. Then, because of feedback, it is likely that the whole merger process can be observed during more than 1 or 2 Gyr \citep[][see also figure \ref{MajorMergerSimulationRob2005}]{2009A&A...507.1313H}. \\

Important progress is needed to disentangle the impact of the different physical mechanisms in the formation of bulges and disks. It seems that the numerical simulations may not be good enough to cover the whole problem. However, if the above hypothesis turns out to be true, the origin of the majority of the present Hubble sequence galaxies could thus be explain by a scenario in which the galaxy evolution is mainly driven by a violent process (major mergers).  \\

  \subsection{An observational scenario: "The Spiral Rebuilding"}
  \label{TheSpiralRebuildingScenarii}

This scenario has been proposed by \citet{2005A&A...430..115H}, and is further detailed by \citet{2009A&A...507.1313H} (see section \ref{Hammer_etal_2009Section}). It can be summarized by 3 major steps (see below for more details): a "pre-merger phase" during which two distant spirals merge, the "LCG phase" where all material from the progenitors falls into the mass barycenter of the system and forms a bulge, and the "disk growing phase" where subsequently accreted material forms a rotating disk. The rebuilding disk scenario suggests a merger origin of the whole Hubble sequence (from ellipticals to late type spirals). It was initially proposed to explain a number of observational pieces of evidence \citep[see table 4 in][]{2005A&A...430..115H}, such as the evolution of the star formation density in the UV and IR, the evolution of the central color and morphology \citep{2004A&A...421..847Z}, as well as the estimated merger rate \citep{2000MNRAS.311..565L}. Moreover, this scenario is in agreement with further observations at z<1 (e.g., the mass and metal evolution of galaxies \citep{2006A&A...447..113L}, see also chapters 5 and 6). \\

\citet{2005A&A...430..115H} used multiwavelength observations of 195, z > 0.4 intermediate mass galaxies (M$_{B} < -20$), and found that 15$\%$ of them are LIRGs (Luminous Infra-Red Galaxies). These galaxies are characterized by a high star formation rate, which could explain by themselves most of the stellar mass formed since z=1. Hammer et al. also showed that in such LIRGs the star formation must be produced in consecutive episodes, because if it is not the case they would evolve into local massive galaxies. This would lead to an over-population of massive elliptical galaxies. Nevertheless, observations show that the number fraction of elliptical galaxies is quite constant since z$\sim$1. Taking these main results as a basis, Hammer et al. then describe all the above evolution trends as due to a recent merger origin of 50-75 $\%$ of spirals. Therefore, a majority of intermediate mass spirals could have experienced their last major merger event (leading to the disruption of the disk) within the last 8 Gyr. \\

The merger scenario is also supported by the merger rate found to be about 10 times more at z $\sim$ 1 than it is today \citep{2000MNRAS.311..565L}. In this theory, the disk angular momentum mostly results from the orbital angular momentum of the last major collision \citep{2007A&A...466...83P,2007ApJ...662..322H}, and, as mentioned in the previous subsection, an increasing number of authors \citep[e.g.,][]{2009ApJ...691.1168H,2006ApJ...645..986R} have successfully tested disk survival after a merger. If the gas fraction is sufficient (about 50$\%$), they predict that the re-formed disk can be the dominant component in the reshaped galaxy. \\

\begin{figure}[!h]
\centering
\includegraphics[width=0.8\textwidth]{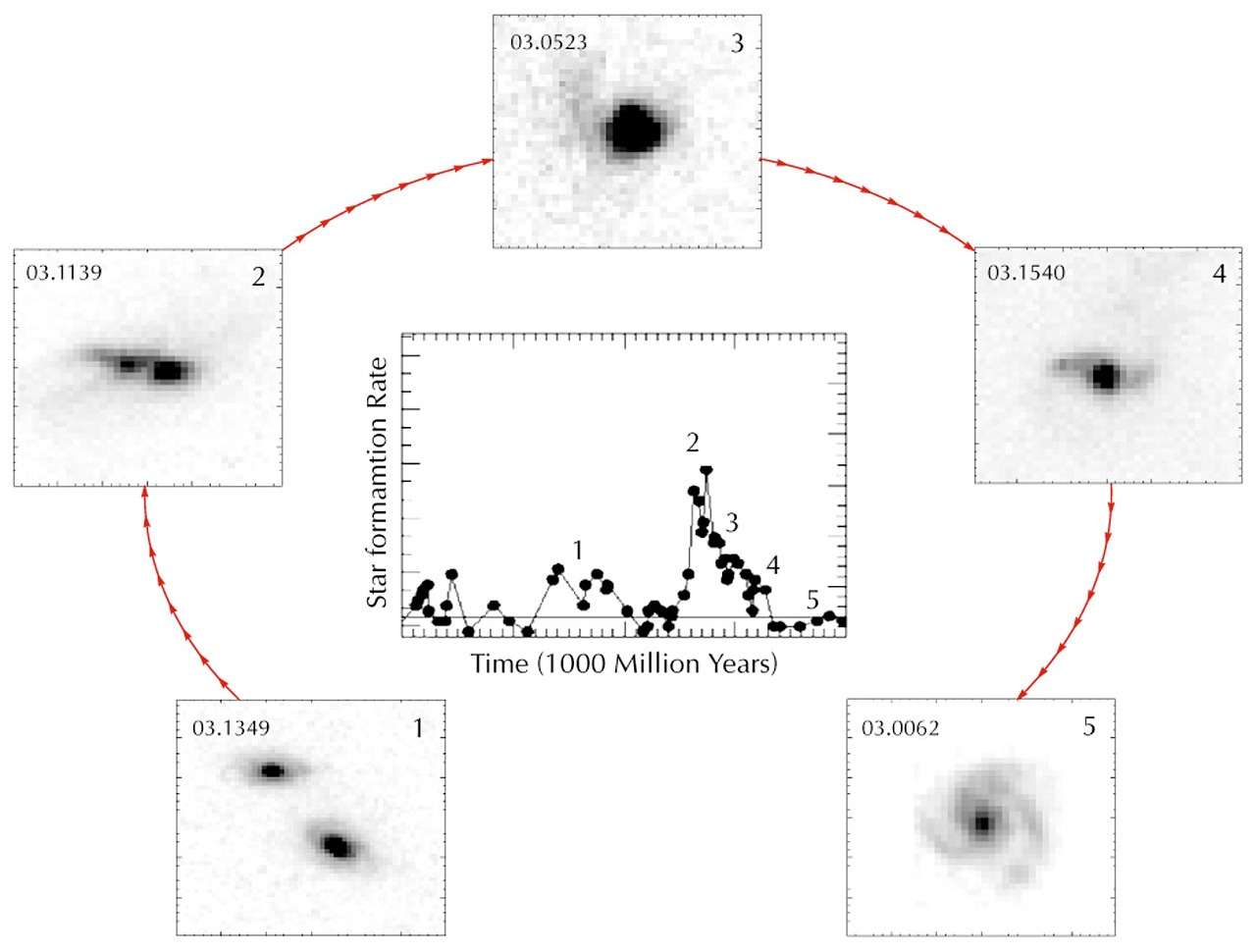}
\caption[Spiral rebuilding scenario diagram]{Illustration of the spiral rebuilding scenario.}
\label{Ill_SpiralRebuildingScenario}
\end{figure}

More recently, \citet{2009A&A...507.1313H} have made a detailed study of the conditions and different phases a major merger can go through to form spiral disks, including also the formation of the bulge, arms, bars and rings, which may mainly be originated from instabilities created during the last major merger of the system. As a result, we can distinguish five phases (see figure \ref{Ill_SpiralRebuildingScenario}):

\begin{itemize}
\item {Phase 1:} or approaching phase, which corresponds to the time until the first passage, 
\item {Phase 2:} or interaction phase. It corresponds to the time elapsed between the first and the second passage. During this phase the dynamical changes in both progenitors are so intense that a lot of matter is ejected into the inter-galactic medium.
\item {Phase 3:} this is the fusion phase, where the disk of the progenitors are totally destroyed. It corresponds to the time after the second passage and before the elaboration of the rebuilt disk. During phase 3 much of the matter falls toward the center of mass of the new system, and the galaxy may take the appearance of a chaotic morphology (see Fig. 3, top) or a central starburst (see Fig. 3, bottom) with an important peak of star formation quite short in time, always accompanied with chaotic velocity field and dispersion peaks clearly offset from the mass center. 
\item {Phase 4:} or disk rebuilding phase. While part of the matter is being compacted to form a bulge, the star formation rate decreases smoothly. At the same time the disk begins to be placed and can be detected. The galaxy would have a blue center associated to the bulge formation, and could be detected as a LCG. The rest of the gas ejected during the first phases (whose density could be too small to be detected) keeps falling into the new system, forming and extending the new disk. At this stage, the star formation through the disk would show a moderate rate. The galaxy rotation is evidenced in a velocity field, but could be offset from the main optical axis, and the dispersion peak(s) is (are) closer to the mass center (see appendix \ref{KinematicStudies}). 
\item {Phase 5:} The last phase corresponds to rotating spirals with regular velocity fields and dispersion peak centred on the centre of mass \citep[e.g.,][]{2006A&A...455..107F}. Notice that, in such a phase, disks are not "a priori" cold, as \citet{2007A&A...466...83P} have shown that most of them have small V/$\sigma$ (thick disks). However, the high dispersion value begins to relax.
\end{itemize}

This scenario is supported by the following numerical simulations based on morphological and kinematic observations of distant galaxies: \\
{\bf a.)} \citet{2009A&A...496...51P} simulated a giant and star-bursting bar induced by a 3:1 merger, and reproduced both the morphology and kinematics of the observed galaxy. \\
{\bf b.)} \citet{2009A&A...501..437Y} showed that a giant disk with two embedded cores results from a 4:1 collision. \\
{\bf c.)} \citet{2009A&A...496..381H} identified a compact and IR-luminous galaxy that is dominated by a dust-enshrouded compact disk\footnote{It is one example of first stage (phase 4) of disk rebuilding.}, itself surrounding a blue, centred "helix" (i.e., a "two arms-plus-bar" structure). This structure and the complex velocity field are predicted in post-merger phases. \\
{\bf d.)} \citet{2009A&A...493..899P} demonstrated that the presence of ionised gas without stars close to a highly asymmetric disk could be understood only as a merger remnant. \\
{\bf e.)} \citet{2010A&A...513A..43F} reproduced another giant blue bar with a flat velocity field which cannot be reproduced by any other mechanism than a merger with a face-on progenitor. By evolving these five models by 6 Gyr in time, they predict the formation of three spirals, one lenticular, and one massive disk that has survived an exceptional centred and polar collision.\\
A final important issue of assuming that major mergers are indeed the dominant process through galaxy evolution is that the number density of dwarf galaxies can also be accounted for if they result from material tidally ejected during such violent interactions \citep{2000ApJ...543..149O}.\\

The main weakness of the disk rebuilding scenario focus on the undetermined amount of gas in distant galaxies and the large uncertainty on the major merger rate. Indeed, on the one hand, this scenario is extremely dependent on the gas availability, which was likely much higher in the past than today. Disk rebuilding is plausible only if the merger progenitors has enough available gas. Recent Kennicutt-Schmidt estimations \citep[][see chapter 7]{2010A&A...510A..68P} are in agreement with the high amount of gas present in distant galaxies. On the other hand, \citet{2005A&A...430..115H} predict that $75\% (\pm 25\%)$ of galaxies must had been involved in a major merger since z$\sim$1. Nevertheless, the estimation of the merger rate is still debated, and this scenario could underestimate the minor mergers importance. \\

  \section{Discussion}
  \label{DiscussionScenariosGalForm}

The improvement and difference between scenarios will be established by the three major issues in modern cosmology, which are the formation of large and thin disks, the acquisition of the disk angular momentum, and the relative importance of each physical process in galaxy formation. \\

As mentioned before, the secular evolution scenario started as the tidal torque theory \citep{1976ApJ...205L.109P,1984ApJ...286...38W}, assuming that the angular momentum of disk galaxies had been acquired by early interactions. It was mainly supported by observations of the Milky Way (MW). Indeed, the past history of the MW appears to be devoid of significant mergers over the past 10 or 11 Gyr. However, compared to other spirals, the MW turns out to be exceptional, with a nearly pristine halo and a particularly small disk scale length and stellar mass \citep{2007ApJ...662..322H}. Then, could this scenario be applied to trace the history of the rest $\sim$70$\%$ of local spiral galaxies, even if they present quite different properties values from the MW? \\

There are many observations that could favor the secular evolution of galaxies (see subsection \ref{Secular_evolutionScenaSubsec}). However, this process predicts a slow evolution of galaxies, so that galaxies could take more than 6 Gyr to form \citep{2004ARA&A..42..603K}. Then, being a non-violent process, during their evolution, galaxies change very slowly without showing drastic or chaotic changes. Therefore, if this were the mechanism that governs the bulk of galaxy formation (since z $\sim$ 2), galaxies were expected to remain in a passive evolution during the past 8 Gyr. Recent studies show that it is not the case \citep{2002PASP..114..797V,2006A&A...455..107F,2008A&A...477..789Y,2005ApJ...632..169L}. These authors show that galaxies, in their majority, have complex morphologies and kinematics, and a high star formation rate, $\sim$6 Gyr ago (see also chapter 6 for a complete and representative comparison of distant and local galaxies). \\

In addition, the secular scenario has to be accommodated with an episodic origin of gas accretions, to explain the observed SFR values in the distant Universe. Nonetheless, in such a case, the difficulty would then be to explain how the gas coming from the intergalactic medium, which ("a priori") does not have a "common" history with the already formed galaxy, and which is coming from different directions (following the filaments, see chapter 1), could have the right angular momentum to form the abundant local thin disks (with the observed galactic angular momentum values). \\

It seems that the disk rebuilding scenario does not present those problems. It is in good agreement with the formation of the high fraction of local young thin disks, and with their angular momentum values. Indeed, \citet{2006AJ....131..226Y} suggest that the properties of thick and thin disks are consistent with gas-rich mergers playing a significant role in their formation where stars formed in these mergers lead to the thick disk, while the settling gas formed much of the thin disk. There is more and more evidence supporting this scenario. Its success comes from the fact that it has been constructed principally from observations. \\



\part{Shape and evolution of galaxies as a witness of the Hubble sequence}

\chapter{Observables in galaxy morphology}
\minitoc

Following the morphological method exposed previously in section 2.1.1, a lot of work has been done during the last two decades. We must remember that such classifications were developed principally before the 1990s, when astronomers only had access to photometric plates. Even if some of them also take advantage of the galaxy spectroscopy and photometry, they are based mainly on the eye inspection of the galaxy images. Nevertheless, with the arrival of the new CCD technology during the 1990s, new tools and techniques have been developed to improve the morphological classifications. I will try to summarize them in this chapter. \\ 

In general we can distinguish between the standard method, and the automatic methods. The classification by eye allows us to examine each galaxy carefully and individually. However, it is affected by a low reproducibility and a strong subjectivity. In the case of the automatic morphological classifications, a few photometric parameters are measured using computing techniques. They can thus be easily applicable to very large galaxy samples, saving a lot of time. Nevertheless, they dare to classify galaxies morphologically by using only a few photometric parameters. Therefore, even if they may be reproducible and subjective, they are "blind". \\

\section{The standard morphological classification method}

\begin{figure}[!h]
\centering
\includegraphics[width=1.0\textwidth]{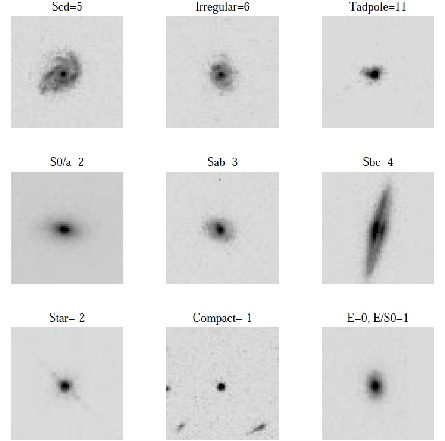}
\caption[Example of images for the morphological classification by eye]{Example of images for the morphological classification \citep{1998ApJ...499..112B}. They use HST images in V band.}
\label{Brinchmann1998MorphoClass}
\end{figure}

The first method used to classify the galaxies according to their morphology consists in observing their optical band images and classifying them following the observer's assessment (see section \ref{MorphologicalClassificationSubsect}). The astronomer disposes of some criteria and images, as a basis, to compare and make his own judgment. The well known criteria for the visual Hubble classification are the size and importance of the bulge in comparison to the disk, the distribution of the spiral arms, and the degree of their resolution. An example of a sample of optical images that was used as a basis for comparison, leading to the morphological classification of other galaxies, is shown in figure \ref{Brinchmann1998MorphoClass}. It is easily imaginable that this methodology can be seriously affected by the resolution of the galaxy images. Furthermore, concerning the distant galaxies, there are some other factors that can affect their visual classification:
\begin{itemize}
\item The farther a galaxy is, the less resolved it appears in the images. Not only because they are intrinsically smaller, but also because the inclination (projection) effect makes them less and less discernible. 
\item The dimming effect affects the galaxy surface brightness by a factor of (1+z)$^{4}$. Then, for a given exposition time, the faint structures of distant galaxies are easily lost. 
\item Finally, we have the spectral shift known as the K-correction. This latter makes galaxies to be observed at longer wavelengths than that in which the light was originally emitted. Thus, for example, if we could observe the same galaxy at different redshifts with the same filter, the galaxy aspect would change because we would be observing different aspects (stellar populations, regions, etc.) of the same galaxy (see subsection \ref{ConsiderationsforClassificationSubsect}).
\end{itemize}
As a consequence of the above effects, it thus becomes more difficult, in function of redshift, to distinguish visually the differences between the different objects and the classification errors happen more easily. "Perhaps the most serious problem with galaxy classification is that it is still largely a subjective visual exercise. The human eye is very good at pattern recognition, and is capable of integrating the information in an image quite quickly. However, a morphological type is not a measured quantity even if it is coded on a numerical scale" \citet{1992pngn.conf....3B}\footnote{with "numerical scale" Buta refers to de Vaucouleurs numerical scale (see figure \ref{HubbledeVaucouleurscom}).}.\\

During the past two decades, new methodologies have been developed to morphologically classify the galaxies, in order to take into account the previous effects. The principal common characteristics of these methods is that they take advantage of the modern computer and observational technologies, e.g., the CCD cameras (for a short introduction to the modern photometric observational techniques, see appendix \ref{IntroImageryPhotometry}).\\

\section{Modern methods for the morphological analysis}
 
These methods are based on quantities linked with the galaxy light distribution. Therefore, by analyzing the light distribution of the objects, they derive such quantities in order to recover the morphological type of each galaxy. In any case, the classification assumes a strong correlation between the derived parameters and the morphological types. We can distinguish between the methods that do not have an "a priori" assumption on the light distribution of the galaxies, called non-parametric methods, and those which use a prior assumption about the light distribution, called parametric methods. \\  

\subsection{The non-parametric methods}
\label{The_non-parametric_methods}

Even if the non-parametric methods do not use a prior assumption about the light distribution of the object, they need a "calibration" of their parameters by using galaxies that are already morphologically classified by other methods. After this calibration, the non-parametric methods extrapolate to recover the morphological type of new galaxies. \\

Amongst the different parameters, we can first mention the mean surface brightness, the radius containing half of the total luminosity, and the light concentration. This last is defined as the ratio between two radii, each one containing a fraction of the total light. This could be also interpreted as the ratio of two fluxes inside two different isophotes. In any case, the internal radius must be large enough to enclose a quite large number of pixels, and the external radius must not be too large in order to avoid any significant contamination from the sky. The definition of the Concentration (C) parameter the most often used is that of \citet{1985ApJS...59..115K} and \citet{1994ApJ...432...75A}:
    \begin{eqnarray}
       C =  \frac{\sum \sum_{i,j\in E(\alpha)} I_{i,j}}{\sum \sum_{i,j\in E(1)} I_{i,j}}
    \end{eqnarray}
where I$_{i,j}$ is the signal at the pixel coordinate $(i,j)$, and E(x) represents a defined area described by isophotal ellipses. However, a more recent definition of the concentration parameter is that of \citet{2000ApJ...529..886C}:
    \begin{eqnarray}
       C =  5 \cdot log \Big(\frac{R_{0.8}}{R_{0.2}}\Big)
    \end{eqnarray}
with R$_{f}$ representing a radius containing a fraction $f$ of the total luminosity of the galaxy. \\

Another parameter has been introduced by \citet{1996ApJS..107....1A}: the asymmetric parameter A \citep[see also][]{1995ApJ...451L...1S}. This parameter measures the degree of galaxy symmetry, and is calculated by obtaining the absolute value of the difference between the galaxy intensity and the intensity of the same galaxy rotated by 180$^{\circ}$ at each pixel (i,j), and normalized by the total intensity of the galaxy. More precisely, it can be expressed as \citep[following][]{2000ApJ...529..886C}:
    \begin{eqnarray}
       A =  min \Big(\frac{\sum \lvert I_{0}(i,j) - I_{180}(i,j) \rvert}{\sum I_{0}(i,j)}\Big) - min \Big(\frac{\sum \lvert B_{0}(i,j) - B_{180}(i,j) \rvert}{\sum B_{0}(i,j)}\Big)
    \end{eqnarray}
where I is the intensity of the galaxy, and B the intensity of the image noise. This last is measured within an image area without any object, and equal in size to the galaxy area, in order to take into account the effects of the sky background. It is very important for recovering the asymmetry A to precisely establish the center of the galaxy. \citet{1996ApJS..107....1A} take the brightest pixel from a smoothed version of the galaxy, and \citet{2000ApJ...529..886C} take the position that minimized the asymmetry. \\

\begin{figure}[!t]
\centering
\includegraphics[width=1.0\textwidth]{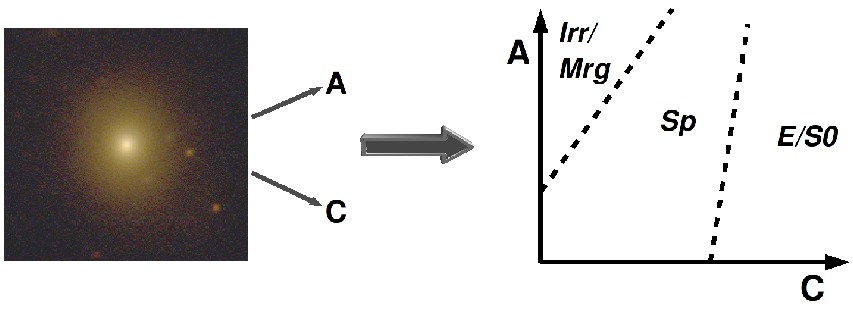}
\caption[Expected distribution of the different morphologies within the A-C plane]{Expected distribution of the different morphologies within the Asymmetry (A) vs central Concentration (C) plane.}
\label{ACplan}
\end{figure}

\begin{figure}[!h]
\centering
\includegraphics[width=0.8\textwidth]{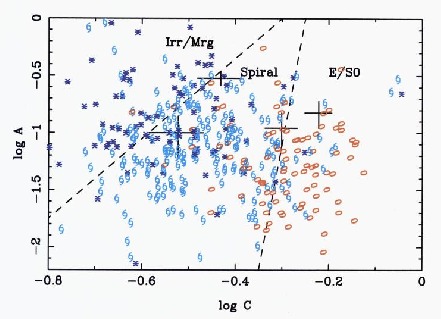}
\caption[Distribution of a galaxy sample in the A-C plane from observations]{Distribution of a galaxy sample in the A-C plane from \citet{1996ApJS..107....1A}. Their plot symbols and colors correspond to the visual classification by an expert (Sidney van den Bergh). Symbols are as follows: compact, S0 and elliptical galaxies are shown as red ellipses; spiral galaxies are shown using blue spiral symbols; and galaxies classified as irregular, peculiar, or mergers are shown as blue asterisks.} 
\label{ACplanDistributionOfgalaxiesVDB}
\end{figure}

On the one hand, the C parameter is essentially a measure of the light concentration of the galaxy, and is correlated to the bulge size. It could then be correlated with the morphological type because elliptical galaxies are generally more concentrated than disk galaxies. On the other hand, the A parameter is a measure of the symmetry of the galaxy, which could be related to the morphological type too. Spiral galaxies, due to their spiral arms (for example), are more asymmetric than elliptical or lenticular galaxies. \citet{1996ApJS..107....1A} proposed to use the A-C plane to morphologically classify the galaxies, assuming that in such a plane the principal morphological types (E/S0, Sp, and Irr/Mrg) are separated in different regions, because of the above reasons (see figure \ref{ACplan}). Nevertheless, as is shown in figure \ref{ACplanDistributionOfgalaxiesVDB}, such a method does not seem to be efficient. The separation of the morphological types presents a contamination higher than 15$\%$ from one type to the others \citep{2006AJ....131..208M}. Nonetheless, ignoring this significant consideration, this method is still used to morphologically classify local and distant (z$\sim$0.6) galaxies (see next chapter)\citep[e.g.,][etc.]{1996MNRAS.279L..47A,1997ApJS..110..213S,1998ApJ...499..112B,2004ApJ...600L.139C,2005ApJ...631..101P,2005sf2a.conf..667L,2006AJ....131..208M,2006A&A...453..809I}. \\

\begin{figure}[!h]
\centering
\includegraphics[width=0.5\textwidth]{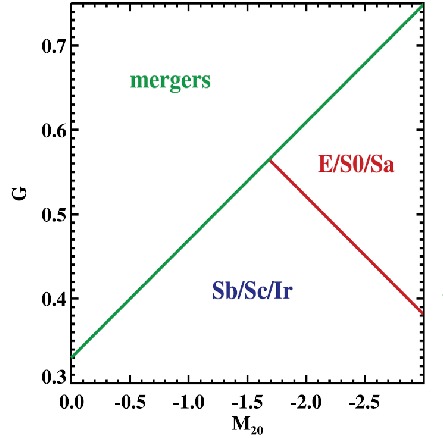}
\includegraphics[width=0.5\textwidth]{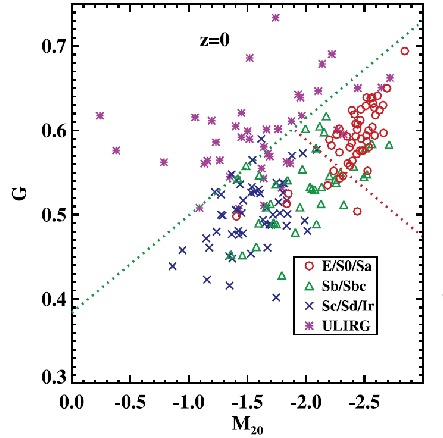}
\caption[Expected distribution of the different morphologies within the G-M$_{20}$ plane and observations]{{\it Top}: Expected distribution of the different morphologies within the Gini (G) vs M$_{20}$ plane. {\it Bottom}: Distribution of a local galaxy sample in the G-M$_{20}$ plane from \citet{2008ApJ...672..177L}. Their symbols and colors correspond to the visual classification from the Carnegie Atlas \citep{1994cagv.book.....S}. Symbols are as follows: elliptical, S0 and Sa galaxies are shown as red circles; Sb and Sbc galaxies are shown using green triangles; Sc, Sd and irregular galaxies are shown as blue crosses. The purple asterisks represent the merger candidates (ULIRGS).} 
\label{GMplanDistributionOfgalaxies}
\end{figure}

More recently, other new parameters have been also introduced: the smoothness (S)\citep{2003ApJS..147....1C}, the Gini (G) coefficient \citep{2003ApJ...588..218A}, the M$_{20}$ coefficient \citep{2004AJ....128..163L}, and the coarseness parameter \citep{2005AJ....130.1545Y}. Similar to the C-A plane, it has been also attempted to propose a G-M$_{20}$ plane to separate the different morphological types. Nonetheless, it is not yet convincing either (see figure \ref{GMplanDistributionOfgalaxies}).\\

Due to the above uncertainties, further steps have been taken. The concentration, the asymmetry, and the smoothness parameters have been combined to form the CAS system. It seems more robust, though it is effective only in separating the early and late type galaxies \citep{2005MNRAS.357..903C}. Then, larger quantities of coefficients (radius, C, A, G, and M$_{20}$, for example) have been combined \citep{2007ApJS..172..406S}. Nonetheless, a significant contamination seems to be always present. \\

As a conclusion, the non-parametric methods or "automatic methods", are easily applied to a large number of galaxies, without human intervention, just by adapting them to a computer code. Nevertheless, even if they could be very promising for the future, at the present day they are not accurate enough to be applied in order to measure precisely and exactly the morphology of galaxies, and quantify accurately the morphological evolution of galaxies. Moreover, \citet{2008ApJS..179..319L} and \citet{2010ApJ...712.1385C} show that the coefficients used in such methods present a strong dependence on S/N ratio. \\

\subsection{The parametric methods}

\begin{figure}[!t]
\centering
\includegraphics[width=0.8\textwidth]{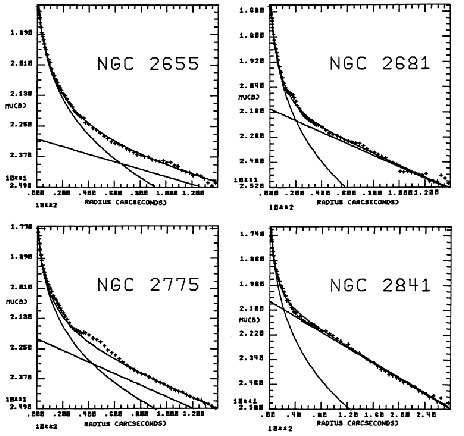}
\caption[Examples of bulge+disk decomposition]{Example of bulge+disk decomposition from \citet{1981ApJS...46..177B}.} 
\label{Bulge+diskDecomposition}
\end{figure}

The main goal of the parametric methods consists in fitting one or different analytical models to the observed light distribution of the galaxy. They can include, e.g., radial multi-Gaussian deconvolution \citep{1991ApJ...366..599B,1998AJ....115.1400F}, or bulge+disk decomposition \citep{1995ApJ...451L...1S,1996ApJ...464...79S,1999AJ....118...86R}(see figure \ref{Bulge+diskDecomposition}). The bulge+disk decomposition has become popular because, as noticed by \citet{2002ApJS..142....1S}, (a) it is rooted in the very first studies of the functional form of galaxy radial surface brightness profiles \citep{1948AnAp...11..247D,1959HDP....53..275D}; (b) it provides a comfortable mental picture of the overall structure of a distant galaxy, i.e., it is conceptually simple to relate a quantitative measurement of galaxy type such as bulge-to-total light ratio to the familiar Hubble types; and (c) photometric entities such as bulges and disks have distinct dynamical counterparts.\\

The bulge can be modeled with a de Vaucouleurs law \citep{1948AnAp...11..247D}, with $f(r)=f_{e}e^{(-r/r_{e})^{1/4}}$, where $r_{e}$ is the effective radius, and $f_{e}$ the flux within the effective radius. The disk can be modeled with an exponential law: $f(r)=f_{0}e^{-(r/r_{d})}$, where $f_{0}$ is the flux in the center of the galaxy, and $r_{d}$ is the disk radius. More recent studies show that it is more advisable to model the bulge, and the disk, with a Sersic law \citep{1968adga.book.....S}. In such a case, the power 1/4 of the bulge profile is replaced by a power 1/n (n is the Sersic index), and the exponential profile describing the disk is similar to the Sersic law when n=1 (see more details in section \ref{StructPara_LightProfiAna}). \\

Taking the best fit\footnote{generally the best fit is determine by using a $\chi^{2}$ minimisation. More details about the $\chi^{2}$ procedure can be found in subsection \ref{StructPara_LightProfiAna}.} to the above models, one can retrieve different physical parameters, such as flux, size, orientation, etc. These methods thus also allow us to measure the flux ratio between the different structures and the total galaxy flux. One example is the bulge to total light ratio (B/T), which is correlated to the morphological types \citep{1986ApJ...302..564S,1985ApJS...59..115K}. \\

The first parametric methods were based on 1D fits, which means that the light distribution of the galaxies was considered to be fixed within a unique value of inclination, position angle (P.A.), and ellipticity along the optical major axis. The first works focused on local galaxies \citep[e.g.,][see figure \ref{Bulge+diskDecomposition}]{1977ApJ...217..406K,1979ApJ...234..829B,1981ApJS...46..177B,1985ApJS...59..115K}\citep[or more recently][]{2006A&A...446..373P}, even though, with the arrival of new technologies and the Hubble Space Telescope (HST), deeper images with good signal-to-noise ratios (S/N) have allowed them to be applied to galaxies in the distant Universe. However, their principal limitation is the fact that they assume a 1D light distribution. Many real galaxies exhibit more than two structural components such as nuclear sources, bars, spiral arms, etc., or even in the presence of only a bulge and a disk, the ellipticity and/or the P.A. of these components might be functions of galactocentric distance \citep{2002ApJS..142....1S}. Another limitation is that they fix a unique center for the whole galaxy structures, which is not necessarily the case.\\

To improve upon these approximations, new techniques, such as isophotal ellipse fitting in which it is customary to let ellipticity, position angle, and centroid vary from one ellipse to another, were developed \citep[e.g.,][]{2002AJ....124..266P,2010AJ....139.2097P}. Thus, the whole galaxy light distribution is now taken into account in a 2D fitting model. Another improvement of such 2D models is that different structures can be modeled at the same time (for example, bulge+disk+bar). These 2D models present a higher number of parameters to be fitted than the 1D models. It is thus necessary to have images where objects are well resolved, and the S/N ratio is high enough to avoid degeneracies \citep[see][for a more detailed discussion on the S/N effects]{2002ApJS..142....1S,2007ApJS..172..615H}. Another solution to reduce the possible degeneracies is to limit the number of free parameters during the fitting process. \\ 

From the retrieval of the physical parameters of the different components, several morphological classifications have been proposed. \citet{2004ApJ...604L...9R} and \citet{2005ApJ...635..959B}, for example, proposed  a classification based only on the Sersic index $n$. In such a case, they distinguish between spheroidal types (n > 2.5), and late type galaxies (n < 2.5). \citet{2005MNRAS.357..903C} and \citet{2007ApJS..172..434S} showed later that a classification based only on $n$ is not as robust as a classification based on the B/T index. Nevertheless, we show in \citet{2008A&A...484..159N} (see also chapter 5 and 6 for more details) that even the B/T value alone is not sufficient to robustly classify morphologically all galaxies. \\

\subsection{The color information}

The color can be defined as the difference between two magnitudes (obtained using two different broad-band filters). By convention, the magnitude derived from the filter having the largest central wavelength is subtracted from the magnitude obtained from the filter having the smallest central wavelength (see subsections \ref{ColorInfo} and \ref{ColordeterminationMyPaper}). A negative value will then means that we observe more light through the filter with smaller wavelength, and it is represented with a blue color. The contrary is represented with a red color. \\

\begin{figure}[!ht]
\centering
\includegraphics[width=1.0\textwidth]{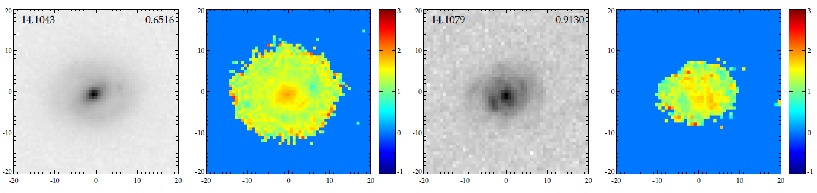}
\caption[Example of color maps]{Example of color maps from \citet{2005A&A...435..507Z}.} 
\label{ColorInformationMaps}
\end{figure}

We can then derive a color for the whole galaxy (integrated color), a color for the different components of the galaxy \citep[e.g., disk, bulge, etc.;][]{2001ApJ...551..111E}, or even pixel by pixel color analysis \citep{1999MNRAS.303..641A,2001MNRAS.322....1M,2004ApJ...612..202M,2004A&A...421..847Z,2005A&A...435..507Z,2005ApJ...631..101P,2007MNRAS.380..571L}(see figure \ref{ColorInformationMaps}, and subsections \ref{ColorInfo} and \ref{ColordeterminationMyPaper}). These last are only possible if the objects are well resolved. An important precaution when working with samples of galaxies at different redshifts is to use different filters which correspond to the same rest-frame band, in order to limit the k-correction effects (more details about the color analysis are shown in chapter 6). \\

In subsection \ref{ConsiderationsforClassificationSubsect}, I show that each filter gives different information about the galaxy\footnote{revealing the different constituents of the galaxy.}. Thus, by comparing the magnitudes from two different bands one can have an idea, e.g., of the stellar population which dominate the whole galaxy or a part of it (old stars are red, and young stars are blue, even though it could be affected by dust). This can also give an idea of the history of the galaxy, or the history of the section of the galaxy we are interested in, by comparing its color with stellar population models \citep[e.g.,][]{2001MNRAS.322....1M,2004A&A...421..847Z,2005A&A...435..507Z,2007A&A...469..483R}. \\

The integrated color is also related to the galaxy morphology \citep{1994ARA&A..32..115R}. Elliptical and lenticular galaxies are composed mainly by old stars, which are red. On the contrary, spiral and irregular galaxies are mainly composed of young stars, which are bluer. However, such a correspondence between integrated colors and morphological types must not be taken too seriously (for a detailed discussion see subsections \ref{DownsizingarevivaloftheprimorColl}, \ref{ColordeterminationMyPaper}, and \ref{ColorDistributionFromMyPaper}). In any case, this correlation has also been adapted to the non-parametric coefficients (see subsection \ref{The_non-parametric_methods}). \citet{2000ApJ...529..886C} proposed a sequence connecting the A and S parameters to the integrated color of galaxies, going from the red and symmetric galaxies (ellipticals) to blue and asymmetric Sc-d. \citet{2005A&A...434...77L} also proposed to use the color information to make a kind of spectro-morphological classification. It is done by constructing A-C planes with the coefficients measured in different wavelength bands (i.e., k-morphological correction)\citep[see also][]{2000ApJS..131..441K,2001A&A...369..421B}. \\


\chapter{Studying the morphological evolution}
\minitoc

\section{Morphological studies in the distant universe}

\subsection{Local universe}

\begin{figure}[!h]
\centering
\includegraphics[width=1.0\textwidth]{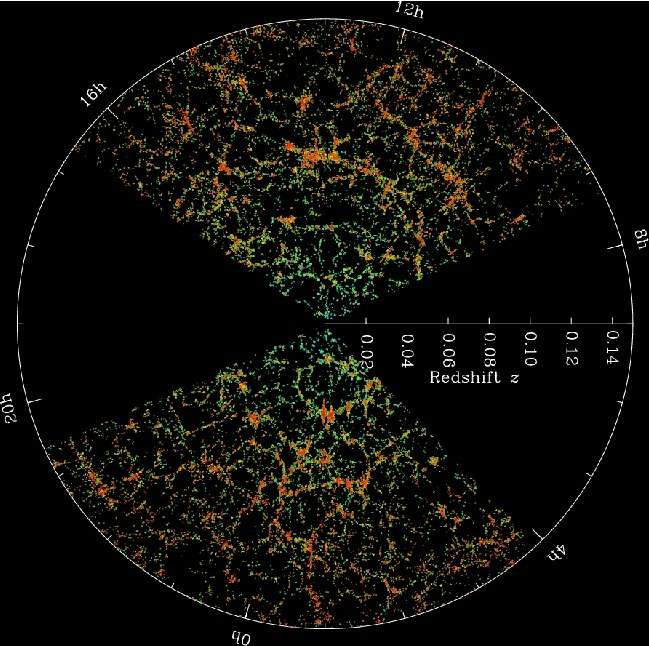}
\caption[SDSS 3-dimensional map of the galaxy redshift distribution]{SDSS 3-dimensional map of the galaxy redshift distribution. Each point represents a galaxy, and colors are according to the ages of the galaxy stars (with the redder points showing galaxies that are made of older stars.). Both slices contain all galaxies within -1.25 and 1.25 degrees declination. (Credit: M. Blanton and the Sloan Digital Sky Survey.)} 
\label{SDSSsurveyPie}
\end{figure}

The local Universe has been studied over the past two centuries (see chapter 1), and it is better known than the distant one (at least in what concerns the fraction of the different galaxy populations). The most recent atlas of local galaxies are, amongst others, \citet{1994cagv.book.....S}, and the Sloan Digital Sky Survey, including almost one million galaxies (see figure \ref{SDSSsurveyPie}).\\

Nowadays, all studies confirm that in the local Universe $\sim$81$\%$ of galaxies are spirals, $\sim$14$\%$ are early-type, and $\sim$5$\%$ are irregulars \citep[][see also chapter 1]{2004AJ....127.2511N,2007AJ....134..579F}. This confirms the composition of the local Hubble sequence discussed in chapter 1. \\

However, no sequence has been derived for the distant Universe. The epoch from z=1 to z=0.4 is crucial for the assembly of galaxies. It still remains to understand how their morphological structures were formed and what physical processes fundamentally drove their assembly. While disks can be built by accreting cooling gas in a dark halo \citep[e.g.,][]{1980MNRAS.193..189F}, the formation of bulges may be regulated by either external or internal processes \citep[see][for a review]{2000bgfp.conf..413C}. The bulges can be formed via either primordial collapses at an early time \citep[e.g.,][]{1962ApJ...136..748E} or hierarchical mergers \citep[e.g.,][]{1996MNRAS.281..487K}. In the later case, violent star formation is expected \citep[ULIRGs are mostly associated with merging systems,][]{1996ARA&A..34..749S}. Comparison of morphological properties between distant and local galaxy populations is essential to, e.g., unveil whether the merging process regulate bulge formation. In the secular processes, bars are mostly suggested to be efficient at driving gas into the galactic center and changing star orbits to form a central bulge-like structure \citep{1990A&A...233...82C,2004ApJ...604L..93D}.

\subsection{Distant Universe (z $\gtrsim$ 0.3)}

Undoubtedly, with the arrival of the HST, over the past two decades a lot of work has been carried out to uncover the morphology of distant galaxies.\\ 

\begin{figure}[!h]
\centering
\includegraphics[width=1.0\textwidth]{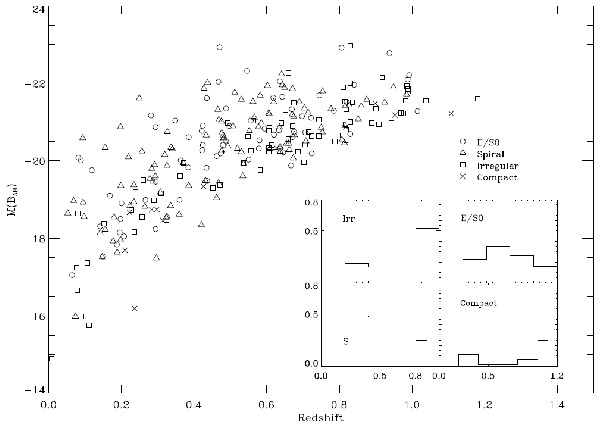}
\caption[Number fraction of different morphologies vs redshift]{Distribution of galaxies in the M$(B_{AB})-z$ plan from \citet{1998ApJ...499..112B}. As it can be noticed in the histograms at the right-bottom, the fraction of irregular galaxies increases with redshift.} 
\label{IrrGalaxEvolutionBrinch1998}
\end{figure}

One of the main contribution of such works has been to address the morphological evolution of field galaxies. \citet{1998ApJ...500...75L} found that massive disk galaxies show a constant number density from z$\sim$1. Elliptical galaxies also exhibit little evolution up to z$\sim$1 in their number density \citep{1999ApJ...525...31S}, although non-negligible star formation at z <1 was detected in a substantial sub-population \citep{2001MNRAS.322....1M,2002ApJ...564L..13T}. Additionally, morphological investigations based on large galaxy samples revealed that the number density of irregular galaxies increases with redshift \citep[][see figure \ref{IrrGalaxEvolutionBrinch1998}]{1998ApJ...499..112B,2000AJ....120.2190V}. Such a significant evolution of irregular galaxies was suggested to be associated with a transformation into regular galaxies \citep{2000ApJ...536L..77B}. \\ 

Indeed, \citet{1996AJ....112..359V,2000AJ....120.2190V} and \citet{1998ApJ...499..112B} found that disk galaxies at redshifts z$\gg$0.3 have morphologies that appear to differ systematically from those of nearby galaxies. Furthermore, while accumulated evidence suggests that most massive elliptical galaxies in the field are in place prior to z =1 \citep[e.g.,][]{2000A&A...362L..45D,2002A&A...381L..68C}\citep[see also][]{2003ApJ...592L..53V,2004ApJ...608..752B,2004ApJ...608..742D}, the formation of spiral galaxies is still debated \citep{2005mmgf.conf..263H}.\\

However, such conclusions must still be taken carefully because they result from different studies, which have been done using different samples with different selection criteria, and not necessarily with equivalent observational data for distant and local galaxies. Moreover, some of these studies have been carried out using different morphological classification methods. It has been noted in the previous chapter that some methods are still not good enough to separate the different morphological types correctly. Therefore, compared with the galaxy distribution along the Hubble sequence in the local universe, a complete morphological census of field galaxies at 0.4 < z < 1 will provide clues to understand their global morphological transformation and evolution. In such a case, a very careful analysis of the observational data available must be done, in order to guarantee the possible comparison between local and distant morphologies.  \\

Concerning objects at z $\gg$ 1, nowadays we do not dispose of morphologically exploitable images for galaxies at more than 8 Gyr in the past. Nonetheless, the futures ELTs will give us a better morphological view of such epochs. \\

\section{Need of representativeness}
\label{NeedOfRepresentativenessSubsect}

In subsection \ref{ConsiderationsforClassificationSubsect}, I made a description of the principal considerations to be taken into account when classifiying a sample of galaxies, assuming that the number of galaxies in the sample being studied is high enough to avoid any morphological type population to be missed (see below). However, beyond the selection criteria (see subsection \ref{ConsiderationsforClassificationSubsect}), the representativeness of a sample is another fundamental consideration to be taken into account. How could we make conclusions, concerning the evolution of a specific galaxy characteristic, by studying a sample that does not statistically represent the whole galaxy populations? It is a difficult task. Indeed, my experience during my thesis showed me that representativeness and selection criteria must be treated carefully. In \citet{2008A&A...484..159N}(see chapter 5), we studied a representative sample of distant galaxies \citep[see also][in chapter 6]{2010A&A...509A..78D}, proving such a representativeness by comparing them to the luminosity function at the respective redshift. Nonetheless, the selection criteria adopted in Neichel et al. limited the morphological types to some specific populations (see subsection \ref{ConsiderationsforClassificationSubsect}).\\

\subsection{Sample representativeness and luminosity function}

Galaxy surveys are gathering large samples of galaxies. Their aims are to minimise statistical uncertainties (for 10000 galaxies, $\sqrt{N}$/N is 1$\%$), and often to overcome the important problem of cosmic variance. One example of the consequence of such statistical uncertainties is evoked in subsection \ref{ConsiderationsforClassificationSubsect}: concerning the different morphological type populations, the smaller is a sample, the larger is the probability to miss one of them (e.g., the less abundant population). However, the study of such large quantities of galaxies poses some time consuming problems, as well as some instrumental availability problems (e.g., if we study the morphology of 1000 galaxies, it will be quite difficult to dispose of kinematic observations for all of them, because the observational techniques are more complicated and time consuming).\\

As a result, a high fraction of works are based on small samples of galaxies to study carefully the different properties of all the galaxies in the sample. Nonetheless, they are facing the small statistics issue (for 100 galaxies, $\sqrt{N}$/N is 10$\%$). Because of this, they have to be very efficient in controlling the selection effects related to the target selection. Having said this, they are quite unique to probe, at the same time, dynamics, stellar populations, gas abundances, extinction and star formation rate. Why is this so important? Because z$\sim$1 is already probing the universe at less than half its present age, where changes in galaxy properties are expected. Indeed, we know that the morphological appearances of distant galaxies are far more complex than those illustrated in the Hubble sequence (see previous subsection). \\ 

To measure the representativeness of a small sample, it is therefore judicious to use the galaxy luminosity functions derived from larger samples \citep[e.g.,][]{2003A&A...402..837P,2007MNRAS.380..585C}. The luminosity function gives the space density of galaxies per luminosity bin. It is hence a representation of the galaxy number distribution with respect to their mass. Since the gravitational force is the principal responsible for the formation of structures (see appendix \ref{CosmologyTheHistoryOfTheUniverse}), the mass can be consider as  the most important parameter that characterized the whole galaxy population. \\

\subsection{What are the galaxies responsible for the evolution?}

The origin of the star formation density decline over the past 8 Gyr is still a matter of debate. Studies based on spectral energy distributions (SEDs) of galaxies predict that 30$\%$ to 50$\%$ of the mass locked into stars in present-day galaxies actually condensed into stars at z<1 \citep{2003ApJ...587...25D,2003A&A...402..837P,2004ApJ...608..742D}. This result is supported by the integration of the star formation density when it accounts for infrared light \citep{1999ApJ...517..148F}. Indeed the rapid density evolution of luminous infrared galaxies (LIRGs, forming more than 20 M$_{\odot} yr^{-1}$) suffices in itself to report for the formation of a high fraction of the total stellar mass found in intermediate mass galaxies (from $3\times 10^{10}$ to $3\times 10^{11}$ M$_{\odot}$)\citep{2005A&A...430..115H}. As opposed to the idea of galaxy "downsizing" \citep{1996AJ....112..839C} - strong evolution only in the faint blue dwarf population - there is a growing consensus that most of the decline of the star formation density is indeed related to intermediate-mass galaxies \citep{2005A&A...430..115H,2005ApJ...625...23B} (see more details in section \ref{Introduction_chapterV}).\\

\begin{figure}[!h]
\centering
\includegraphics[width=0.8\textwidth]{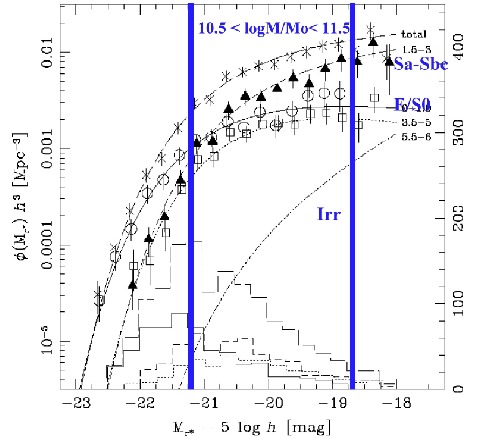}
\caption[Intermediate mass galaxies definition from a MDLFs plot]{Intermediate mass galaxies \citep{2005astro.ph..9907H}. This is an adaptation of figure 3 of \citet{2004AJ....127.2511N}, which presents the morphological-dependent luminosity functions (MDLFs) from the SDSS. The two vertical blue lines distinguish the area of intermediate mass galaxies. For this purpose \citet{2005astro.ph..9907H} have assumed a ratio M/L$_{r^{*}}$=5, by averaging results from \citet{2003ApJ...587...55G} fits of the spectral energy distribution of SDSS galaxies.}
\label{IntermediateMassGalaxiesMDLFs}
\end{figure}

It is more and more convincing that massive E/S0s and dwarves\footnote{\citet{2005astro.ph..9907H} shows how unclear are the words "massive" or "dwarves". Indeed, let's define massive galaxies as those having M$_{star}$ > $3\times 10^{11}$ M$\odot$ (i.e. bigger than the Milky Way), and dwarves as those having M$_{star}$ < 3 10$^{10}$ M$\odot$. This is somewhat arbitrary (as any definition), although with these criteria, E/S0 dominate the massive regime, spirals dominate the intermediate mass regime, and irregulars become a significant population in the dwarf regime (see figure \ref{IntermediateMassGalaxiesMDLFs}).} could not be responsible of the star formation density decline, because the first ones were mostly in place at z=1 (see previous subsection), and the second ones contribute marginally to the global stellar mass or metal content. It is therefore interesting to carry out a detailed study of the morphological and kinematic properties of the population of intermediate mass galaxies, in which at least 2/3 of the present-day stellar mass is locked \citep{2000ApJ...536L..77B,2004Natur.428..625H}. \\


\part{Our approach: Looking for the evolution at z $<$ 1}

\chapter{IMAGES}

\begin{figure}[!h]
\centering
\includegraphics[width=0.8\textwidth]{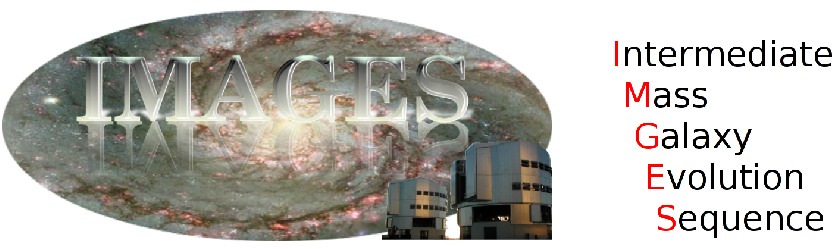}
\end{figure}

\minitoc

\section{Introduction}
\label{Introduction_chapterV}

Over the years before our survey IMAGES (Intermediate MAss Galaxies Evolution Sequence), a lot of efforts had been made to study the properties of galaxies in the distant universe (at z $\sim$ 1), showing growing evidence of a significant evolution with cosmic time. Examples of such evidence are the rapid time decrease of the cosmic SFR (star formation rate) density, the role of merging in the evolution of galaxies, the importance of studying in details the morphology of high-redshift galaxies, and the possible evolution of the TFR.  \\

\begin{figure}[!h]
\centering
\includegraphics[width=0.8\textwidth]{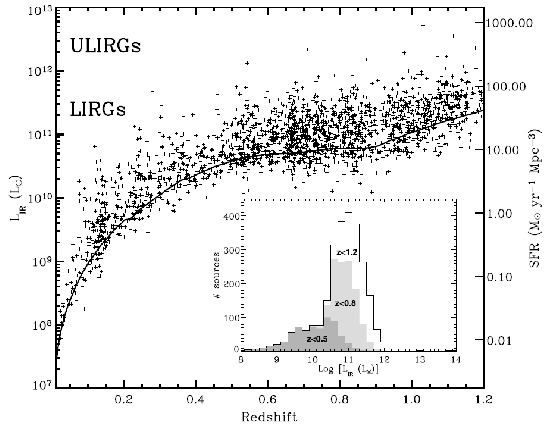}
\caption[SFR evolution with redshift]{SFR evolution with redshift showed by \citet{2005ApJ...632..169L}Le Floc'h et al. (2005).}
\label{SFRevolution}
\end{figure}

Concerning the SFR history, it had been found a decline of the cosmic SFR by a high factor $\sim$10 from z $\sim$ 1 to the present \citep[see figure \ref{SFRevolution};][]{2005ApJ...632..169L,2000ApJ...544..641H,2002AJ....124.1258W,1999AJ....118..603C,1999ApJ...517..148F,1997ApJ...481...49H,1996ApJ...460L...1L,1996MNRAS.283.1388M}. Although the conclusions are reached from different data, they are consistent within a factor of 3 \citep{2004ApJ...615..209H}. Furthermore, recent results from Herschel observations, which give a more precise measurement of the SFR\footnote{The SFRs are calculated by assuming that they are proportional to the integrated infrared luminosity (LIR, typically 8-1000 $\mu$m, Kennicutt 1998). Then, the flux through an IR broad-band range is converted into LIR by fitting the flux using a SED library \citep[e.g.,][]{2001ApJ...556..562C}, in which for a given redshift and band flux, a unique solution to LIR exists. The previous works had only available observations in the 12-18 $\mu$m \citep[e.g.,][]{1999ApJ...517..148F} and 24 $\mu$m (Spitzer observations). These wavelengths are relatively far from the IR-SED peak (produced principally by the emission of the dust heated by young stars), which makes them sensitive to extrapolation errors. In addition, at these wavelengths, polycyclic aromatic hydrocarbon (PAH) emissions, as well as the AGN-heated dust emissions, are present and the ratio of their fluxes to LIR may create significant scatter. With Herschel observations at 160 $\mu$m (PACS), fluxes probe the emission much closer to the SED peak and are unaffected by the above issues \citep[more details in][]{2010A&A...518L..24N}. It allows Herschel observations to provide the most reliable and least biased SFR measurements to date, avoiding large SED extrapolation errors.}, consolidate such a strong evolution of the cosmic SFR \citep{2010A&A...518L..25R}. \citet{2004Natur.428..625H}, using the SDSS survey, suggest that the co-mobile star formation density may have reached its peak as late as z $\sim$ 0.6, and this bulk of star formation was mostly dominated by star formation in galaxies in the range 3-30x10$^{10}$ M$_{\odot}$ \citep[see also][]{2005A&A...430..115H}. Even if such a peak could be at z$\sim$2 due to dust effects and the larger measure uncertainties at z$\geq$1 \citep{2005ApJ...630...82P,2008ApJ...675..234P}, the high SFR values at z$<$1 led argue that about half of the local stellar mass density has been formed since z = 1, mostly in luminous infrared galaxies \citep{2005ApJ...625...23B,2005A&A...430..115H,2003ApJ...587...25D}.\\

As a consequence of these facts, galaxy interactions and merging are mechanisms that could play a significantly larger role for star-formation and mass assembly in the distant universe than today \citep{2009A&A...507.1313H,2010MNRAS.402.1693H}. \citet{2000MNRAS.311..565L} found that the merger rate in the distant universe (z$\sim$1) was about 10 times higher than at low redshift \citep[see also][]{2008ApJ...681.1089R,2006ApJ...652..270B,2006ApJ...636..592L,2003AJ....126.1183C}. The high merger rate detected in the distant universe raises a challenge to the standard scenario of disk formation\footnote{Early studies of the MilkyWay (MW) have led to a general description of the formation of a disk galaxy embedded in a halo \citep{1962ApJ...136..748E}. \citet{1980MNRAS.193..189F} set out a model of galaxy formation assuming that disks form from gas cooling condensing in dark halos. Protogalactic disks are assumed to be made of gas containing a substantial amount of angular momentum, which condenses into stars to form thin disks \citep{1976MNRAS.176...31L}. These disks then evolve only through secular processes. This so-called standard model successfully reproduces the flat rotation curves and the size of spiral galaxies \citep[e.g.,][and references therein]{1998MNRAS.295..319M}, even though it does not reproduce the observed local angular momentum values (see subsection \ref{Secular_evolutionScenaSubsec}).}, since till today most of the simulations predict that major mergers form essentially elliptical galaxies (see subsection \ref{Secular_evolutionScenaSubsec}). Given the number of pairs detected in the past, if major mergers always generate elliptical galaxies inevitably, we would find a large fraction of elliptical galaxies in the local Universe rather than the $\sim$70$\%$ of observed spiral galaxies. Similar problems appear with the observed fraction of luminous compact blue galaxies (LCBGs), which increases with redshift by about an order of magnitude out to z $\sim$ 1 \citep{2007A&A...469..483R,2004ApJ...617.1004W}. First, LCBGs were associated to progenitors of local spheroidal or irregular galaxies at low redshift \citep[e.g.,][]{1995ApJ...440L..49K,1999ASPC..187..271G}. Later, \citet{2001ApJ...550..570H} proposed that these galaxies were the progenitors of the bulges of intermediate-mass spirals \citep[admitted later by][]{2006ApJ...640L.143N}. The last authors, using deep (HST/UDF) images, detect disk around distant LCBGs. Further 3D spectroscopic studies of the internal kinematics of LCBGs suggest that they are likely merger remnants \citep[see][for a distant Universe sample]{2006A&A...455..119P}\citep[see][for a local example]{2001A&A...374..800O}. Interestingly, just a quick look of the ACS/HST images of most of the galaxies showed in \citet{2006A&A...455..119P} reveals quite peculiar morphologies.  \\

Indeed, in the past, morphological investigations have brought observational evidence that a large fraction of galaxies at intermediate redshift have peculiar morphologies that do not fit into the standard Hubble sequence \citep[e.g.,][]{1996ApJS..107....1A,1998ApJ...499..112B,2001AJ....122..611V,2005A&A...435..507Z}. This could be related to the merger rate and SFR evolution cited above \citep{2005A&A...430..115H}. Furthermore, different observations suggest that most of the massive elliptical galaxies were in place prior to z = 1 \citep{2007ApJ...669..947J,2006AJ....131.1288B}. While the formation and evolution of spiral galaxies still pose some problems and different physical processes can modify the galaxy properties over cosmic time (mergers have always been associated to the destruction of the disk), a robust morphological analysis of distant galaxies is crucial to understand their component structures and the eventual relation between morphology and kinematics in the distant Universe. \\

\begin{figure}[!h]
\centering
\includegraphics[width=0.85\textwidth]{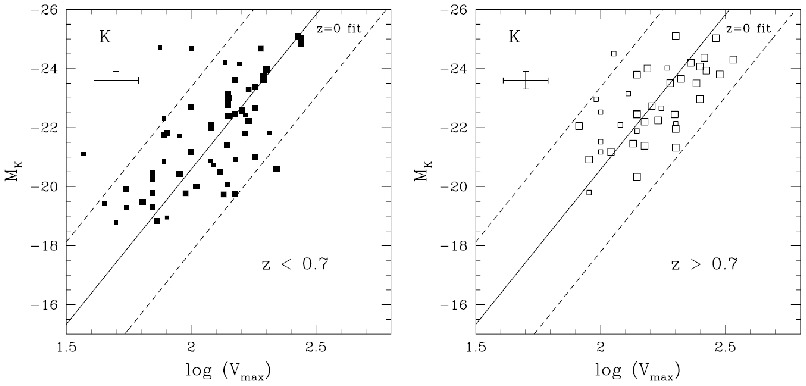}
\includegraphics[width=0.4\textwidth]{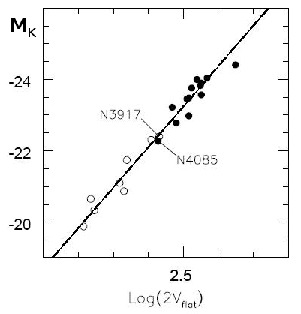}
\includegraphics[width=0.45\textwidth]{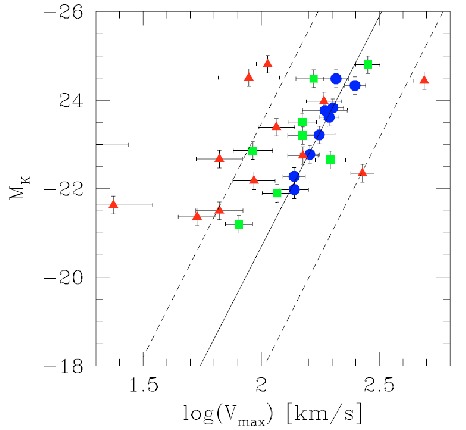}
\caption[TFR scatter at low and high redshift]{{\it Top}: TFR scatter at high redshift from \citet{2005ApJ...628..160C}. {\it Bottom} (from left to right): 1.) TFR scatter in the local Universe from \citet{1997PhDT........13V,1997ASPC..117..190V}\citep[see also][]{2001ApJ...550..212B}. 2.) TFR scatter in the distant Universe using integral field spectroscopy from \citet{2006A&A...455..107F}. Here, the blue circles are the galaxies classified as rotating disk, red triangles are galaxies with complex kinematics, and green squares are galaxies presenting a perturbed rotation. If we take into account only the rotating disk galaxies the significant scatter observed in the distant TFR (bottom-left) disappears.}
\label{TFRscatterHighZ}
\end{figure}

\begin{figure}[!h]
\centering
\includegraphics[width=0.6\textwidth]{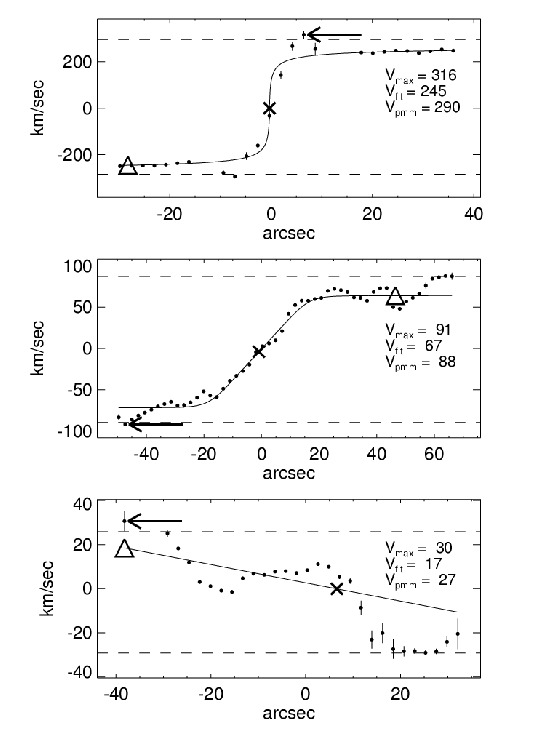}
\caption[Detection of anomalous rotating curves in local galaxies]{Detection of anomalous rotating curves in local galaxies by \citet{2002AJ....123.2358K}. We can see the waves in the part corresponding to the "plateau" of the rotating curves.}
\label{Kannappan_2002rotating}
\end{figure}

In addition to the study of properties of individual galaxies with the IMAGES survey, we have also studied fundamental relations such as the Tully-Fisher relationship \citep[TFR,][]{1977A&A....54..661T}\citep[see][for a cosmological view]{1997ApJ...477L...1G}, which relates the luminosity and the rotation velocity of disk galaxies (see figure \ref{TFRscatterHighZ}). Pioneering studies \citep{1996ApJ...465L..15V,1997ApJ...479L.121V}, using slit spectroscopy, have claimed a strong evolution of the TFR as a function of cosmic time . Out to z $\sim$ 1, the B-band TFR has been found to evolve by $\sim$0.2-2 mag \citep[e.g.,][and references therein]{2007MNRAS.375..913P}. This brightening of the B-band TFR can be explained by the enhanced star-formation rates at higher redshifts \citep{2001ApJ...557..165F,2004MNRAS.355...64F}, but it is still a matter of debate. \citet{2005ApJ...628..160C}, instead of B-band, used the K-band to retrieve the TF relationship. The importance of the K-band is that it is less affected by the dust and is a better tracer of the stellar mass than the B-band. Conselice et al. do not find significant evolution in either the stellar mass or K-band TFR's slope or zero point. However, the most striking evolution of the TFR is provided by its large scatter at high redshifts \citep[][see figure \ref{TFRscatterHighZ}]{2005ApJ...628..160C}, which may be related to disturbed kinematics in distant galaxies that can be infer by extrapolating the observed kinematics of local galaxies \citep[e.g.,][see figure \ref{Kannappan_2002rotating}]{2004AJ....127.2694K}. Another possible cause of this scatter at high z is the sample selection (see subsection \ref{ConsiderationsforClassificationSubsect}). In any case, it is crucial to find an explanation to the discrepancy between the previous studies, in order to understand whether the TF relationship is preserved at higher redshift, or if it is only a property of local galaxies, as well as better understant the morpho-kinematic evolution of galaxies. \\

\begin{figure}[!h]
\centering
\includegraphics[width=1.0\textwidth]{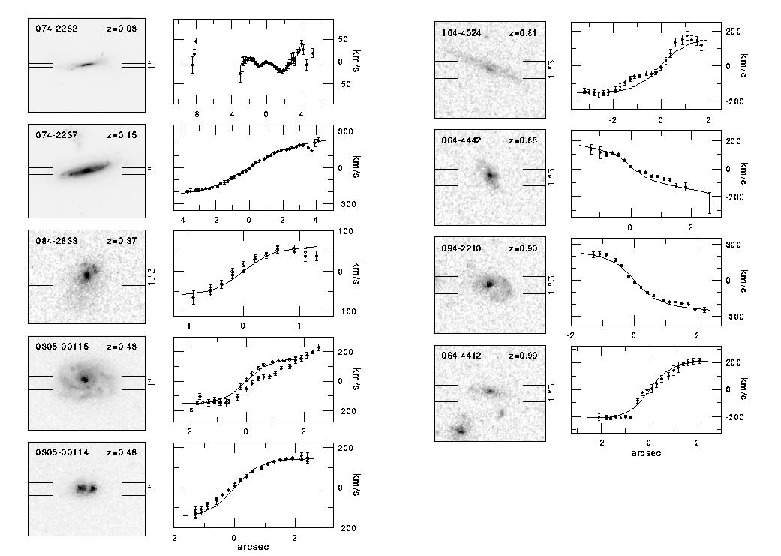}
\caption[Rotating curves with the respective galaxy images and its long-slit arrangement]{Rotating curves obtained by \citet{1996ApJ...465L..15V}. One column shows the spectroscopic long-slit arrangement overplotted on the galaxy image, while the other column shows the respective rotating curves.}
\label{Vogt1997rotating}
\end{figure}

As the above kinematic investigations at intermediate redshift were based on long (or multi) slit spectroscopy, the large dispersion (e.g., in the TFR) may thus be related to instrumental effects. Indeed, measuring the kinematics of distant galaxies precisely and robustly is necessary to study galaxy evolution. However, this is often beyond what can be achieved with classical long-slit spectroscopy (see figure \ref{Vogt1997rotating}). The morphologies and kinematics of distant galaxies are often complex, and their small sizes make it very difficult to precisely position and align the slit. Both limitations can be overcome with integral-field spectroscopy. 3D spectroscopy appears to be a pre-requisite to sample the whole velocity field of individual galaxies in order to limit the uncertainties related to the major axis determination and those related to galaxy interactions. Such effects have been already tested for nearby spirals by comparing Fabry-Perot to slit measurements which show that slit measurements can easily provide under or over estimates of the maximum velocity by factors reaching 50$\%$ \citep{1995A&AS..113...35A}. Worse yet is that many of the high redshift studies do not even place the slits along the major axis in order to preserve the multiplex advantage of slitlet MOS. Physically speaking, much larger uncertainties are expected for distant galaxies which are actively forming stars (up to rates larger than 100M$_{\odot}$/yr) and for which the frequency of interactions is very common. Only velocity fields from 3D spectroscopy can disentangle the pure circular rotation from chaotic/non circular motions \citep{2006A&A...455..107F}. Even if rotation curve anomalies, such as those detected by \citet{2002AJ....123.2358K} (see figure \ref{Kannappan_2002rotating}) using long slit spectroscopy on nearby galaxies, can be explained by velocity field distortions due to minor or major mergers taking place during the history of each galaxy, it is unclear whether such details can be identified on distant galaxies by using long slit spectroscopy. These last years, it has then become urgent to investigate the full 3D kinematics of distant emission line galaxies. \\

\begin{figure}[!h]
\centering
\includegraphics[width=1.0\textwidth]{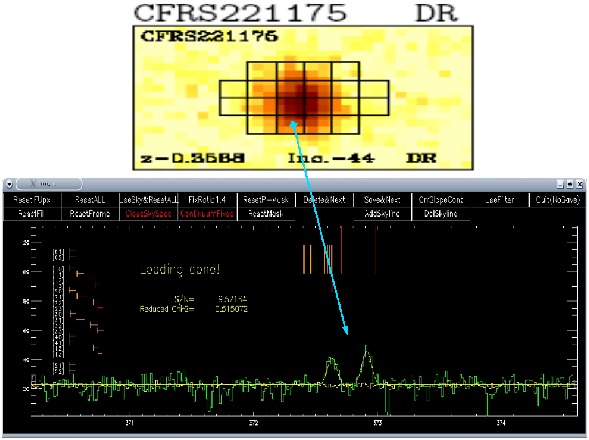}
\caption[GIRAFFE line analysis interface to measure the line wavelength and its FWHM]{Measurement of the line wavelength and its full width at half maximum (FWHM) using the GIRAFFE line analysis interface.}
\label{Line_interface2}
\end{figure}

\begin{figure}[!h]
\centering
\includegraphics[width=0.6\textwidth]{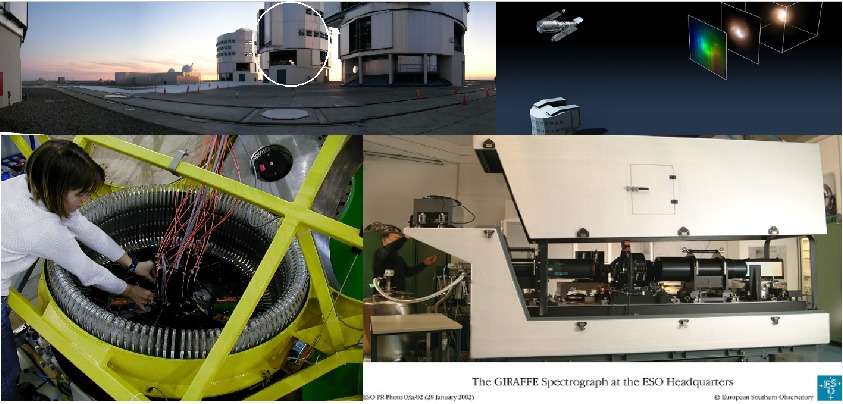}
\includegraphics[width=0.6\textwidth]{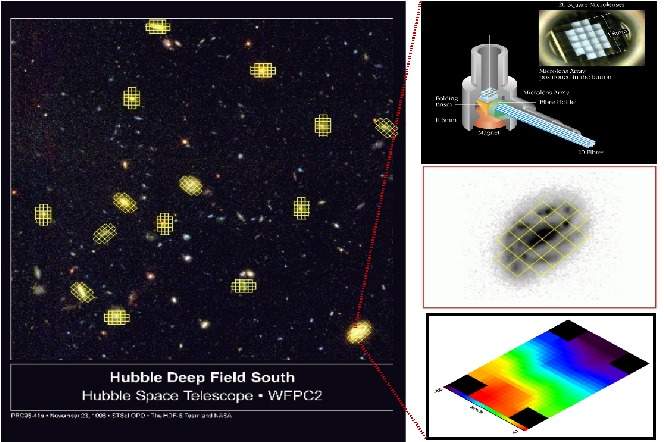}
\caption[Illustration of FLAMES/GIRAFFE, installed on the UT2 at VLT]{FLAMES/GIRAFFE, installed on the UT2 at VLT, can get the integral field spectroscopy of 15 objects at the same time. For this, in its IFU mode, it is equipped with 15 IFUs. Each one composed of 6x4 micro-lens (0.52 arsec/lens) which spatially cut the light of the galaxy, allowing to obtain 20 spectrum of different parts of the galaxy. Like this, the internal kinematics of distant galaxies can be decrypted.}
\label{FlameGirrafe_Ifus}
\end{figure}

3D spectroscopic observations can provide us with detailed kinematics, including accurate measurements of velocities in order to study physical relations such as the Tully-Fisher one \citep{1977A&A....54..661T} \citep[see also][]{2006A&A...455..107F}. VLT/GIRAFFE is one of the instrument which provides us with such an IFU mode. This special mode, thanks to the high spectral resolution (R $\sim$ 9 000) of GIRAFFE, allows us to disentangle the [OII]$\lambda \lambda$3726,3729 doublet (see figure \ref{Line_interface2}). For these reasons, the kinematics observations of galaxies by IMAGES are made with the integral field spectrograph FLAMES/GIRAFFE. This spectrograph is installed on the UT2 at VLT, and has a wavelength range from 370 nm to 950 nm. The Integral Field Unit (IFU) mode of this instrument is composed of 15 IFUs with R=9 000. Each IFU is made of 6x4 micro-lens (0.52 arsec/lens) that can be displayed in the telescope focal plan. This makes it the only instrument in the world able to get the integral field spectroscopy of 15 objects at the same time (see figure \ref{FlameGirrafe_Ifus}).\\

Therefore, given the above tantalizing hints for galaxy evolution, the obvious complexity of the problem, and the contradictory results, the IMAGES survey has applied a different approach, using the state of the art in instrumentation, in order to study the kinematic of distant galaxies. We then consider that the most significant way forward is to conduct a dynamical study of a large sample of galaxies with excellent integral spectrophotometry (VLT/FORS2 observations) and photometric data (in terms of depth and angular resolution). The combination of the integral field spectrograph FLAMES/GIRAFFE with high resolution images from HST/ACS allow us to take advantage of estimating the evolution in both the morphological properties and kinematics of galaxies of a large sample of galaxies. As a result, with IMAGES, we will be in a position to evaluate the relative importance of the different physical processes (merging, dynamical friction between sub-components, monolithic collapse) driving the evolution of the ensemble of galaxies spanning a wide range of morphologies.\\

\section{The survey}

The Intermediate MAss Galaxies Evolution Sequence (IMAGES) survey has been constructed to get a maximum of information in a representative sample of distant galaxies (see below, and figure \ref{IMAGESinstruments}). Through the ESO large programme IMAGES, it has been intended to (1) establish the kinematic and morphological evolution of galaxies; (2) test the different physical processes leading to the present day Hubble sequence; (3) explain the star formation history of individual galaxy; and (4) test the evolution of mass-metallicity relation, angular momentum, size, and mass. The analysis involves combining multi-instrumental (principally GIRAFFE, FORS2 and HST) observations being carried out of a sample of intermediate mass galaxies, selected mainly by their rest-frame J$_{AB}$ magnitude, derived from the multi-wavelength photometric data available. \\

The galaxies in the IMAGES survey are situated in the Chandra Deep Field South (see figure \ref{Chandra_DFS}). The principal reason to chose this field is that it has been observed with a large number of instrument telescopes (WFI/MPI-2.2, EMMI/NTT, MIPS/Spitzer, RGS and EPIC/XMM-Newton, et.). Thus, photometry and morphology of galaxies in the visible comes from deep HST observations with the ACS camera (0.03"/pix). The coverage in the mid-IR at 3.6 $\mu$m, 8 $\mu$m, 12 $\mu$m and 24 $\mu$m is provided by the deep survey of the satellite Spitzer (with its instruments IRAC and MIPS). Observations in the radio range (3 mm, 12 mm, 3 cm, 6 cm, 13 cm and 20 cm) from the Australian Telescope Compact Array and the Very Large Array (VLA), and X observations (0.1 to 15 keV, which correspond to $\sim$ 0.1 nm - 12.4 nm) from the Chandra X Ray Observatory and XMM-Newton telescope are also public. Finally, many spectroscopic observations from ground telescopes provide the redshift of objects (e.g. VVDS, FORS2, K20, and our survey IMAGES). This field is then one of the largest deep field of sky observed by such a multitude of instruments. \\

\begin{figure}[!t]
\centering
\includegraphics[width=1.0\textwidth]{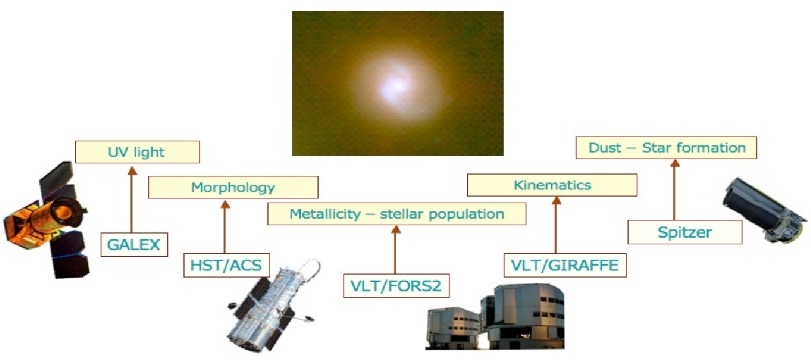}
\caption[Instruments used by IMAGES and the related observed quantities]{Instruments used by IMAGES and the related observed quantities. These observations allow us to study the kinematics, the morphology, emission/absorption line properties, and the integral chemical properties of galaxies. }
\label{IMAGESinstruments}
\end{figure}

\begin{figure}[!t]
\centering
\includegraphics[width=0.4\textwidth]{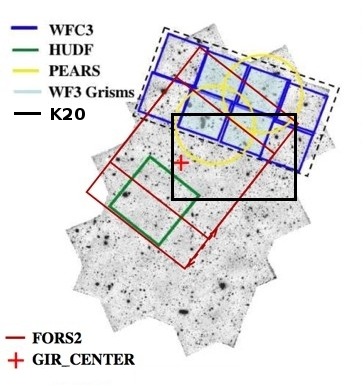}
\includegraphics[width=0.5\textwidth]{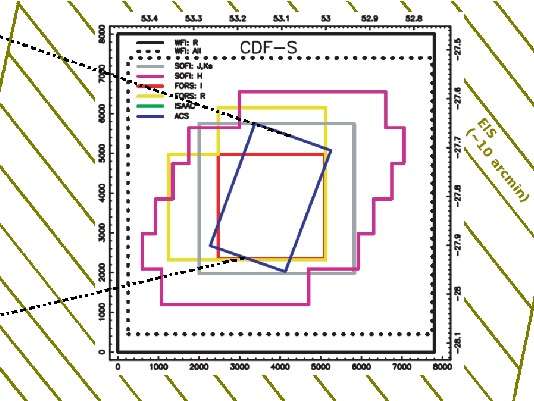}
\caption[Schematic view of the Chandra Deep Field-South survey]{{\it Right:} schematic view of the CDF-S (Chandra Deep Field - South) amongst other surveys. {\it Left:} a zoom of the HST/ACS observation.}
\label{Chandra_DFS}
\end{figure}

\section{Sample selection and representativeness}

The IMAGES sample has been selected with the aim of containing a representative sample of galaxies having an absolute magnitude in J band brighter than -20.3. This corresponds to galaxies with a mass roughly > 1.5x10$^{10}$ M$_{\odot}$. Galaxies with stellar masses from 3x10$^{10}$ to 3x10$^{11}$ M$_{\odot}$ are also known as intermediate mass galaxies, and at intermediate redshift (0.4 $\leqslant$ z $\leqslant$ 1), where the cosmic SFR density reaches its peak \citep{2004Natur.428..625H}, they host most of the star formation \citep{2005A&A...430..115H}. Furthermore, these galaxies include most (from 65 to 85$\%$) of the present-day stellar mass, which implies that about half of the stellar mass in intermediate-mass galaxies was formed since z = 1 \citep{2003ApJ...587...25D}. Therefore, in the frame of the IMAGES survey, and complementary to the existing data from low resolution spectroscopic follows up (VVDS, K20, GOODS/FORS2/VIMOS), \citet{2007A&A...465.1099R} observed a sample of galaxies with VIMOS/VLT to define the full IMAGES sample. This includes galaxies with 0.4 $\leqslant$ z $\leqslant$ 0.9, and M$_{J}$ $\leqslant$ -20.3. \\

\begin{figure}[!h]
\centering
\includegraphics[width=0.5\textwidth]{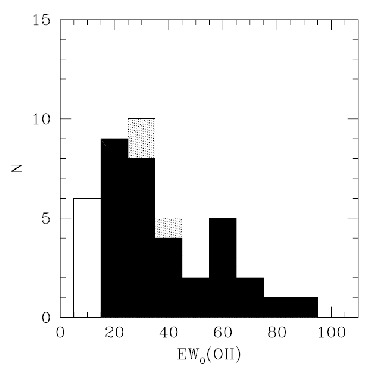}
\caption[EW$\lbrack OII \rbrack$ histogram of the GTO galaxy sample]{EW$\lbrack OII \rbrack$ histogram for GTO sample galaxies \citep{2006A&A...455..107F}. For galaxies with EW($\lbrack OII \rbrack$)$\geqslant$ 15 $\mathring{A}$, GIRAFFE can retrieve their kinematics after 8 h of integration time (black region). The major difficulty studying the detected galaxies is then related to the flux strength. The three galaxies with EW$[OII] \geqslant$ 15 $\mathring{A}$, but with too spatially concentrated emission are displayed in grey. The open histogram includes objects with EW0($\lbrack OII \rbrack$) < 15 $\mathring{A}$.}
\label{GTOsampleEW}
\end{figure}

Concerning the kinematics, galaxies have been also selected by having enough [OII]$\lambda$3727 line emission, to be detected by GIRAFFE in a reasonable integration time. This line is inside the spectral range of GIRAFFE observations for galaxies at intermediate redshifts (0.4 $\leqslant$ z $\leqslant$ 0.9). Based on the GIRAFFE GTO by \citet{2006A&A...455..107F}, intermediate redshift galaxies with an EW[OII]$\lambda$3727 > 15 $\mathring{A}$ can be detected by GIRAFFE (see figure \ref{GTOsampleEW}). These galaxies having a significant line emission flux represent 60$\%$ of intermediate mass galaxies at intermediate redshifts \citep{1997ApJ...481...49H}. Then, the kinematic observations of the IMAGES sample are based on the analysis of emission lines, mainly the [OII]$\lambda$3727 emission line \citep[][and appendix \ref{KinematicStudies}]{2008A&A...477..789Y}.  \\

\begin{figure}[!t]
\centering
\includegraphics[width=0.6\textwidth]{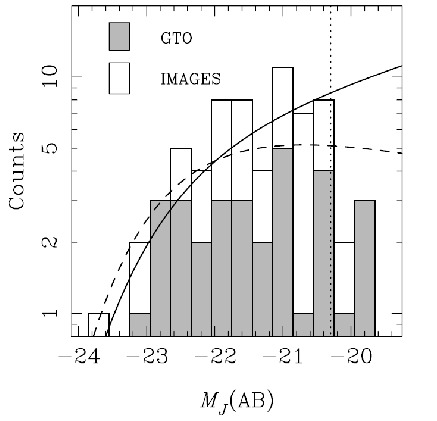}
\caption[IMAGES galaxy sample representativeness]{Number counts (in logarithmic scale) of selected galaxies versus AB absolute magnitude in J-band. In this figure, the GTO sample refers to \citet{2006A&A...455..107F}; the IMAGES sample refers to this paper. The vertical dotted line indicates the limit of the IMAGES program. Two luminosity functions derived from \citet{2003A&A...402..837P} are shown (full line: z = 0.5; dashed line: z = 1). The galaxies of IMAGES sample have redshifts ranging from z = 0.4 to z = 0.75. This implies that the combined sample of 63 galaxies with MJ(AB) $\leqslant$ -20.3 is representative of galaxies with stellar masses larger than $1.5x10^{10} M_{\odot}$ at z $\sim$ 0.6.}
\label{IMAGESrepresentative}
\end{figure}

In order to test the representativeness of the sample we have compared the distribution of the J-band absolute magnitudes of the IMAGES sample, combined with the GTO sample of 35 galaxies \citep{2006A&A...455..107F}, with the luminosity function (LF) from \citep{2003A&A...402..837P}. Both samples (GTO+IMAGES) can be merged because the selection of Flores et al. is equivalent to that of IMAGES and because both studies use the same instrumental set-ups and integration times \citep{2008A&A...477..789Y}. Applying our criteria of MJ(AB) $\leqslant$ -20.3, 0.4 $\leqslant$ z $\leqslant$ 0.9, and EW([OII]) $\geqslant$ 15 $\AA$, the IMAGES+GTO sample is left with 63 galaxies with high quality data to carry out our analysis. The luminosity distribution of the combined sample and the luminosity function at redshift of 0.5 and 1 are compared in figure \ref{IMAGESrepresentative}. Kolmogorov-Smirnov tests support that the complete sample follows the luminosity function in the z = 0.4 - 0.75 range at >99.9$\%$ confidence level \citep{2008A&A...477..789Y}. Furthermore, the combined two samples include galaxies from 4 different fields, namely the CDFS, HDFS, CFRS03h and CFRS22h. It is then unlikely that our conclusions are affected by cosmological variance due to field-to-field fluctuations, and related to large scale structures \citep[see][]{2007A&A...465.1099R,2004ApJ...600L.171S}. Our sample is thus representative of MJ(AB) $\leqslant$ -20.3 emission line galaxies at z = 0.4 - 0.75. At this redshift range, it is the only representative sample of distant galaxies with measured 3D kinematics that exist nowadays.

\section{Results of the IMAGES survey}

IMAGES broad results are described within a series of four papers. The first one \citep[][see subsection \ref{YangSubsection}]{2008A&A...477..789Y} presents the kinematic study of distant galaxies, the second \citep{2008A&A...484..159N} makes the morpho-kinematic comparison in galaxies 6 Gyr ago, the third \citep{2008A&A...484..173P} studies the Tully-Fisher relation and its evolution since z$\sim$0.6, and the last one \citep{2008A&A...492..371R} investigates the chemical abundance evolution and the mass/metallicity relation of galaxies over the past 8 Gyr. I participated in two of them (IMAGES II and IMAGES IV, see sections \ref{Neichelsection} and \ref{Rodriguessection}). In these studies, my contribution is related to the morphological analysis of galaxies and its relation with their kinematics, the measure of R$_{1/2}$, inclinations, disk radius, EW[OII], among others. Disk radius and inclinations are also important to establish the rotation velocities of galaxies. More details about the morphological analysis can be found in chapter \ref{MorphoAnalysisChapter}. I have also studied the morphology of distant and local galaxies \citep{2010A&A...509A..78D}, using this survey and the SDSS (Sloan Digital Sky Survey). This last study is detailed in chapter \ref{MorphoAnalysisChapter}. \\

To better understand our results in subsections \ref{Neichelsection} and \ref{Rodriguessection},  I must introduce the kinematic studies of the IMAGES survey and its conclusions. The first study of intermediate redshift galaxy kinematics using FLAMES/GIRAFFE was carried out by \citet{2006A&A...455..107F}. They presented the first study of a statistically meaningful sample of 35 intermediate-mass galaxies at z = 0.4-0.7, using the integral-field multi-object spectrograph GIRAFFE at the ESO-VLT (see figure \ref{FlameGirrafe_Ifus}). They defined a classification scheme to distinguish between rotation and kinematic perturbations, which may stem from interactions and mergers, based on the 3D kinematics and high-resolution HST imaging (see appendix \ref{KinematicStudies}). \citet{2008A&A...477..789Y} enlarges this first sample to finally count 63 distant galaxies kinematically studied with FLAMES/GIRAFFE. As mentioned in the previous section, this final sample is representative of intermediate redshift galaxies with MJ(AB) $\leqslant$ -20.3. In addition, I have been analyzing new galaxies (see appendix \ref{KinematicStudies}). The methodology we use to retrieve the galaxy kinematic through FLAMES/GIRAFFE is well described in \citet{2006A&A...455..107F} and \citet{2008A&A...477..789Y} (see also appendix \ref{KinematicStudies} for more details about the 3D kinematic analysis). \\

    \subsection{Kinematic evolution of galaxies since z=1}
    \label{YangSubsection}

\begin{figure}[!h]
\centering
\includegraphics[width=1.0\textwidth]{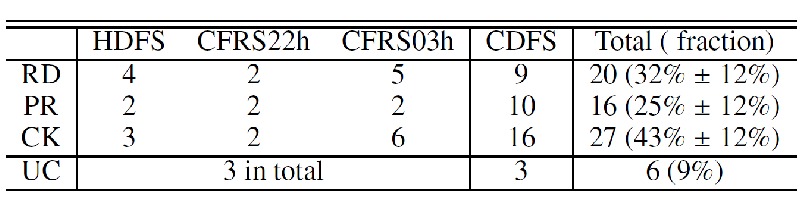}
\caption[Kinematic evolution of galaxies over the past 6 Gyr]{Results from \citet{2008A&A...477..789Y} showing the large percentage of galaxies with perturbed or complex kinematics 6 Gyr ago.}
\label{KineEvolGal}
\end{figure}

Applying the same kinematic analysis than \citet{2006A&A...455..107F} to a larger and representative sample of 63 distant galaxies from the IMAGES survey, \citet{2008A&A...477..789Y} find 32$\%$ of rotating disk, 25$\%$ of perturbed rotation galaxies, and 43$\%$ of galaxies with complex kinematics (see figure \ref{KineEvolGal}).

Considering that galaxies with significant emission line flux (EW([OII]$\lambda$3727) > 15 $\mathring{A}$) represent 60$\%$ of intermediate mass galaxies at 0.4 < z < 1.0 \citep{1997ApJ...481...49H}, and assuming the rest 40$\%$ of galaxies to be dynamically relax, they conclude that at this redshift range at least 41$\pm$7$\%$ of galaxies has perturbed or complex kinematics. This kinematic state of galaxies at z$\sim$0.6 is very different from that observed 6 Gyr later (in the local Universe). Indeed, in the local Universe, roughly 70$\%$ of intermediate mass galaxies are dynamically relax spiral galaxies, while a few percents correspond to irregular, compact, or merger galaxies \citep[][and references therein]{2007AJ....134..579F}. Therefore, galaxy kinematics are among the most rapidly evolving properties between z$\sim$0.6 and today. Which is the physical process driving to such fast evolution?, Is this kinematic evolution corroborated by a morphological evolution?

\subsection{The morpho-kinematic correlation 6 Gyr ago}
\label{Neichelsection}

First, we have studied the correlation between morphology and kinematics at z$\sim$0.6, as well as compared different morphological classification methods (Concentration-Asymmetry, Gini-M20, our own).

This was the principal objective in the second IMAGES paper \citep{2008A&A...484..159N}, in which we defined a robust morphological classification methodology. By applying this new classification to a kinematically studied sample of distant galaxies (see previous subsection), we show that the retrieved morphologies are well correlated to their kinematic classification. We find that, at z$\sim$0.6, 4/5 spiral galaxies are rotating disk, and more than 4/5 peculiar galaxies are not in a dynamical equilibrium. Therefore, using this new method to morphologically classify a galaxy guarantees, in a high percentage (see figures 5 and 7 in the paper), that the morphology is representative of the dynamical state of that galaxy. This correlation represents a new important tool to understand the galaxy formation and evolution. However, such a correlation is only guaranteed when our classification methodology is applied. Indeed, we further show that automatic morphological classifications over-estimate the number of spiral rotating disks by a large fraction \citep[see also][]{2003AJ....126.1183C,2005ApJ...620..564C}.\\

\fbox{\textbf{ See: \itshape{Neichel et al. 2008, A$\&$A, 484, 159N}}}

\subsection{Chemical evolution of intermediate-mass galaxies since z$\sim$0.7}
\label{Rodriguessection}

Since the previous studies reveal the agitated history of galaxies over the past 6 Gyr, one could expect a chemical abundance evolution of galaxies since that epoch. This is the principal goal of our IMAGES IV paper, where we find that galaxies at z$\sim$0.6 are, on average, two times less metal-rich than local galaxies at a given stellar mass, confirming the earlier results of \citet{2006A&A...447..113L}. Combining our results with other studies, we also find that the metal abundance of the gaseous phase of galaxies is evolving linearly with time since z$\sim$3 until now. Such result implies that $\sim$30$\%$ of the stellar mass of local galaxies have been formed through a gas supply. It could be interesting to compare this result to the evolution of the baryonic TFR (see section \ref{Hammer_etal_2009Section}). \\

\fbox{\textbf{ See: \itshape{Rodrigues et al. 2008, A$\&$A, 492, 371R}}}

\section{Beyond the IMAGES survey}

The IMAGES survey has then allowed us to develop new techniques and methodologies in order to preserve a well correlation between kinematics and morphology of distant galaxies, as well as unravel the evolution of some of their physical properties. \\

At this stage, it becomes naturally necessary to undertake a more complete study of the galaxy morphology, from the distant Universe to the local one, in order to estimate its evolution. Different reasons support it. First, morphology is the result of the different physical process that form a galaxy. Second, we dispose now of a classification method that guarantee the correlation between kinematics and morphology. Third, this method must be extended to include all the morphological types amongst the Hubble sequence. Fourth, there is still no work to date that has studied representative samples of local and distant galaxies at the same time, applying the same methodology to equivalent local and distant observation data. Fifth, there is still no work to date able to determine the past Hubble sequence. \\

\chapter{Reconstructing the distant Hubble sequence}
\label{MorphoAnalysisChapter}
\begin{flushright}
"El mundo era tan reciente que muchas cosas carec{\'i}an de nombre, y para mencionarlas hab{\'i}a que se{\~n}alarlas con el dedo." \\ Gabriel Garc{\'i}a M{\'a}rquez, Cien a{\~n}os de soledad.
\end{flushright}

\minitoc

This chapter represents the heart of my work during my thesis. Here, I present a complete description of the morphological analysis and methodology we have developed during the past three years, as well as our results concerning the morphological evolution of galaxies since the past 6 Gyr. Further results of my work are also described. Then, I finish by a discussion about the different morphological classification methods, and I argue why our method is more reliable than the standard ones to retrieve the morphological properties of galaxies (see also chapters 3 and 4.).\\

As mentioned in chapter 2, the Hubble sequence was created in 1936, and has been used during the last century by different authors to classify the Schmidt plates and the first CCD images of local galaxies. In the 1990s, with the arrival of the HST, astronomers began to obtain good resolution images of distant galaxies. Only high resolution images allow us to quantify the structural parameters such as, e.g.,  B/T \citep[][see subsection \ref{The_structural_parameters}]{1998ApJ...500...75L}\footnote{Other kind of parameters have also been created (see chapter 3 for more information and details). Briefly, we can mention the Concentration (C) parameter \citep{1994ApJ...432...75A} which roughly correlates with the bulge over disk ratio (B/D), and the Asymmetry (A) parameter \citep{1996ApJS..107....1A} which divides the sample between irregulars and more symmetric objects. More recently two other parameters have been introduced: the Gini coefficient \citep{2003ApJ...588..218A} which is a measure of the relative distribution of galaxy pixel flux values, and the M20 parameter \citep{2004AJ....128..163L}, which is the relative second-order moment of the brightest 20$\%$ of a galaxy's pixels.}, because only a high spatial resolution allow us to deblend the information coming from different regions of the galaxy, in particular the bulge and the disk. In the case of WFPC images, it was considered that only in galaxies with an R$_{1/2}$ (see below) larger than 3 kpc \citep[known as the compactness limit;][]{2005ApJ...632L..65M}, it could be possible to differentiate such structures. It is interesting to note that one of the best ground images at that time were those from the CFRS, having a 0.8" FWHM with a pixel size of 0.207" \citep{2001ApJ...550..570H}. Such FWHM (and pixel size) represents 5.5 kpc (and 1.43 kpc) at z=0.65, implying that the classification (compactness) limit of 3 kpc would be strictly and largely smaller than the smallest spatial resolution element imposed by the PSF (limit$_{R_{1/2}} \sim$ FWHM/2), and thus the structural deblending would not take into account the scatter induced by the measurement process. Here, the importance of HST images. Astronomers then started to classify samples of such galaxies using the traditional methods. Nevertheless, authors also started to create new methodologies to recover the structural parameters of distant and local galaxies, and study their evolution \citep[][and reference therein]{1998ApJ...499..112B,1996AJ....111..174F}. In this context, during the past decade, the structural parameters are usually recovered using "deconvolution"\footnote{what is really done is a convolution of the model with the PSF. This convolution focus on reproducing the different effects caused by the telescope optics and/or the atmosphere.} softwares, such as GIM2D and Galfit \citep[][respectively]{2002ApJS..142....1S,2010AJ....139.2097P}. \\

Our methodology to classify galaxies according to their morphology began to be developped in 2004 by \citet{2004A&A...421..847Z}, introducing, for the first time, the color information (color maps) to classify galaxies (see subsection \ref{ColorInfo}). It was later improved by \citet{2008A&A...484..159N}, where we introduce a decision tree to schematise our morphological classification process. Finally, I have extended this methodology in order to account for all the morphological types amongst the Hubble sequence, to include the analysis of bulge and disk center positions during the determination of the morphology, and to allow the study of galaxies with smaller sizes. This last, thanks to the spatial resolution of the images at which we have access. Indeed, with the arrival of the HST/ACS images we have been able to push the compactness limit to 1 kpc (see subsection \ref{The_structural_parameters}), thanks to their 0.108" FWHM and their pixel size of 0.03". Such FWHM (and pixel size) represents 0.75 kpc (and 0.2 kpc) at z=0.65. Our classification (compactness) limit is thus larger than the smallest spatial resolution element (limit$_{R_{1/2}} \sim$ 1.33$\times$FWHM), with a well sampling of the PSF, and consequently takes also into account the scatter due to the measurement process. For a further comparison with observations of the local Universe, the SDSS images present a 1.4" FWHM (and a pixel size of 0.396). In the local Universe, e.g. at z=0.025, such value corresponds to 0.74 kpc (and 0.2 kpc), which matches perfectly with HST/ACS observations of distant galaxies (see above). \\

\begin{figure}[!h]
\centering
\includegraphics[width=0.8\textwidth]{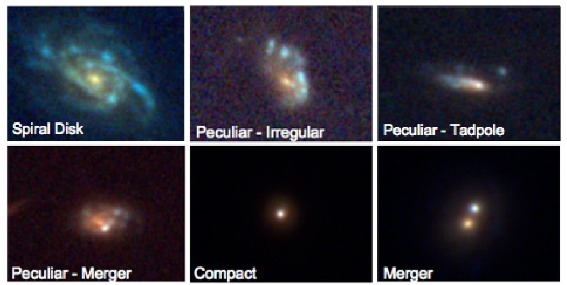}
\includegraphics[width=0.8\textwidth]{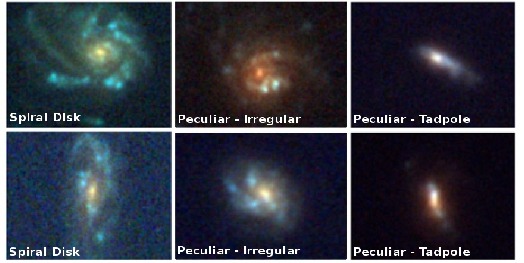}
\caption[Color images of distant galaxies showing their huge variety of shapes and physical properties]{Distant galaxies. They present a huge variety of shapes and physical properties.}
\label{Distant_galaxiesNei2008}
\end{figure}

As mentioned in section \ref{PhyHubbleSeq}, the morphological classification by eye, which has been used till recently \citep[e.g.,][see chapter 3]{1998ApJ...499..112B}, can generate an important scatter when different astronomers do it \citep{1995MNRAS.274.1107N}. It will depend a lot on "{\it the experience or pleasure of who is doing it}". Then, a less subjective method must be constructed to guide the classifier, and let the morphological classification process be reproducible with a very little scatter\footnote{This is related to the precision of the method.}. On the other hand, the method must not let the process be totally automatic ("blind") because it could affect its accuracy, and introduces systematic errors, given the different shapes and independent physical properties a galaxy can have (especially, those that can be found in distant galaxies, see figure \ref{Distant_galaxiesNei2008}). As we cannot, e.g., describe the appearance of a person by only measuring the size of his nose, it is similarly difficult to determine the morphology of a galaxy only using one automatic measurement. Even if this last could be an easier and faster method, we lose the goal in the study of distant galaxies which can be cited as: "understanding the evolution and state of each galaxy".\\

During my thesis, I worked in order to improve the method on how galaxies are morphologically classified. In the next sections I will describe the sample used in our study (local and distant), the different parameters used to classify the galaxies, and the methodology to recover them. \\

\section{Our galaxy sample}

We have worked with two samples of galaxies. One made up with local galaxies (0.0207 $\leqslant$ z $\leqslant$ 0.030) observed by the SDSS, and the other one composed of distant galaxies (0.4 $\leqslant$ z $\leqslant$ 0.8) located in the CDFS field and observed by the HST/ACS \citep[see][at the end of this chapter for more details]{2010A&A...509A..78D}. Both samples have been set up following one simple selection criterion: the absolute magnitude in J band brighter than -20.3. Furthermore, during the construction of our samples we have ensured that each galaxy have the information necessary for our analysis: good quality spectrum which includes the [OII]$\lambda$3727, and high resolution images in at least three optical bands. The first would allow us to distinguish between starburst and quiescent galaxies. The second, to construct color maps and three color images. In this way, we got a local sample of 116 galaxies, and a distant sample of 143 galaxies.\\

Finally, we show that both samples are representative of the galaxies at each respective epoch. To do so, we compare our samples with the luminosity functions obtained for local and distant galaxies. Kolmogorov-Smirnov tests indicate probabilities larger than 94$\%$ that our samples and the corresponding luminosity functions are drawn from the same distribution. \\

\section{The parameters retrieval: analysis of the observables}

\begin{figure}[!h]
\centering
\includegraphics[width=0.8\textwidth]{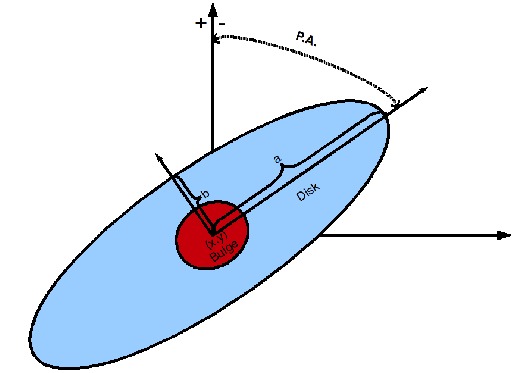}
\caption[Illustration of different galaxy parameters]{Galaxy parameters. They are define as follow: a = semi-major axis, b = semi-minor axis, (x,y) = center coordinates, P.A. = position angle (which goes from 0 to 90 degrees. Positive within the left of the grand {\it y} axis, and negative in the other direction), ellipticity = 1-(b/a), inclination = arccos(b/a).}
\label{GalaxyParameters}
\end{figure}

As discussed in section \ref{PhyHubbleSeq}, one of the basis of any morphological classification method is the definition of the observables to be taken into account (images bands, color images, etc.), as well as fixing the importance of the different morphological details (bulge, disk, bar, ring, etc.). This is important in order to extract the underlying physical processes (kinematics, composition, physical events, etc.). In this context, we have developed a method that allows us to correlate morphology and kinematics of distant galaxies (see section \ref{Neichelsection}), and with other physical properties and events (e.g., mergers). This method is summarized by a decision tree illustrated later in section \ref{MorphoMethod}. In our methodology we use different tools and all the available information to morphologically classify each galaxy:

\begin{itemize}
\item GALFIT parameters: these include the bulge and disk centers, magnitudes, radius, inclinations, P.A.s, and sersic index (see figure \ref{GalaxyParameters}).
\item GALFIT model and residual images: to gauge the fitting process, and carefully examine the different structures.
\item Error image: to weight the different structures detection.
\item disk-bulge-galaxy profile: to analysis the light profiles of different components.
\item bulge-to-total light ratio (B/T): to distinguish the different morphological types.
\item half light radius (R$_{1/2}$): to study the light concentration and define the compactness.
\item color-map: to measure the color of individual features. It is constructed using two images in different filter bands. 
\item color image: to examine the small-scale structures. It is constructed using three images in different filter bands. 
\item Original images in three different bands: together with the color map and the color image, to identified dusty or star forming regions, as well as examine the different structures.
\end{itemize}

In the next sections I describe in more details the extraction of the parameters, the construction of the different images (or maps), and their utility inside our decision tree.\\

       \subsection{The light profile analysis}
       \label{StructPara_LightProfiAna}

This step in our methodology is essential to determine the majority of all the parameters we use to classify each galaxy (the five first items above). Therefore, special attention must be given to it for a good understanding of what we are retrieving. \\

We use the software galfit \citep{2010AJ....139.2097P} to carry out two-dimensional modeling of each galaxy in order to obtain their structural parameters. For this we use optical high resolution images: HST/ACS z band images in the case of distant galaxies, and SDSS r band for the local galaxies. These two bands are equivalent due to the K-correction: r band $\sim$ (z band)/(1+($z$=0.65)). Galfit models the galaxy image as a linear combination of a bulge and a disk, using a well established analytic model for each of the two components (see below), and convolves them with a PSF in order to extract the parameters free of instrumental effects. An important caveat in this approach is the underlying assumption that a galaxy can be represented as a linear combination of simple, smooth analytic functions \citep{2007A&A...469..483R}. Real galaxies are known to be much more complex, with the presence of spiral arms, bars and central point sources, etc. Despite these reservations, galaxy fitting algorithms have been demonstrated to be successful in the past for the purpose of quantifying the galaxy morphological parameters \citep{2002AJ....124..266P,2002ApJS..142....1S}. \\

To model the galaxy bulge+disk intensity profile we use the combination of two Sersic laws \citep{1968adga.book.....S}\footnote{The elegance of the Sersic profile is that it forms a continuous sequence from a Gaussian (n = 0.5) to an exponential (n = 1) and a de Vaucouleurs (n = 4) profiles simply by varying the exponent. It is then very useful for modeling bulge, pseudo-bulges, ovals, bars and flat disks; the smaller the index n is, the faster the core flattens within r < r$_{e}$, and the steeper the intensity drop beyond r > r$_{e}$.},

   \begin{eqnarray}
       \Sigma (r) = \Sigma_{e} exp[-\kappa[({\frac{r}{r_{e}}})^{\frac{1}{n}}-1]]
       \label{SersicProfileEq}
    \end{eqnarray}

with r$_{e}$ being the half light radius, $\Sigma_{e}$ the surface brightness at r$_{e}$, $\kappa$ a constant\footnote{$\kappa$ is coupled to {\it n} such that half of the total flux is always within r$_{e}$. Then, for {\it n} $\gtrsim$ 2, $\kappa \thickapprox$ 2{\it n} - 0.331. At low {\it n}, $\kappa$ flattens out toward 0 and is obtained by interpolation \citep{2010AJ....139.2097P}.},  and {\it n} being the sersic index. The difference between the bulge and the disk lies then in the sersic index. For the bulge the sersic index is let free\footnote{the classic de Vaucouleurs profile \citep{1948AnAp...11..247D} that describes a number of galaxy bulges is a special case of the Sersic profile when n = 4 ($\kappa$ = 7.67). Nevertheless, some bulges may be better represented by other Sersic profiles \citep{2009ApJS..182..216K,2008ASPC..396..297K,2007MNRAS.381..401L,2004ARA&A..42..603K,1994MNRAS.267..283A,1991ApJ...378..131K,1989MNRAS.237..903S,1978ApJ...223L..63K}.}, while for the disk the sersic index is fixed to 1. This last case is equivalent to the exponential function \citep{1970ApJ...160..811F}\footnote{Freeman showed that dynamically "hot" stars in galaxies make up spheroidal bulges having a roughly de Vaucouleurs light profile, whereas "cold" stellar components make up the more flattened, rotationally supported, exponential disk region \citep[see also][and the references cited in the previous footnote]{1996A&AS..118..557D,1996A&A...313...45D}.},

   \begin{eqnarray}
       \Sigma (r) = \Sigma_{0} exp[- \frac{r}{r_{d}}]
    \end{eqnarray}

with $\Sigma_{0}$ being the central surface brightness and r$_{d}$ the e-folding scale length. \\

\begin{figure}[!h]
\centering
\includegraphics[width=0.8\textwidth]{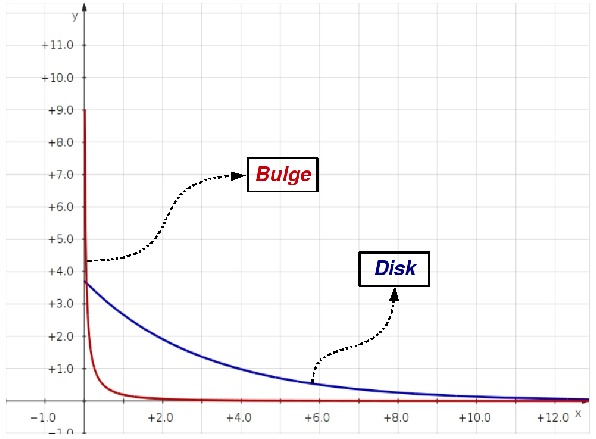}
\caption[Examples of bulge+disk Sersic profiles]{Examples of Sersic profiles. In red a Sersic profile with n=4, which correspond to a bulge. In blue a Sersic profile with n=1, modeling a disk.}
\label{SersicProfiles}
\end{figure}

The galaxy model is then constructed as a linear combination of these two components which includes some free\footnote{All parameters like r$_{e}$, r$_{d}$, {\it n}, position angle etc., with the exception of the sky background, are allowed to be free in the fitting process, so that galfit can explore the full range of parameter space to find the global minimum. In such a case, we avoid galfit to pull into a local minimum due to constraints put in by hand.} parameters such as:
\begin{itemize}
\item[-] the sky background value $bkg$, which is assumed to be constant across the galaxy image. This is a good approximation because the images we use are 15{\it x}15 arsec$^{2}$ large for distant galaxies ($\sim$ 103{\it x}103 kpc$^{2}$ at z=0.65), and 198{\it x}198 arsec$^{2}$ large for local galaxies($\sim$ 100{\it x}100 kpc$^{2}$ at z=0.025). 
\item[-] the ellipticities $\epsilon_{bulge}$ and $\epsilon_{disk}$ of the bulge and the disk isophotes respectively,
\item[-] and their position angles $\theta_{bulge}$ and $\theta_{disk}$.
\end{itemize}

Galfit needs a first guess of such parameters, as well as the magnitude, radii and position angle (PA) of each component, to run the fitting. The SExtractor software \citep{1996A&AS..117..393B} applied to our galaxies allows us to recover these initial guesses\footnote{We use SExtractor software for this task because it performs source detection and photometry, and it is able to deblend sources using flux multithresholding, as well as estimate a sky value using the complete image and not just an specific area.}. Then, galfit iterates the best analytic model of the galaxy using the PSF of the observation (see below) and comparing systematically with the observed galaxy image. The values of the free parameters are determined iteratively by minimizing the reduced $\chi_{red}^{2}$ defined as usual \citep[see][for more details]{2010AJ....139.2097P},

   \begin{eqnarray}
       \chi_{red}^{2} = \frac{1}{N_{d.o.f}} \sum_{x=1}^{nx} \sum_{y=1}^{ny} \frac{(galaxy_{x,y}-model_{x,y})^{2}}{\sigma_{x,y}^{2}}
    \label{EqChi2}
    \end{eqnarray}

where N$_{d.o.f.}$ is the number of degrees of freedom, galaxy$_{x,y}$ and model$_{x,y}$ are the counts in the pixel {\it (x, y)} of the galaxy and the model image respectively, $\sigma_{x,y}$ is the noise in the pixel {\it (x, y)} and {\it nx},{\it ny} are the number of pixels in the galaxy image in the {\it x} and {\it y} direction respectively.\\

\begin{figure}[!h]
\centering
\includegraphics[width=0.75\textwidth]{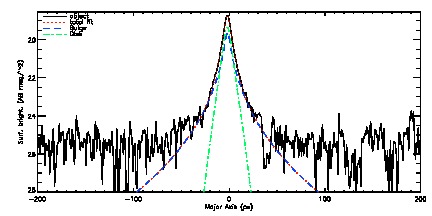}
\includegraphics[width=0.75\textwidth]{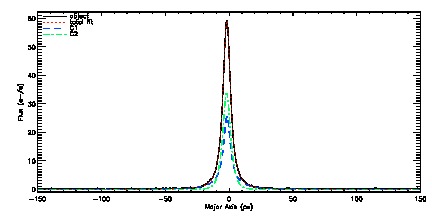} \\
\includegraphics[width=0.215\textwidth]{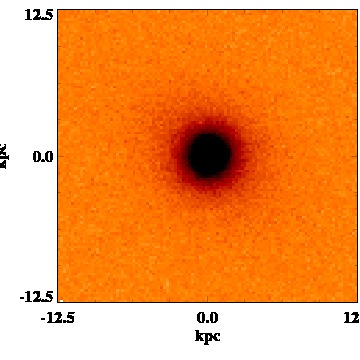}
\includegraphics[width=0.215\textwidth]{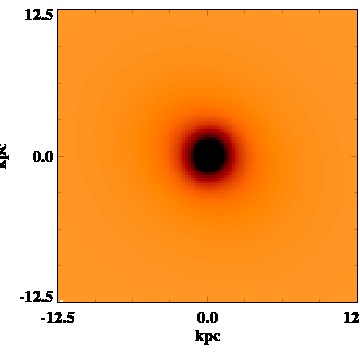}
\includegraphics[width=0.215\textwidth]{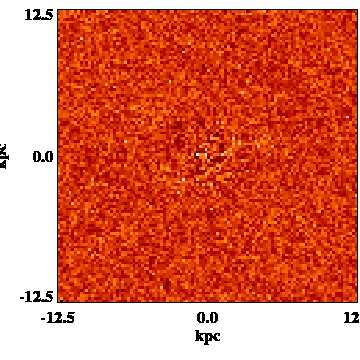}
\includegraphics[width=0.27\textwidth]{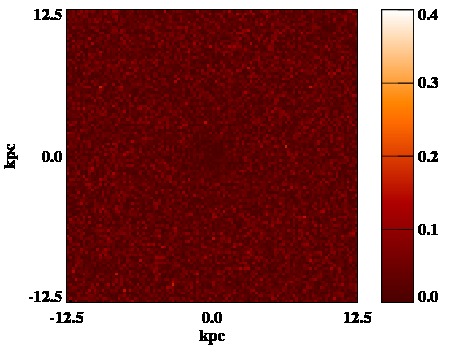}
\caption[Example of a galfit simulation for one galaxy in our sample]{Example of a galfit simulation for one galaxy in our sample. {\it Top}: display of the components profiles over-plotted to the original galaxy profile in magnitude units.  {\it Middle}: same as before, but the profile are in flux units. {\it Bottom}: from left to right we have (a) the observed galaxy image; (b) the best model; (c) the residual image, constructed by taking the difference between the real and the modeled light distribution; (d) the "error" image.}
\label{GalfitSimulation}
\end{figure}

{\textbf{\itshape The point-spread-function (PSF)}}\\

The PSF is a mathematical function describing the response of an imaging system to a point source. \\

\begin{figure}[!h]
\centering
\includegraphics[width=0.6\textwidth]{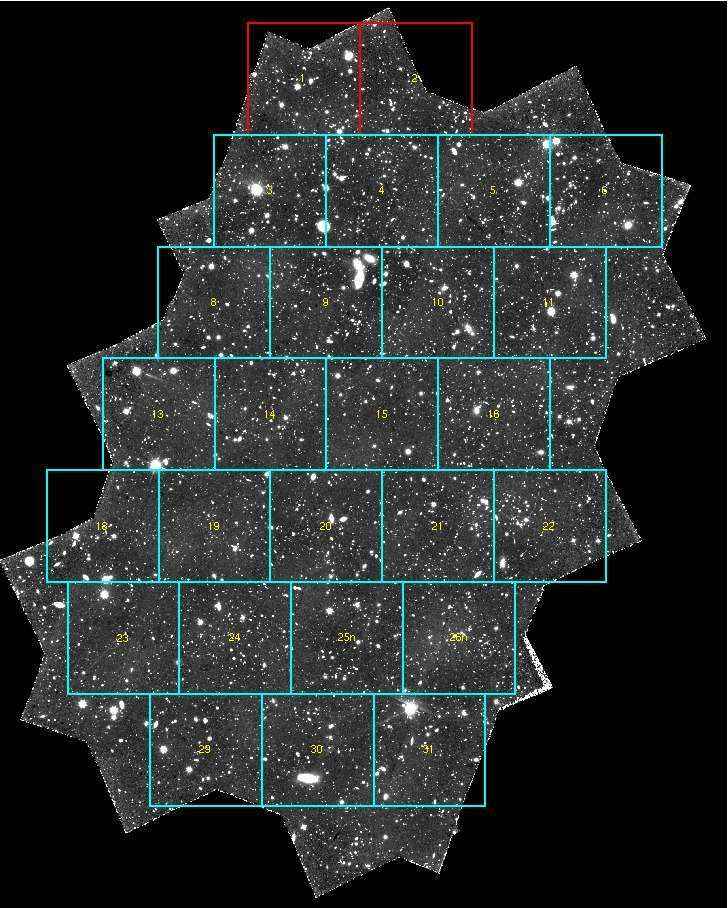}
\caption[HST/ACS GOODS image from the ESO/GOODS project]{HST/ACS GOODS image from the ESO/GOODS project \citep{2010A&A...511A..50R} (Credit: http://archive.eso.org/cms/).}
\label{HST-ACS_GOODS_image}
\end{figure}

Therefore, a special attention is paid when determining the PSF to be used during the fitting of each galaxy because it is fundamental to well simulate the original galaxy image. As it has been mentioned before, the convolution with a PSF is necesary to reproduce the blurring caused by the telescope optics and/or the atmosphere. In our work, the star closest to a given galaxy is used to determine the PSF for the purpose of bulge+disk decomposition. Stars are the most pristine characterization of the PSF in any given astronomical observation, as they have gone through the same optics as well as the reduction pipeline as the studied galaxies. In this purpose, especially in the case of HST/ACS drizzled images, real stars are better than simulated ones in order to take into account all the optics and reduction effects. It must be remembered that during the drizzling process a linear reconstruction of the individual astronomical images is done before their combination. This implies that, even if the algorithm preserves photometry and resolution, each input images is weighted according to the statistical significance of each pixel. The effects of geometric distortion on both image shape and photometry are also removed. In addition, individual images must be rotated according to their orientation, and it is possible that, in the presence of cosmic rays, the images are combined by a dithering procedure. As a consequence, the simulated PSFs may not be able to reproduce all these steps, which could become complicated enough. Only bright (non saturated) and well behaved stars, free of neighbours and other contamination were used as candidate PSF stars in our work following the study carried out by \citet{2007A&A...469..483R}.\\

{\textbf{\itshape The $\sigma$-image and the $\chi_{red}^{2}$ role to retrieve the best model}}\\

In the above scheme, an accurate estimation of the noise model $\sigma_{x,y}$ to be used in Eq. \ref{EqChi2} is also crucial for a robust determination of the derived structural parameters of a galaxy. As the $\chi_{red}^{2}$ "measures" how far the modeled image is compared to the original light distribution, it is generally used to estimate the quality of the fit. Then, the role of the $\sigma$-image is to give relative weights to the pixels during the fit, because, even with the absence of other eventualities (cosmic ray hits, dead/hot pixels, flatfielding problems) in the galaxy image, all pixels are not equal due to the counting uncertainties of the photon (Poisson) statistics. In an image, pixels with high signal have smaller uncertainties (fractionally speaking) than pixels with fewer counts. Therefore, the fit can not be weighted by the uncertain data at the same degree as the more certain ones. Thus, in Eq. \ref{EqChi2} each point is normalized by the uncertainty at each pixel (the $\sigma$-image), following the Poisson weighting as it is the natural scheme for number counting. \\

Due to the fact that Poisson statistics applies to electrons, when constructing the $\sigma$-image, pixel values must be in electrons. Thus, we use the GAIN, NCOMBINE, and EXP\_TIME parameters of the image, to convert all the pixel values into electrons. Then, following the Galfit manual\footnote{http://users.obs.carnegiescience.edu/peng/work/galfit/galfit.html} and \citet{2002AJ....124..266P,2010AJ....139.2097P}, we do the noise calculation by making the sum of the Poisson uncertainty at each pixel with the other noise sources in quadrature (e.g. detector readnoise). Finally, we convert everything back to the same units as the input data (counts/second). \\

Statistically, if the model would be a perfect fit of the data, the difference (galaxy$_{x,y}$ - model$_{x,y}$) in Eq. \ref{EqChi2} would be roughly the size of $\sigma_{x,y}^{2}$, on average, over infinite number of observations at each pixel. Therefore, when the sum is computed over all the pixels, and then divided by the number of pixels $\chi_{red}^{2}$ will be roughly equal to 1. In galaxy fitting, you almost never get $\chi_{red}^{2}$ equal to 1 because galaxies are not perfect ellipsoids with the profiles we try to force on them. Nevertheless, it is still important to get the right sigma image because if not it might be hard to interpret the profile parameters such as the Sersic index (n).\footnote{http://users.obs.carnegiescience.edu/peng/work/galfit/CHI2.html}\\

\begin{figure}[!h]
\centering
\includegraphics[width=0.5\textwidth]{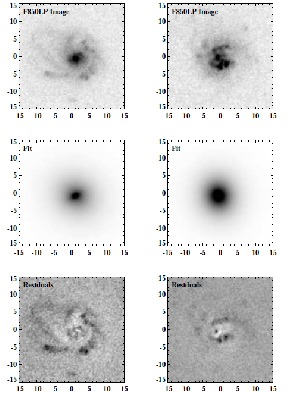}
\caption[Comparison of the $\chi_{red}^{2}$ parameter of two different galaxies]{$\chi_{red}^{2}$ parameter of two different galaxies \citep{2008A&A...484..159N}. From top to bottom: original image, Galfit model image, and residual image. For these two objects the software gives a similar confidence level in the fitting process ($\chi_{red}^{2} \thickapprox$ 1.4). For the galaxy on the left, the $\chi_{red}^{2}$ is high due to spiral arms, whereas it is due to the two clumps located at the southeast of the center for the galaxy on the right.}
\label{Chi2_parameterAlone}
\end{figure}

As a result, we found that the $\chi_{red}^{2}$ parameter alone is not sufficient for deciding whether a fit is reliable or not (see figure \ref{Chi2_parameterAlone}). In fact, the main limitation of parametric methods is that they use an "a priori" assumption on the light distribution, making them unsuitable for treating patchy light distribution often found in spiral galaxies, for example, spiral arms. A spiral galaxy with prominent arms unavoidably produces a $\chi_{red}^{2}$ > 1 because the spiral pattern is taken as if it was noise during the fitting. However, we can gauge the appropriateness of the fit by carefully examining the residual map, the "error" map, and the flux and magnitude profiles shown in figure \ref{GalfitSimulation}. Then, the combination of the $\chi_{red}^{2}$ value and the careful examination of the aboves allows us to adequately recognise the successful fits.\\

       \subsection{Structural parameters}
       \label{The_structural_parameters}

\begin{figure}[!h]
\centering
\includegraphics[width=1.0\textwidth]{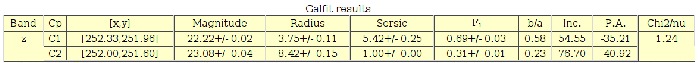}
\caption[Example of Galfit result parameters from our morphological analysis]{Example of Galfit result parameters from our morphological analysis. C1 represents the bulge and C2 the disk. $[x,y]$ indicates the center position in pixel values, $Sersic$ is the sersic index value ($n$ in equation \ref{SersicProfileEq}), F$_{i}$ is the component flux to the total flux ratio (B/T and D/T respectively), $b/a$ is the semi-minor axis to the semi-major axis ratio, and $Inc.$ is the inclination in degrees.}
\label{GalfitResultParameters}
\end{figure}

Figure \ref{GalfitResultParameters} shows an example of the structural parameters values resulting from the fitting procedure detailed in the previous subsection. Furthermore, a series of images and plots, also resulting from the galfit fitting procedure, are shown in figure \ref{GalfitSimulation}. \\

{\textbf{\itshape Bulge-to-total light ratio (B/T)}}\\

Once the best fit galaxy model has been obtained, we calculate the bulge fraction B/T, defined as:

   \begin{eqnarray}
       \frac{B}{T} = \frac{flux_{bulge}}{flux_{bulge}+flux_{disk}}
    \end{eqnarray}

where the flux$_{bulge}$ and flux$_{disk}$ are calculated by integrating the bulge and the disk profiles respectively over the major axis using the best fit parameters obtained by galfit (see figures \ref{GalfitResultParameters} and \ref{GalaxyParameters}). This bulge fraction B/T is found to be broadly correlated with the traditional Hubble type of a galaxy in the sense that an early type galaxy has a high B/T ratio and a late type spiral has a low value \citep{1985ApJS...59..115K,1986ApJ...302..564S,2007MNRAS.381..401L}. It is this ratio that we use for characterizing the morphology of our galaxy sample (see section \ref{MorphoMethod}).\\

Another parameter not included in figure \ref{GalfitResultParameters}, because it does not result from the galfit fitting, is the half light radius. Nonetheless, our morphological classification method also takes advantage of this parameter (see decision tree in figure \ref{DecisionTree}).\\

{\textbf{\itshape Half light radius (R$_{1/2}$)}}\\

This is the radius within which half the light (luminosity or flux) of the galaxy is contained.\\

\begin{figure}[!h]
\centering
\includegraphics[width=0.4\textwidth]{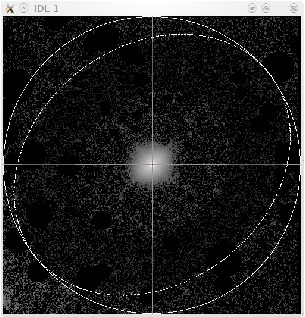}
\includegraphics[width=0.4\textwidth]{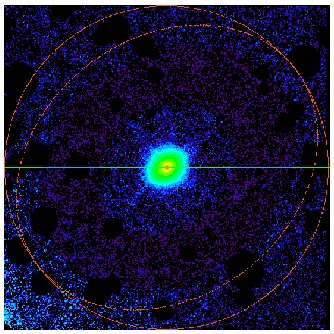}\\
\includegraphics[width=0.8\textwidth]{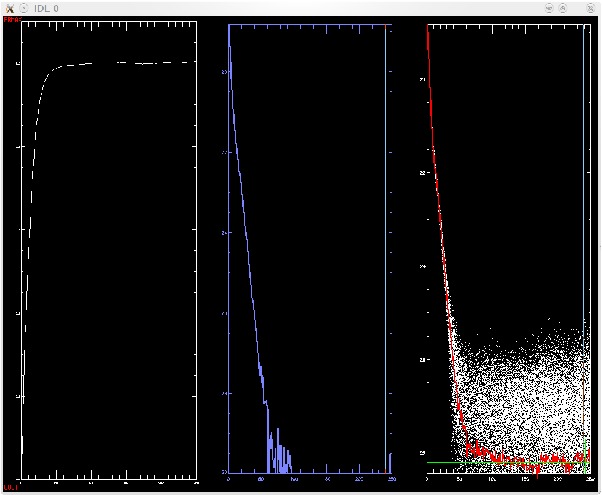}
\caption[Display of our methodology to measure R$_{1/2}$]{Measuring R$_{1/2}$. {\it Top}: display of the progression of the ellipses with different radii overplotted on the galaxy image. These animations are used to better visualize, in real time, the ellipses regions being mesured, as well as the selected region to derive the sky background value. {\it Bottom}: from left to right we have (a) the plot of the flux content within ellipses of different radii with a step of one pixel, (b) the magnitude profile, (c) the flux profile.}
\label{RhalfCode}
\end{figure}

To measure the half-light radius, we have developed our own IDL procedure, which allows us to visually analyze different profiles of the galaxy at the same time: (1) the flux and magnitude profiles; and (2) the flux content within ellipses of different radii with a step of one pixel (the PA used here is the one retrieve by galfit). The first profiles let us determine a reasonable sky value and the last one gives us the half light-radius when a real "plateau" is visualized (see figure \ref{RhalfCode}). \\

       \subsection{The color information}
       \label{ColorInfo}

The galaxy color is information of great significance, as it is the comparison of the light in different wavelengths coming from the galaxy. It can be obtained by comparing its magnitude in two or three different bands. It is even possible to get a color of each galaxy features (similar, for example, to a picture of someone taken by a modern photographic camera), if the images have a high spatial resolution and their alignment is better than 0.16 pixels \citep{2004A&A...421..847Z}. We made use of the available multi-band high-resolution images to construct two sets of color maps, used in our morphological classification methodology.\\

{\textbf{\itshape Color maps}}\\

We substract, pixel by pixel, the magnitude in two observed bands using our own algorithm that allows us to estimate colors and their uncertainties \citep{2004A&A...421..847Z}.\\

A color map is a difference of two images in a logarithm scale. Here, we can ignore the possible scaling constant which will not affect the final results. Given one image in the blue rest-frame band with flux F$_{B}$ and another one in the red rest-frame band with flux F$_{R}$, the color image is defined as

   \begin{eqnarray}
       Y = log F_{B}-log F_{R}
   \label{ColorMapEquation}
   \end{eqnarray}

\begin{figure}[!h]
\centering
\includegraphics[width=0.32\textwidth]{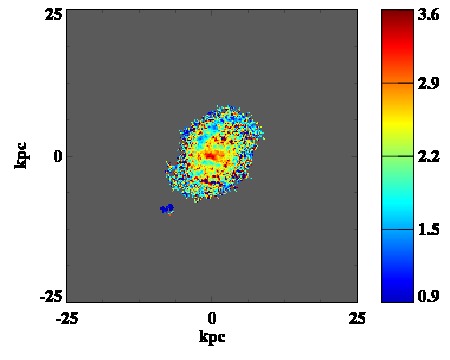}
\includegraphics[width=0.32\textwidth]{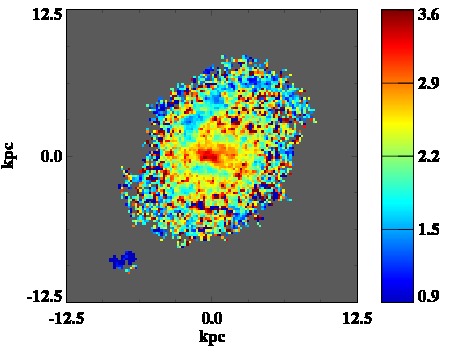}
\includegraphics[width=0.32\textwidth]{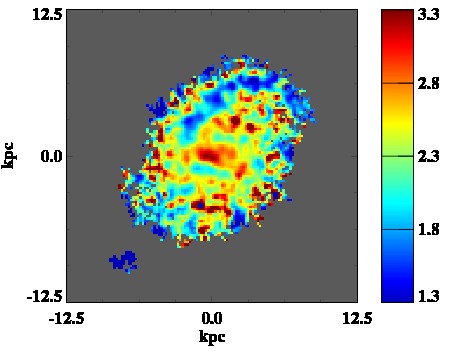}\\
\includegraphics[width=0.32\textwidth]{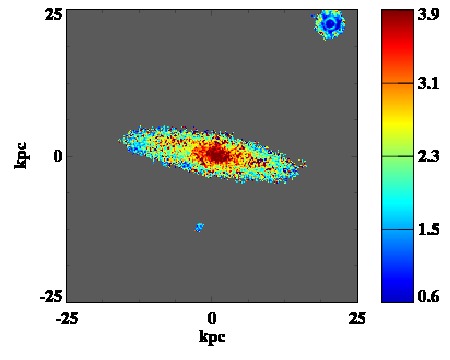}
\includegraphics[width=0.32\textwidth]{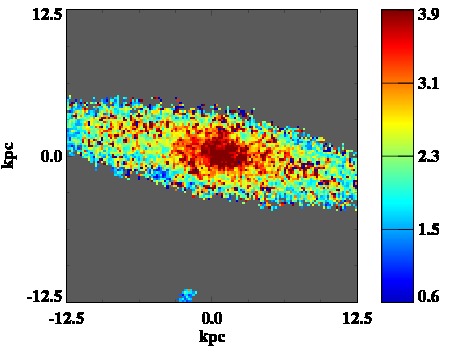}
\includegraphics[width=0.32\textwidth]{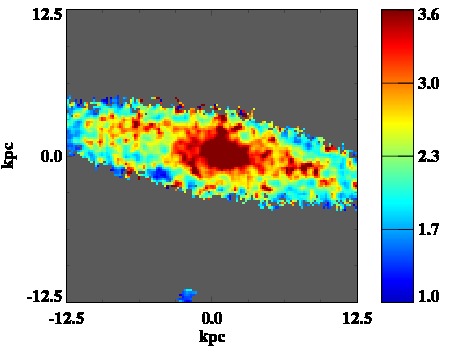}
\caption[Example of color maps for two different galaxies in our morphological analysis methodology]{Example of color maps for two different galaxies in our morphological analysis methodology. {\it Left}: 50x50 kpc color map. {\it Middle}: 25x25 kpc color map. {\it Right}: Smoothed color map. This last is created from the 25x25 kpc color map, applying a 2x2 boxcar average. The scales are adjusted to have a nice view of the whole galaxy.}
\label{ColorMap}
\end{figure}

In this relation, the flux is a function of the position in an image, including signals of background and object. Then, applying SExtractor to the red rest-frame image and taking only the detected pixels through a 3-$\sigma$ threshold,  we obtain a color map similar to figure \ref{ColorMap}.\\

It is important to note that the color information in the color map is normalized, pixel by pixel, by the S/N ratio. It allows to create a real and homogeneous color information of the galaxy. If the color map does not take into account the S/N ratio, it could give incorrect information. One example is the extreme case where one structure, within a resolution element, is detected with the B band, but it is not detected with the R band. In such a case, without a S/N normalization, equation \ref{ColorMapEquation} will be equal to infinity (-$\infty$, if we do not ignore the scaling constant).\\

{\textbf{\itshape Color images}}\\

\begin{figure}[!h]
\centering
\includegraphics[width=0.45\textwidth]{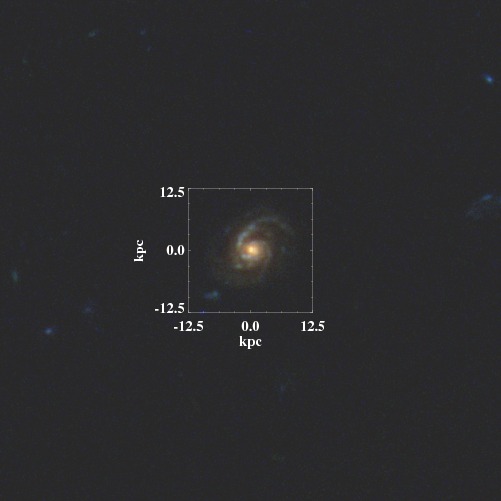}
\includegraphics[width=0.45\textwidth]{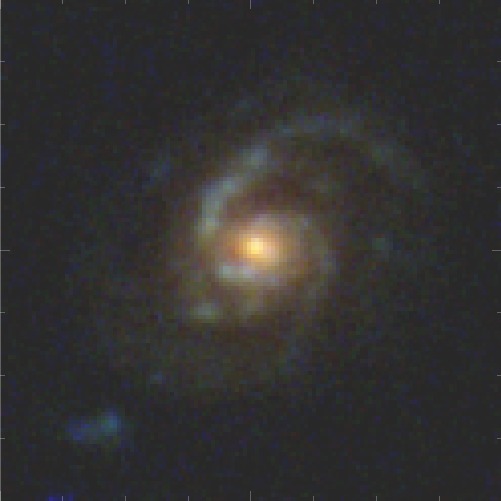} \\
\includegraphics[width=0.45\textwidth]{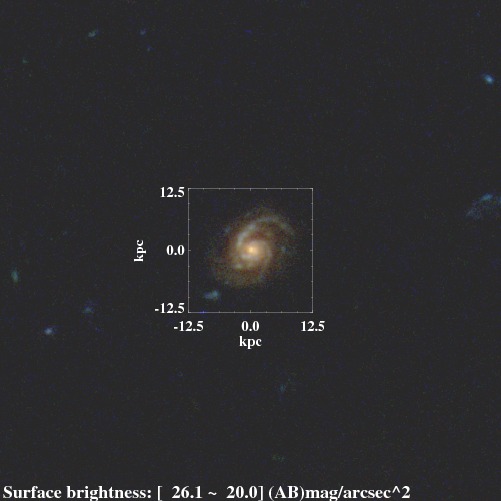}
\includegraphics[width=0.45\textwidth]{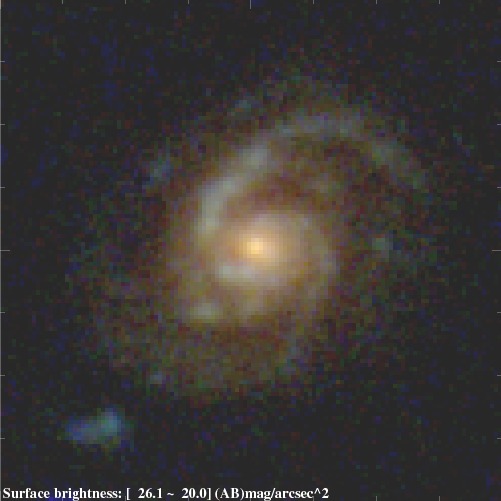}
\caption[Examples of color images we use in our morphological analysis methodology]{Examples of color images we use in our morphological analysis methodology. {\it Top}: linear intensity scale. {\it Bottom}: Logarithm intensity scale. {\it Left}: image with a spatial scale of 50x50 kpc. {\it Right}: image with a spatial scale of 25x25 kpc.}
\label{3ColorImages}
\end{figure}

With "color image" I refer to the combination of images in three different bands. For a better visualisation of the difference scale structures of each galaxy, we create three color images with different intensity and spatial scales. In the intensity space, linear and logarithm scales are used. For the spatial case, we compute 25x25 kpc and 50x50 kpc images, as shown in figure \ref{3ColorImages}.

       \subsection{Bar inspection}

Bars are primarily stellar structures in the center of galaxies and it has been shown that they play an important role in the dynamic and structure formation of galaxies. Models show that the nonaxisymmetric bar potential induces large-scale streaming motions in the stars and gas \citep[][and references therein]{1993RPPh...56..173S,1992MNRAS.259..345A,1992MNRAS.259..328A}. They are mostly suggested to be efficient at driving gas into the galactic center and changing star orbits to form a central bulge-like structure \citep{1990A&A...233...82C,2004ApJ...604L..93D}, as well as inducing the formation of spiral arms. \\

A bar structure can be recognized by a characteristic change of ellipticity ($\epsilon$) and position angle (P.A.) from an analysis of isophotes over the entire image of the galaxy. The existence of a bar will cause a monotonically increase of the isophote ellipticity, while the P.A. remains relatively constant.  At the end of the bar, the ellipticity drops sharply, and the P.A. changes as the isophotes belonging to the underlying disk are fitted \citep{2005A&A...435..507Z,2003ApJ...592L..13S,1997AJ....114..965R}.\\

\begin{figure}[!h]
\centering
\includegraphics[width=0.48\textwidth]{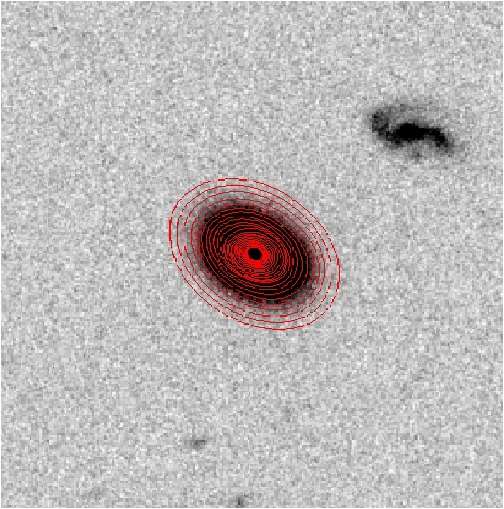}\\
\includegraphics[width=0.47\textwidth]{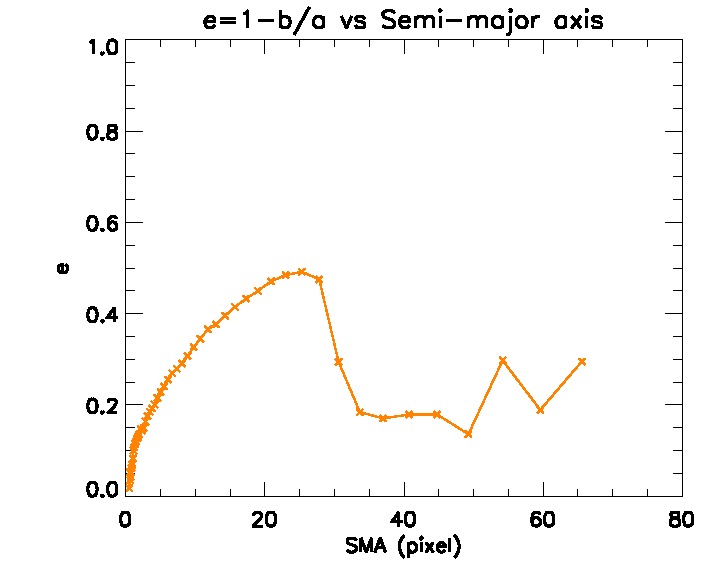}
\includegraphics[width=0.47\textwidth]{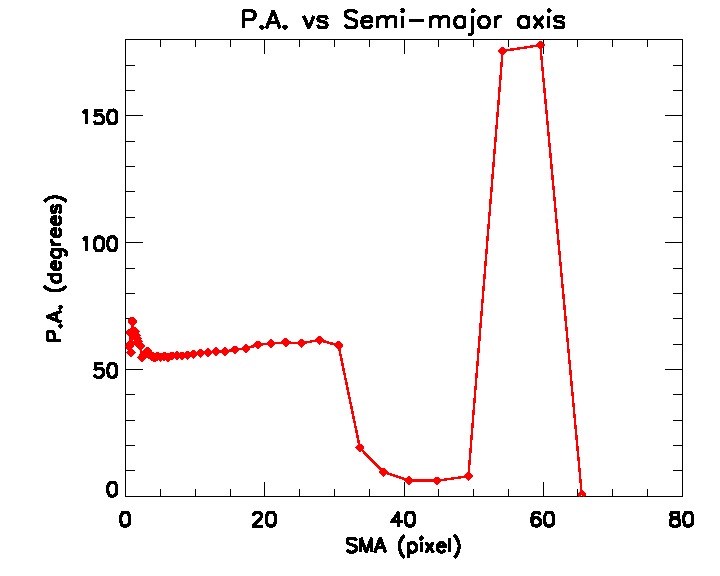}
\caption[IRAF/Ellipse analysis showing a bar detection]{Example of the isophotal fitting using the tool IRAF/ELLIPSE, and the respective ellipticity and P.A. analysis showing a bar detection.}
\label{Ellipse_analysisBarDetection}
\end{figure}

We use the tool IRAF/ELLIPSE to perform the analysis of the surface brightness distribution via fitting the isophotes with ellipses as shown in figure \ref{Ellipse_analysisBarDetection}. \\

Plotting the ellipticity and P.A. of the isophotal ellipses can thus allow us to detect a bar as in figure \ref{Ellipse_analysisBarDetection}. A bar detection is validated when the increase of ellipticity exceeds 0.2, and then drops by at least $\delta \epsilon$ > 0.1 \citep[][and references therein]{2008ApJ...675.1141S}. Moreover, the position angle profile, after keeping a relatively constant value over the bar region, must change by at least 10$^{\circ}$ as the isophotes enter the disk. \\

It is easy to suspect that this technique is quite ineffective in detecting bars in highly inclined spiral galaxies, as well as in detecting bars having a similar P.A. than the galactic disk, and/or if the underlying galactic disk is too faint to be adequately imaged, and/or if the data have inadequate resolution to resolve bars.  \\

For these reasons, a further confirmation of the bar existence is carry out by carefully examining the color maps and color images of the galaxy (see subsection \ref{ColorInfo}), the galfit simulations maps, and the flux and magnitude profiles (see subsection \ref{StructPara_LightProfiAna}). In any case, we can infer that the ellipse fitting technique is unlikely to overestimate the fraction of barred galaxies.\\

      \subsection{Color determination}
      \label{ColordeterminationMyPaper}

I argued in subsection \ref{PhyHubbleSeq} that even if large number of physical properties/parameters are well correlated with the galaxy morphological types through the Hubble sequence, some of them, as it is the case for the integrated color \citep[see end of subsection \ref{DownsizingarevivaloftheprimorColl}; see also section 6 in][]{2007MNRAS.380..585C}, present considerable scatter. Is this particular scatter due to the applied morphological classification method, and/or is it caused by real intrinsic physical properties? Having representative samples of distant and local galaxies, I have tried to get new insights on this issue.\\

As remarked at the end of subsection \ref{DownsizingarevivaloftheprimorColl}, some recent studies show quite high contaminations of late-type galaxies in the so called "red-sequence" ($\sim$20$\%$ in the local Universe, and $\sim$35$\%$ in the distant one), which is expected to be composed of just early-type galaxies. This is a controversial topic, which needs to be clarify to better understand the evolution of galaxies and the color-luminosity functions. In this context, I have thus compared the integrated color vs the morphological type of \citet{2010A&A...509A..78D} representative samples.  \\

To extract the integrated color of the galaxies, I have subtracted the rest-frame U and B absolute magnitudes. The rest-frame (U-B) color of each galaxy allowed me then to separate blue from red galaxies by applying a separation limit. I used the limit define by \citet{2007MNRAS.380..585C}. Following \citet{2004ApJ...608..752B}, "it is defensible and perhaps more natural to define early types in terms of colors". This seems not to be the case. Results are shown in section \ref{ResultsOfOurGalaxiesSamples}.\\

\section{The morphological classification method}
\label{MorphoMethod}

\begin{figure}[!h]
\centering
\includegraphics[width=0.65\textwidth]{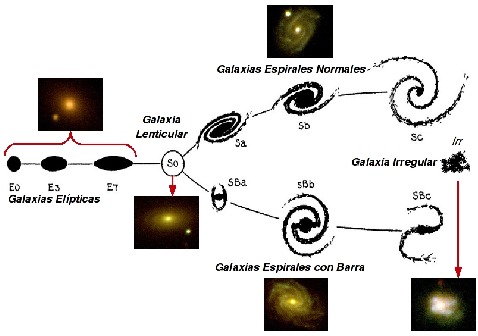}
\caption[Scheme of the local Hubble sequence]{Scheme of the local Hubble sequence.}
\label{HubbleSequenceBase}
\end{figure}

\begin{figure}[!h]
\centering
\includegraphics[width=1.0\textwidth]{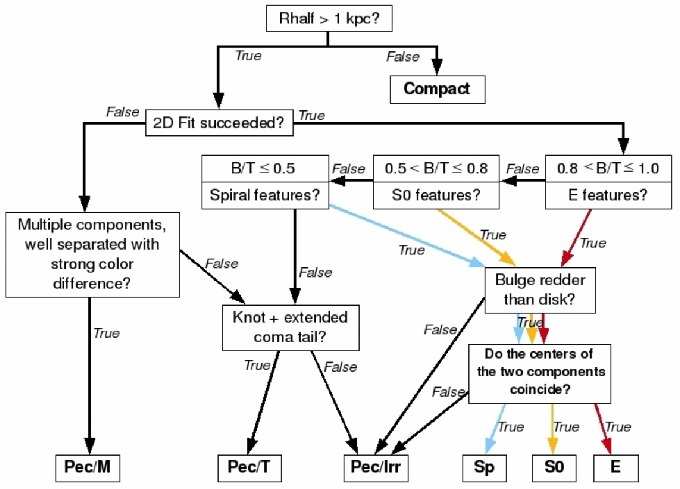}
\caption[Our morphological classification decision tree]{Our morphological classification decision tree.}
\label{DecisionTree}
\end{figure}

We thus adopt a methodology that systematically compares morphologies of distant galaxies to those of local galaxies. Identification of peculiarities is directly assessed by the discrepancy between a given galaxy and the galaxies that populate the local Hubble sequence (see figure \ref{HubbleSequenceBase}). Such a technique is based on a simple and reproducible decision tree (see figure \ref{DecisionTree}). \\

First, we identify those galaxies that are compact, and can thus not be decomposed in a bulge and a disk due to their small size. Second, we model the galaxy as detailed in subsection \ref{StructPara_LightProfiAna}. If the fit does not succeed, the galaxy is classified as peculiar according to its color map and image. If the fit succeeds, we derive a B/T value to distinguish between spiral, lenticular and elliptical galaxies. Even though, to be classified as spiral, lenticular or elliptical, the color map must show a center redder than the disk, the centers of the bulge and the disk must be in agreement, and all the images (color images, galfit residual images, etc.) must show compatible features \citep[see section 3.4 in][]{2010A&A...509A..78D}. \\

\begin{figure}[!h]
\centering
\includegraphics[width=1.0\textwidth]{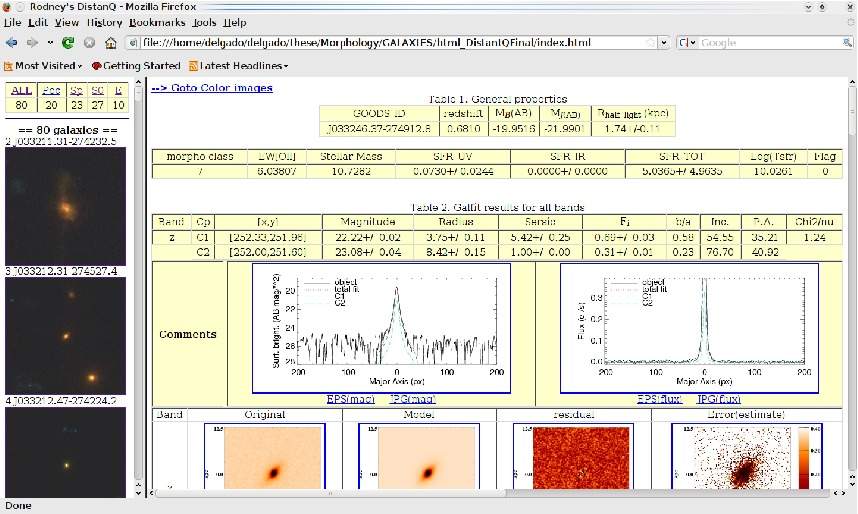}
\caption[Galaxies database created to gather all the galaxies information used during our classification methodology]{Galaxies database (webpage format) created to gather all the galaxies information used during our classification methodology. It makes the application of the decision tree by any astronomer a lot much easier.}
\label{Galaxies_Webpage}
\end{figure}

I have then created a data base, as shown in figure \ref{Galaxies_Webpage}, in order to have a fast and easy access to all the information concerning each galaxy (absolute magnitudes, SFR, redshift, galfit parameters, B/T, R$_{1/2}$, bulge-disk-galaxy profiles, galfit images, color maps, color images, etc.). Finally, the morphological classification was made individually by three of us (RD, FH, YY) and compared. The agreement between the individual classification was excellent (only 2 galaxies of 143 led to a discussion, arguing between peculiar or spiral galaxy. It was due to their relativily low surface brightness).\\

\section{Results}
\label{ResultsOfOurGalaxiesSamples}

       \subsection{Galaxy number density}

In short, for our morphological study we have gathered a distant representative sample of galaxies observed by the HST/ACS in the GOODS survey area. We have further  selected a representative sample of nearby galaxies from the SDSS. In applying the same morphological classification method, we have been able to derive a past and a present Hubble sequence. We also estimated the uncertainties of our reliable methodology, finding unprecedented results. Moreover, we showed that the observational conditions of the distant and local galaxies in our samples, needed to retrieve their morphology, are similar in an unbiased way. This has allowed us to link the past Hubble sequence to the present-day one. We do find that spiral galaxies were 2.3 times less abundant in the past, which is compensated exactly by the strong decrease by a factor 5 of peculiar galaxies, while the fraction number of elliptical and lenticular galaxies remains constant. It shows that more than half of the present-day spirals had peculiar morphologies, 6 Gyr ago (see figure \ref{HubbleSequencesFinal}).

       \subsection{Bar fractions}

\begin{figure}[!h]
\centering
\includegraphics[width=0.8\textwidth]{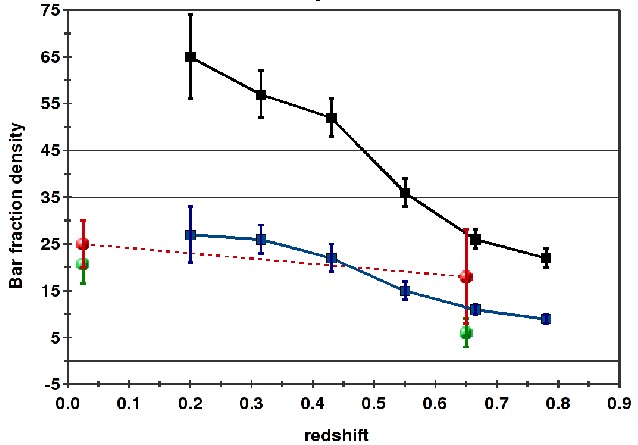}
\caption[Bar fraction density results from our local and distant representative samples]{\citet{2008ApJ...675.1141S} results are presented by straight lines. The blue straight line takes only into account galaxies with strong bars. Such bars are defined arbitrarily as having an ellipticity higher than 0.4. The black straight line includes all bar detections (weak + strong bars). Results from our representative local and distant samples are shown as red and green circles. The red circles can be compared to Sheth et al. as they are calculated using a similar criteria (bar detections within spiral galaxies). The green circles represent the fraction of barred Sp and S0 galaxies within the whole population. Error bars are calculated using a Poisson statistics.}
\label{Bar_fraction_densities}
\end{figure}

Taking into account only bar detection in Sp and S0 galaxies\footnote{It means I do not take into account those detections in galaxies that have been classified as Peculiar.}, I find $\sim$6$\% \pm 3\%$ of galaxies with bars in the distant sample, and $\sim$21$\% \pm 4\%$ in the local sample (see green circles in figure \ref{Bar_fraction_densities}). In the distant sample there are less than $10\%$ of Sp or S0 galaxies having an inclination ({\it i}) higher than 65$^{\circ}$, while in the local sample $\sim 24\%$ of Sp or S0 galaxies have the same property. In the literature it has been argued that galactic structures such as bars are difficult to identify and quantify in galaxies with {\it i} $>$ 65$^{\circ}$ \citep{2008ApJ...675.1141S}. Thus these results give a lower limit to the fraction density of barred galaxies at each epoch.\\

To compare my results to \citep{2008ApJ...675.1141S}, I quantified the fraction of barred galaxies only within those galaxies classified as spirals, as Sheth et al. eliminated all elliptical, lenticular, and peculiar galaxies when constructing their sample. In such a case, I find 18$\% \pm 10\%$ and 25$\% \pm 5\%$ of barred galaxies within the distant and local spirals, respectively (see red circles in figure \ref{Bar_fraction_densities}). These results are in good agreement with Sheth et al. \citep[see also][]{2008ASPC..396..333A}.\\

Finally, I want to remark that the morphological classification used by Sheth et al. is very different from the accurate methodology I have presented in this chapter. Moreover, even if the present results are limited by the Poisson statistic, I have showed that our samples are representative of local and distant galaxies. Being also limited by the fraction of galaxies with {\it i} $>$ 65$^{\circ}$ cited above, my result may be consider as a lower limit.

       \subsection{Color distribution}
       \label{ColorDistributionFromMyPaper}

\begin{figure}[!ht]
\centering
\includegraphics[width=0.67\textwidth]{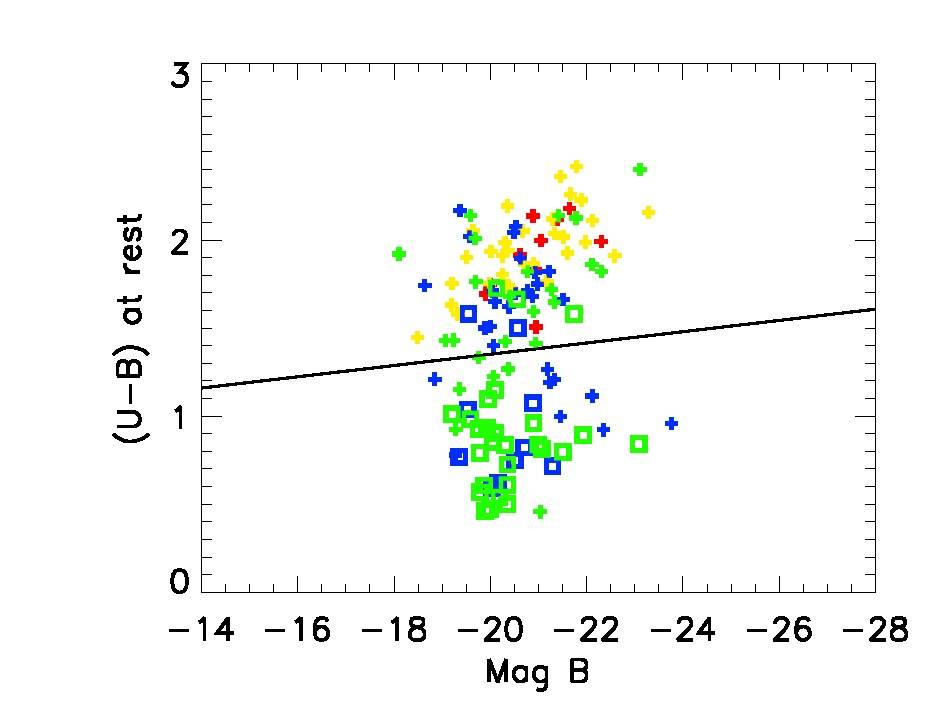}
\includegraphics[width=0.67\textwidth]{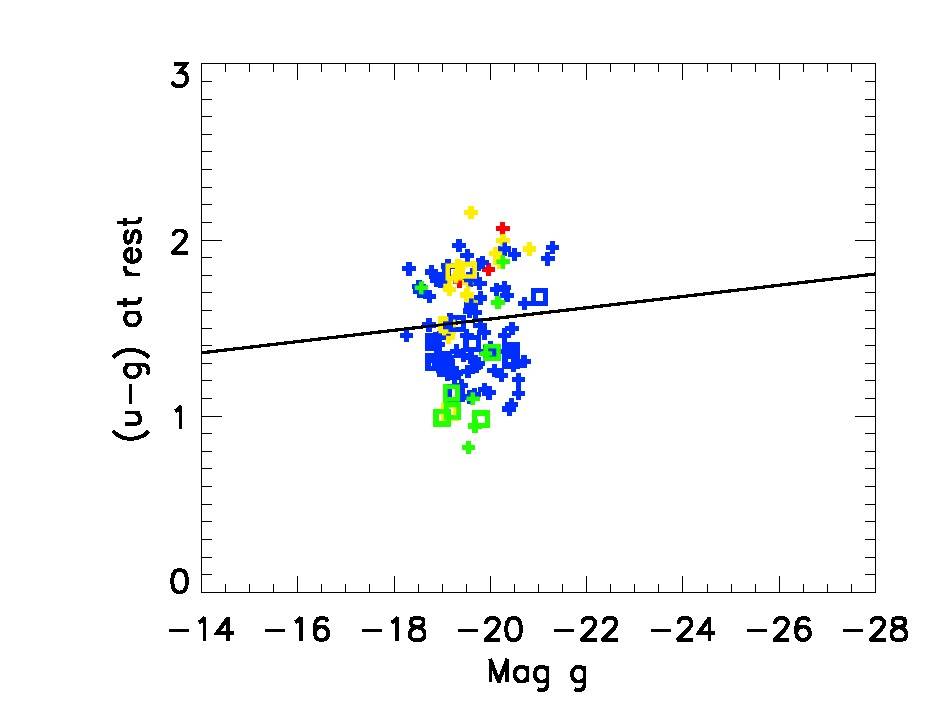}
\caption[Morphological contamination in the so called "red sequence" shown by a 3D plot of the color distribution of our representative distant and local samples]{Morphological contamination in the so called "red sequence" shown by a 3D plot of the color distribution of our representative distant (at {\it top}) and local (at {\it bottom}) samples. Red/yellow crosses represent E/S0 galaxies, blue symbols represent spiral galaxies, and green symbols represent peculiar galaxies. Crosses show galaxies classified as quiescent (EW[OII]$\lambda 3727 < 15\mathring{A}$), and squares galaxies classified as starburst. The straight line gives the limit between red and blue galaxies from \citet{2007MNRAS.380..585C} (in the local plot, I made a correction for the straight line of +0.2 mag to roughly fall in the "separation" between "red" and "blue" galaxies). All E/S0 galaxies of our representative samples are well in the red region. However, the "contamination" of the red region by other morphologies (Sp and Pec) is quite large. Noteworthy, I have found that using a starbusrt-quiescent limit smaller than 11$\mathring{A}$ there is a better correlation between starbusrt-quiescent galaxies and bleu-red galaxies, respectively.}
\label{ColorDistributionOurSamples}
\end{figure}

\begin{figure}[!ht]
\centering
\includegraphics[width=0.95\textwidth]{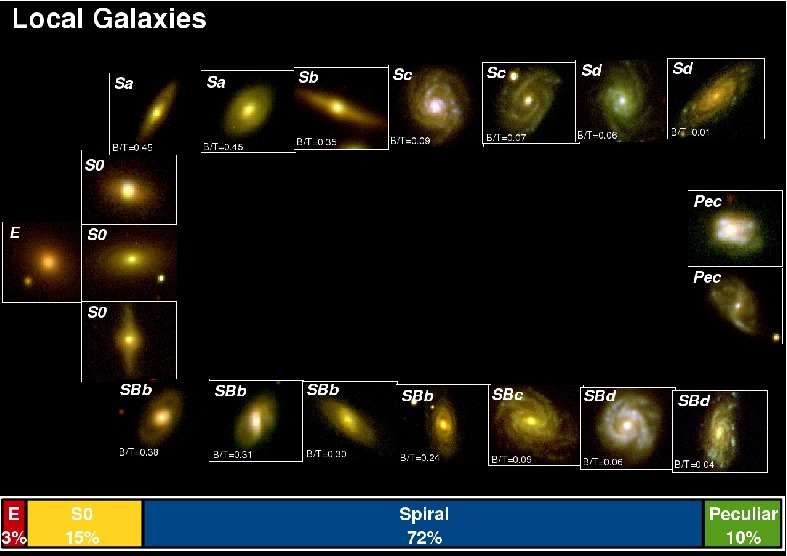}\\
\vspace{0.5 cm}
\includegraphics[width=0.95\textwidth]{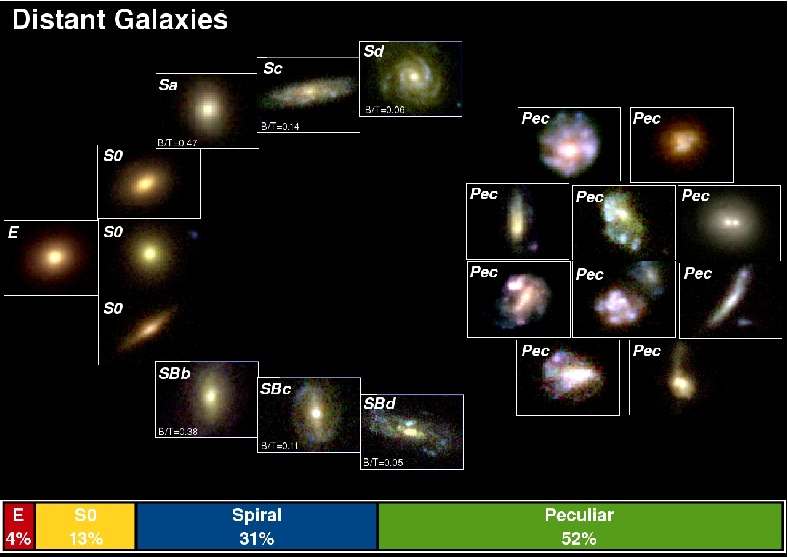}
\caption[Distant and Local Hubble sequences derived from our morphological study]{Local and Distant Hubble sequences. Each galaxy image represents $\sim 5\%$ of the whole galaxy population at each respective epoch.}
\label{HubbleSequencesFinal}
\end{figure}

Figure \ref{ColorDistributionOurSamples} shows the color distribution of our local and distant samples. Yellow and red symbols represents the S0 and elliptical galaxies, respectively. Thus the well known "red sequence"\footnote{all galaxies above the black straight line.} is not only composed by such galaxy types. I find that in our local sample the red sequence is composed by: 33$\% \pm 8\%$ of E/S0, 61$\% \pm 11\%$ of Sp, and 6$\% \pm 3\%$ of peculiar galaxies. In the case of the distant sample the red sequence includes 49$\% \pm 8\%$ of E/S0, 25$\% \pm 6\%$ of Sp, and 26$\% \pm 6\%$ of peculiar galaxies. \\

These results imply that more than 50$\%$ of galaxies in the local and distant red sequence are not elliptical or lenticular galaxies. This is a lot more than what has been published previously \citep[e.g.,][and references therein; see also subsection \ref{ColordeterminationMyPaper}]{2007MNRAS.380..585C}. It may be due to our morphological classification methodology, which we have shown to be more reliable than other methods. Indeed, the use of colors to estimate morphological evolution of galaxies has been done erroneously, as galaxy color alone is not a good indicative of the galaxy morphology. Different reasons can explain this. One of them is that U-B color are very sensitive to extinction, which is more important in spiral and peculiar galaxies because of their large amount of gas. \\

In short, colors are well used to classify stars. Nevertheless, galaxies are much more complex systems, composed of gas and millions of stars. It is therefore expected that extinction and geometric effects (edge-on galaxies) affect their color with a significant deviation. \\

\newpage

\newpage

\fbox{\textbf{ See: \itshape{Delgado-Serrano et al. 2010, A$\&$A,509A, 78D}}}

\chapter{Conclusion and Discussion}
\minitoc

During my PhD thesis I have confronted serious problems of methodology concerning the morphological and kinematic classification of distant galaxies. This has forced us to create a new simple and effective morphological classification methodology in order to guarantee a morpho-kinematic correlation, make the reproducibility easier and restrict its subjectivity. This is not the case for the rest of morphological classification proposed in the past decades. Such classification methodologies\footnote{concerning the automated ones.} suppose that the high number of galaxies they can classify in a relative short time compensate for the errors they make in the morphological classification of a quite important fraction of galaxies. It is impudent to believe that methodology has a second role when classifying galaxies. How could a limited methodology not affect the results?\\

Giving the characteristic of our morphological classification, we have thus been able to apply the same methodology, using equivalent observations, to local and distant galaxies. It has allowed us to determine a morphological evolution of galaxies over the last 6 Gyr. Our results present just a small statistical error bar, which could be improved by adding more galaxies to the studied samples.\\

Indeed, the IMAGES survey and subsequents studies, in which I participated, provide us with a complete description of galaxy properties 6 Gyr ago (see figure \ref{IMAGESresults}). In short, this thesis has thus allowed me to work on the morphology and kinematics of galaxies. By making a careful analysis of each galaxy, we have been able to determine the evolution of their morphology and kinematics. The conclusions are the following:

\begin{figure}[!h]
\centering
\includegraphics[width=1.0\textwidth]{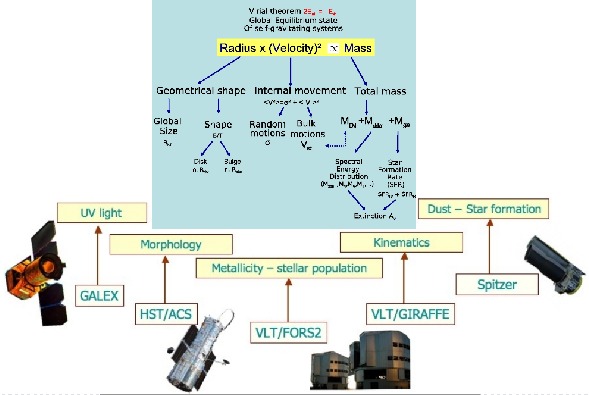}
\caption[IMAGES results diagram from different instrument observations]{IMAGES results diagram from different instrument observations (Courtesy of M. Puech).}
\label{IMAGESresults}
\end{figure}

\begin{itemize}
\item The morphological classification method presented here represents a very promising tool, as (1) it provides a strong correlation with the kinematic stage of the galaxies, (2) it is easily reproducible, and (3) it has a considerably limited subjectivity. It could be easily extended to higher redshifts when deeper and better resolution images will be available (e.g., from the ELT).  

\item We find that E/S0 galaxy populations show no evidence of number density evolution during the past 6 Gyr, while slightly more than half of the distant galaxies had peculiar morphologies. These last are associated to anomalous kinematics.

\item The fraction of regular spirals was 2.3 times lower 6 Gyr ago than in the present epoch. This shows that more than half of the present-day spirals had peculiar morphologies, 6 Gyr ago. Thus almost all the evolution is caused by the transformation of galaxies with peculiar morphologies into regular spiral galaxies at present epoch.

\item The number fraction of barred galaxies in the local Universe is higher than at z$\sim$0.65.
 
\item Integrated color determination to separate early and late type galaxies is not an adequate tool. A similar result has been found for the parametric methods (A-C plane, Gini and M$_{20}$ coefficients, etc.) 

\item Further results, concerning the kinematics of $M_{B_{Vega magnitude}}< -16.4$ galaxies at z$\sim$ 0.25 (see appendix \ref{KinematicStudies}), allow us to state that galaxies with complex kinematics present the same signature in emission lines of the same ionization (e.g., [OIII]$\lambda$4959 and [OIII]$\lambda$5007), while emission lines of different species (e.g., H$\beta$ and [OIII]) show different complex kinematics. In contrast, rotating disks have the same kinematic signature using any emission line. This indicates that our methodology is effective at separating spiral rotating disks from galaxies having complex kinematics and peculiar morphology. While spiral rotating disks are dynamically stable, peculiar galaxies with complex kinematics undergo chaotic internal motion. 
\end{itemize}

\section{Merger events and the baryonic Tully-Fisher relation}
\label{Hammer_etal_2009Section}

The previous results seem to favor a "spiral rebuilding" scenario \citep{2005A&A...430..115H,2009A&A...507.1313H}. In this context, we can add the following conclusions to \citet{2009A&A...507.1313H}, and \citet{2010A&A...510A..68P}:

\begin{itemize}
\item Analysis of the Tully-Fisher relation at z = 0.65 \citep{2006A&A...455..107F,2008A&A...484..173P,2010A&A...510A..68P} are more consistent with merger processes happening at that time \citep[see also][]{2010ApJ...710..279C}; \citet{2009A&A...507.1313H} reproduced the morphology and the kinematics of anomalous galaxies at z=0.65 assuming major merger processes, including instabilities in the remnant phase;

\item The merger rate required to explain that half of the z=0.65 galaxies are in a merger process is in agreement with the most recent models of halo occupation \citep[see][]{2009MNRAS.397..802H}. This is because the whole duration of the merger process is long, and can reach up to 2, 3 or even more Gyr;

\item The comparison between the distant and the local Hubble sequences implies that most peculiar galaxies should be the progenitors of present-day spirals. This is because the fraction of early-type galaxies (E/S0) does not evolve. \citet{2009A&A...507.1313H}, and \citet{2010A&A...510A..68P} have shown that the progenitors z $\sim$ 0.6 galaxies had large enough gas fractions ($>$50$\%$), implying that the progenitors of the assumed mergers were very gas-rich, thereby allowing a significant rebuilding of the disk of the merger remnant \citep[see][]{2009ApJ...691.1168H,2009ApJ...702..307S};

\item Other mechanisms such as minor mergers or outflows are unlikely to be dominant because they have a considerable lower efficiency \citep[see][]{2008ApJ...688..757H}, and because their observational signatures are not detected in the galaxy spectra \citep{2009A&A...507.1313H}.

\item The non evolution of the baryonic TF relation, and the previous conclusion concerning the higher content of gas in distant galaxies, implies that such gas is gravitationally bound to the system, which suggest that galaxies evolve on average as in a close box model since z$\sim$0.65. This could explain the evolution of the stellar mass-metallicity relation reported by \citet{2008A&A...492..371R} (see subsection \ref{Rodriguessection}). This is in agreement with a "spiral rebuilding scenario" (see subsection \ref{TheSpiralRebuildingScenarii}). This last explains that gas-rich major mergers can expel gas, which can be later re-accreted to re-build a new disk.

\item The fact that the gas is exhausted with time would unavoidably limit star formation during the post-merger phase at lower redshift. The lack of available gas with which to form a new disk would naturally lead to no new disks forming with strong star formation at low redshift, as reported by \citet{2005ApJ...632L..65M}.

\item The relative abundance of E/S0 has not changed in the past 6 Gyr, which does not imply that E/S0 galaxies did not undergo mergers since then \citep[possible dry mergers;][]{2006ApJ...640..241B}. Nonetheless, it makes clear that the process of mergers favors the formation of spirals since z$\sim$0.65.
\end{itemize}

\fbox{\textbf{ See: \itshape{Hammer et al. 2009, A$\&$A, 507, 1313H}}} \\ \\

\fbox{\textbf{ See: \itshape{Puech et al. 2010, A$\&$A, 510A, 68P}}}

\section{Prospectives}

The previous results open several interesting prospectives:

\begin{itemize}

\item {1.)} {\it Searching for dynamic substructures in the core of galaxies in fusion}. The morphological analysis developed in this thesis, and observations from the HST/UDF, will allow us to trace the different morphological signatures of the abundant merger events at z$\sim$0.65. Such morphological properties may allow us to understand how the system move (and evolve), as for example the detection of inner spiral structures. I have already created a data base concerning such galaxies, similar to that created for the galaxies I studied during my PhD, making galfit fitting in four different bands to analyze the fit and residual image of each galaxy. A new type of error map is also in perspective, which could allow us to make a real measure of the galfit pixel detections. Thus, the work is in progress. 

\item {2.)} {\it Stellar tidal streams in nearby galaxies}. The detection of such streams \citep[(e.g.,][]{2001Natur.412...49I,2008ApJ...689..184M,2009ApJ...692..955M} will contribute to constrain the hydrodynamical simulations, and will give us a better idea of the past merger history of the local Universe. I paticipated in three night observations at the ESO-NTT searching for such streams in six local MW like galaxies. This data is being exploited. Furthermore, one of the project that will be developed in the Panama observatory (see appendix \ref{PanamaObservatory}) concerns the study of tidal stream in local galaxies. It has been already shown \citep{2008ApJ...689..184M,2009ApJ...692..955M} that, with enough exposition time, it is possible to detect such structures with small diameter telescopes. However, it is necessary to develop a project which allow to do the photometry of such faint structures with small telescopes.   

\item {3.)} {\it Morpho-kinematic Study of higher redshift galaxies}. There might be a transition epoch where the main driver for star formation shifts from cold gas streams to major mergers, around z $\sim$ 0.8-1.4, since preliminary results at z $\sim$ 1.5 suggest that cold flows were still important at this epoch \citep[][and references therein]{2008A&A...486..741B,2009A&A...504..789E}. Indeed, numerical simulations, at z$\geqslant$1.5, suggest that cold-gas streams from inter-galactic filaments can feed galaxies in fresh gas \citep[e.g.,][]{2009MNRAS.397L..64A,2010MNRAS.404.2151C}, which could explain their observed very high star formation rate and clumpiness \citep[e.g.,][]{2008ApJ...687...59G,2009ApJ...697.2057L,2009ApJ...694L.158B}. In this scenario, major mergers have a much lower duty cycle and therefore are not expected to be the main driver for star formation in z$\sim$2 galaxies \citep{2009Natur.457..451D}, as the study of individual galaxies apparently show no signs of major merging events \citep{2006ApJ...645.1062F,2007ApJ...658...78W,2009ApJ...699..421W,2007ApJ...669..929L,2009ApJ...697.2057L,2006Natur.442..786G,2008ApJ...687...59G}. However, what we currently know at high redshift is simply too fragmentary to tell us exactly what is going on. Thus, alternatively, there may not be such cold flows, as the number of kinematic studied galaxies at these redshift is still very small to have precised conclusions. Moreover, the morphology of very high redshift galaxies has been very difficult to exploit, because of the need of very deep data. Noteworthy, it is precisely at these epochs that the co-mobile star formation density reaches its peak, which makes this range particularly valuable for further 3D-kinematic and deep morphological studies. This is the reason why we are applying to ESO with a large program to observe a representative sample of 120 z=0.6-1.4 galaxies with GIRRAFFE CDD and FORS2 in the HUDF and WFC3-ERS (Early Release Science) fields (R. Delgado-Serrano: CoI). The available deep NIR imaging (i.e., rest-frame V-band) in these fields is essential to correctly image the distribution of stellar mass, an information which is currently lacking in all 3D surveys at z$\geqslant$1.

\item {4.)} {\it Kinematic studies of quiescent intermediate redshift galaxies}. Since  kinematic studies in our sample are based on star-forming galaxies (arbitrary defined as galaxies having an EW[OII]$\lambda 3727 \geq 15\mathring{A}$), a complementary study of our quiescent galaxies (EW[OII]$\lambda 3727 < 15\mathring{A}$) is needed. \citet{2009A&A...501..437Y} has shown it is possible, as the z$\sim$0.4 galaxy they kinetically studied with Flames/Giraffe has an EW[OII]$\lambda 3727 = 2\mathring{A}$. We may take advantage of the much larger sensitivity of X-SHOOTER to observe a set of emission lines in these objects. Indeed, using the exceptionally large spectral bandwidth of X-SHOOTER, it is now possible to obtain spatially resolved line maps of distant galaxies, using 5 prominent emissions lines (i.e., [OII]$\lambda$3726,3729, H$\beta$, [OIII]$\lambda$5007, H$\alpha$). These lines are the most important tracers of the gaseous phase, which fuels star formation and disk growth in distant galaxies. As an example we have that for z$\sim$0.6 galaxies, under good seeing conditions ($\leq$0.8 arcsec), FLAMES can spatially resolve the [OII] emission line in 8 hours of integration time in z$\sim$0.6 intermediate-mass galaxies. X-SHOOTER will roughly provide a five to six times better throughput compared with GIRAFFE at a smaller spectral resolution, which means that the same signal-to-noise ratio will be reachable in only $\sim$1 hours (on-source). As a result, within the team GTO, there is an on-going proposal to ESO (R. Delgado-Serrano: CoI).

\item {5.)} {\it Hydrodynamical simulations}. The comparison of the observations with such simulations will allow us to constrain the different dynamical process that play an important role during the merger phases, as well as the progenitors stage. For each distant galaxy, we have at our disposal a complete description of its physical properties, including morphology, kinematics, and mass. As has been demonstrated in five explicit examples (see subsection \ref{TheSpiralRebuildingScenarii}), this allows us to model distant galaxies in a way similar to what is currently done for nearby galaxies. Restricting the simulations with all the information we dispose for each galaxy through the IMAGES survey (especially the morphological and kinematic information) is a promising way to model, with high accuracy, the formation of structures and substructures within discs.

\end{itemize}

A longer term prospective is the application of the morphological classification method developed here to observations with the future ELTs. Such next generation of telescopes are a very promising tool for the astronomical community, as they will improve the observation deepness in some order of magnitude compared with the present instruments.


\chapter*{Publication list}
\addstarredchapter{Publication list}

\section*{Publications}

\begin{enumerate}
\item Neichel B., Hammer F., Puech M., Flores H., Lehnert M., Rawat A., Yang Y., {\bf Delgado-Serrano R.}, Amram P., Balkowski C., Cesarsky C., Dannerbauer H., Fuentes-Carrera I., Guiderdoni B., Kembhavi A., Liang Y.~C., Nesvadba N., {\"O}stlin G., Pozzetti L., Ravikumar C.~D., di Serego Alighieri S., Vergani D., Vernet J. and Wozniak H., {\it IMAGES. II. A surprisingly low fraction of undisturbed rotating spiral disks at z $\sim$ 0.6 The morpho-kinematical relation 6 Gyr ago}, A$\&$A, vol. 484, pages 159-172, June 2008, 484, 159 

\item Rodrigues M., Hammer F., Flores H., Puech M., Liang Y.~C., Fuentes-Carrera I., Nesvadba N., Lehnert M., Yang Y., Amram P., Balkowski C., Cesarsky C., Dannerbauer H., {\bf Delgado-Serrano R.}, Guiderdoni B., Kembhavi A., Neichel B., {\"O}stlin G., Pozzetti L., Ravikumar C.~D., Rawat A., di Serego Alighieri S., Vergani D., Vernet J. and Wozniak H., {\it IMAGES IV: strong evolution of the oxygen abundance in gaseous phases of intermediate mass galaxies from z $\sim$ 0.8}, A$\&$A, vol. 492, pages 371-388, December 2008, 73, 105 

\item Hammer F., Flores H., Puech M., Yang Y.~B., Athanassoula E., Rodrigues M. and {\bf Delgado-Serrano R.}, {\it The Hubble sequence: just a vestige of merger events?}, A$\&$A, vol. 507, pages 1313-1326, December 2009, 507, 1313 \\

\item {\bf Delgado-Serrano} R., Hammer F., Yang Y.~B., Puech M., Flores H. and Rodrigues M., {\it How was the Hubble sequence 6 Gyr ago?}, A$\&$A, vol. 509, pages A78+, January 2010, 509, A78 

\item Puech M., Hammer F., Flores H., {\bf Delgado-Serrano R.}, Rodrigues M., $\&$ Yang Y., {\it The baryonic content and Tully-Fisher relation at z $\sim$ 0.6}, A$\&$A, 2010, 510, 68 

\end{enumerate}

\section*{Press release}

\begin{enumerate}
\item ESA/NASA/HST: {\it Forming the present-day spiral galaxies}, February 2010,  \\

\item CNRS/INSU/Observatoire de Paris: {\it L'origine des galaxies spirales actuelles}, February 2010,  \\
\end{enumerate}


\appendix

\part*{Appendix}
\addcontentsline{toc}{part}{Appendix}

\chapter{Cosmology: The history of the Universe and the formation of structures}
\label{CosmologyTheHistoryOfTheUniverse}


\begin{flushright}
"...You need a lot of smart peoples, and faith that with enough smart people, thinking about difficult problems in enough different ways, someone can transform our view about how the world works. It's happened many times before, and it will surely happen again." \\ Liddle and Loveday, The Oxford Companion to Cosmology.
\end{flushright}

\minitoc

In the observable Universe, the gravitational force is the principal responsible for the formation of structures. One little mass overdensity can result in an accretion process. It is the beginning of the formation of structures. Then, other forces show up, while the accretion process is in action, and they could finally drive the system to equilibrium. The final spatial scale of the object will depend on the amount of mass accreted or that could be accreted. Two examples of forces that could counteract gravity, in the case of small scales, are the thermal radiation and the nuclear forces. For large scales, the conservation of angular momentum is crucial in the way to the equilibrium. \\

In any case, the new structure can be considered as an isolated {\it set of components} which do not follow the general Universe expansion, but its own evolution. However, the first thought, before studying those sets as isolated, is how the Universe has evolved to form all the different objects or structures at different scales as we can see today through the best telescopes in the world. This is not so well known until now. Nevertheless, two theories confront each other to try to explain the Universe evolution, and find an answer to how the different structures at different scales have been formed. Both of them are based on observational and theoretical results, but they present two different ways of the Universe evolution.

\section{The primordial collapse model}

\begin{figure}[!h]
\centering
\includegraphics[width=0.5\textwidth]{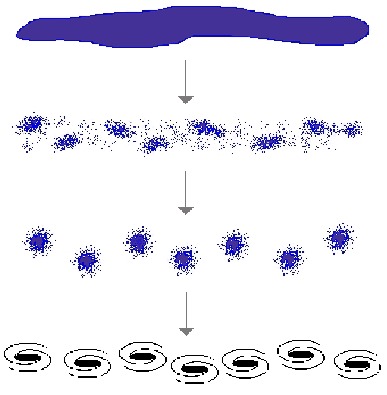}
\caption[Illustration of the Top-Down model]{Illustration of the Top-Down model.}
\label{TopDownModel}
\end{figure}

The primordial collapse model, also known as the adiabatic scenario, proposes the formation of smaller structures after the formation of the largest ones. This is the reason why it is known as the {\it top-down} model (see figure \ref{TopDownModel}). As a consequence, this model predicts the formation of massive galaxies very soon after the big-bang (z $\geqslant$ 3) when a primordial collapse of matter is produced.\\

\citet{1970Ap&SS...9..368S} give the principle \citep[see also][]{1967SvPhU...9..602Z,1970A&A.....5...84Z,1967SvA....11..233D,1970PhRvD...1..397P,1970ApJ...162..815P,1989RvMP...61..185S}. First, little adiabatic perturbations take place in the beginning of the Universe. Then, only the most important perturbations survive the Recombination\footnote{It is the moment when the decoupling of photons and matter take place (see more details in section \ref{The_hierarchical_model}).} because those smaller than a limit mass were smoothed by the photon diffusion. Finally, large structures are formed with ellipsoid shape (z $\sim$ 5), to be then flattened following the minor axis and forming what Zeldovich called "pancakes". These pancakes segment and form the galaxies. One of the principal results of this theory is that it is consistent, as a natural consequence, with the scale of the large structures.\\

\begin{figure}[!h]
\centering
\includegraphics[width=0.9\textwidth]{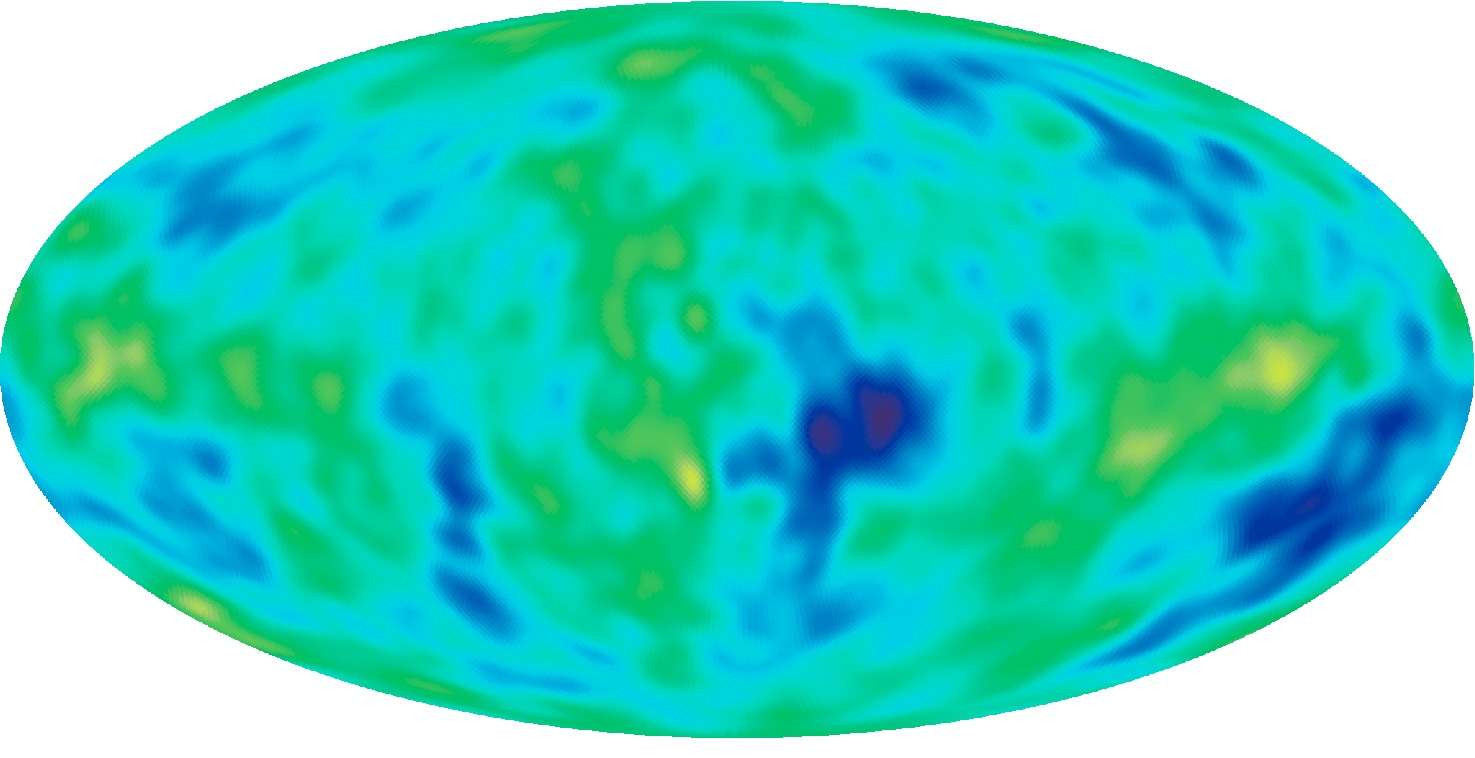}
\caption[CMB map observed by COBE]{Cosmic Microwave Background (CMB) map observed by COBE \citep[Credit: NASA/WMAP Science Team; see also][]{1996ApJ...464L...1B}. The spectrum of the CMB is well described by a blackbody function with T = 2.725K.}
\label{COBE_CMB}
\end{figure}

So far, the model takes care only of the baryonic matter assuming that the baryons dominate the Universe, and thus predicts that the perturbations during the Recombination are of the order of 3x10$^{-3}$. Nevertheless, with satellite COBE it has been found that those perturbations could not be larger than 2x10$^{-5}$ \citep[see figure \ref{COBE_CMB};][]{1992ApJ...396L...1S}\citep[see also][for a review]{2006astro.ph..1307S}. Here there was then a problem to solve. Moreover, astronomers had been finding evidence about a disagreement between the mass of galaxies derived from their luminosity and that from their dynamics \citep[e.g.,][]{1937ApJ....86..217Z,1983MNRAS.202P..21G,1986AJ.....91.1058D,1990SvAL...16..454N,1988A&A...201...51R,1987AJ.....93.1114L}\citep[see also][for a review]{1983MitAG..60...23L,2009EAS....36..113F,2008ASPC..395..283R,2009AIPC.1192..101P}\footnote{Very few studies were against this idea \citep[e.g.,][]{1982ApJ...260...65G,1983IAUS..100...93G}.}. The mass calculated after their dynamics was much bigger than the mass estimated after the total luminosity we receive from them. Consequently, a logical hypothesis to emerge was the idea of the presence of a certain kind of mass we can not see, and which differ in some characteristics from the "normal" baryonic mass \footnote{For example, it does not interact with electromagnetic radiation, neither with the ordinary matter via electromagnetic forces.}. The non-baryonic "dark matter" was thus born.\\

\begin{figure}[!h]
\centering
\includegraphics[width=0.7\textwidth]{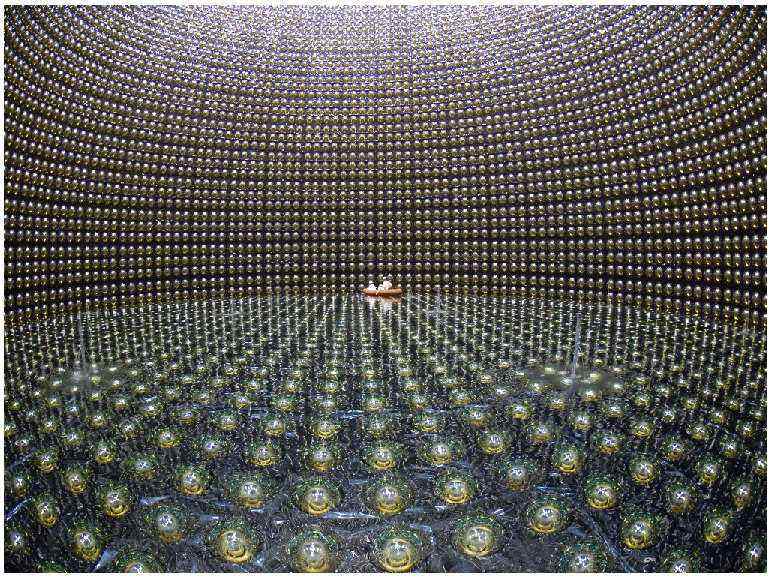}
\caption[Photo of one part of the Super-Kamiokande]{Photo of one part of the Super-Kamiokande. For scale, in the center one can distinguish a rowboat with two scientists inside. The Super-Kamiokande is surrounded by 50 000 tons of pure water, and is composed of 11 200 photomultiplier detectors. Its dimensions are: 41.4 m high, and 39.3 m across (Credit: Kamioka Observatory, ICRR-Institute for Cosmic Ray Research, The University of Tokyo).}
\label{SuperKamiokande}
\end{figure}

With the idea of the existence of dark matter, the adiabatic model becames the Hot Dark Matter (HDM) model. Within a few years, this renewed model was developed to be in agreement with the results mentioned above. The HDM model predicts, similar to the original adiabatic model, the formation of large structures at the beginning which break up in smaller structures through time. The principal candidates for the hot dark matter are the super-massive neutrinos \citep[e.g.,][]{1988PASP..100.1364S,1996NuPhS..51..254C,2010arXiv1003.0459N,1966JETPL...4..120G,1972PhRvL..29..669C}. Neutrinos travel with ultrarelativistic velocities, and are very difficult to detect because of their way to interact with baryonic matter. Neutrinos only interact with baryonic mass by the weak interaction and gravity, which are well known not to be strong forces. One of the most famous projects to detect neutrinos is the Super-Kamiokande neutrino observatory in Gifu-Japan (see figure \ref{SuperKamiokande}). \\

During the first three minutes of the Bing Bang, while the Universe goes from its spectacular 10$^{32}$ kelvin to roughly 10$^{9}$ Kelvin, takes place the production of neutrinos among the reactions which produce photons. The theory also predicts a decoupling of neutrinos from matter, similar to the photon-matter decoupling which creates the today well known CMB (Cosmic Microwave Background). However, the neutrino-matter decoupling happens before the Recombination. Therefore, a cosmic neutrino background as dense as the photons of the CMB is also predicted.\\

One of the problems of the HDM model is the inability to account for the galaxy distribution observed in the Universe \citep[e.g.,][]{1983ApJ...274L...1W}. The neutrino density was very high in the early Universe, and added to their greatly relativistic velocities, they could then smooth out any fluctuation in matter density and make the formation of any overdense region impossible. This implies that those overdense regions could be formed only later when neutrinos notably slow down (called neutrino "cooling") as a consequence of the decreasing temperature of the expanding Universe. Then, combining this result with those observed by the satellite COBE on large scales, scientists have found that a Universe dominated by neutrinos would not contain enough power on small scales to fit the observed Universe properties.\\ 

As a result, any purely HDM model of the Universe is excluded. For this reason, when scientists talk about HDM model today, they mostly refer to a "Mixed Dark Matter" model, which combines "Cold" and "Hot" dark matter \citep[CHDM, for Cold+Hot Dark Matter; see for example][]{1998astro.ph.10204P,1998MNRAS.296..109C}\footnote{There are also other combinations giving for example a $\Lambda$CHDM model \citep[e.g.,][]{2001ApJ...558...10P}, which includes a cosmological constant different from 0, or even we could also have a Warm DM model \citep[e.g.,][]{2009JCAP...07..037A,2009MNRAS.399.1611T}\citep[see][for a review]{2001canp.book..287P}.}. A huge fraction of the dark matter is cold, while the hot fraction is tiny. Experiments predict that the amount of hot dark matter in the Universe would be a few percent\footnote{some experiments even predict an amount of HDM which could be unmeasurable.}, while the cold dark matter component represents roughly 85$\%$ of the total mass ($\sim 33\%$ of the total energy) in the Universe.

\section{The hierarchical model}
\label{The_hierarchical_model}

This model, also known as the isothermal model, proposes the formation of small structures at the beginning and then, by accretion of these ones, the largest structures develop. Therefore, it is an hierarchical model with a {\it bottom-up} evolution.\\

It was developed almost parallely to the primordial collapse model \citep{1968ApJ...154..891P,1969ApJ...157.1075P}\citep[see also][]{1967SvPhU...9..602Z,1967SvA....11..233D}, and could be seen as the opponent of it if we consider their scale evolution. Its principal idea states that almost each fluctuation before the Recombination could give birth to one structure. In this context, the formation of structures has begun just after the Recombination. At such an epoch, the Universe was smaller, and thus its density was a lot much higher. This implies that the probability of objects to collide was much more important. Hence the first little objects interact and begin to form larger and larger structures until the formation of galaxies, galaxy clusters and so on.\\

Two important results of this model are the following. The first one is that it predicts that between the first baryonic objects to be formed some of them have sizes roughly similar to the known globular clusters. Curiously enough, these last are also known to be amongst the oldest objects in the Universe. The second result is that this model can explain very well the fast metal increase of the Universe by the action of the first massive stars, or more exactly by the explosion of these ones in supernovae.\\

Similar to the adiabatic model, the hierarchical model needed to be adapted to the new results coming from COBE and those about the dynamic of galaxies, among others. It then became the Cold Dark Matter (CDM) model \citep{1984fegl.proc..163P}. The dark (non baryonic) matter, in this case, travels relatively slow, and this is the reason why it has been called "cold". In the CDM model, very early in the Universe history (before the Recombination), the dark matter follows the little fluctuations and began to form structures of very small mass, then by coalescence the bigger structures form. \\

One weakness of the cold dark matter theory is that it does not make any predictions about exactly what the cold dark matter particles are. Nevertheles, there are some candidates to constitute the "cold dark matter". They have very curious names: (1) WIMPs, for Weakly Interacting Massive Particles; (2) MACHOs, for Massive Astronomical Compact Halo Objects; and (3) Axions, coming from the quantum chromodynamics (QCD) theory\footnote{Just for curiosity, the name "Axion" was adopted from a brand of detergent, making allusion to the fact that one of the QCD problems had been "cleaned" with their assumption.}. The first ones (WIMPs) are supposed to be a kind of massive particles. There is no known particle within the standard model of particle physics with the required properties (similar to HDM particles, but with higher masses and much slower velocities)\footnote{Nontheless, in supersymmetry theories, the neutralino is a possible WIMP \citep[e.g.,][and references therein]{2009arXiv0906.1615B}\citep[see also][for a review]{2004PhR...405..279B}.}. Researches continue in the particle accelerators. The second ones (MACHOs) would be composed of very compact objects like black holes, neutron stars, white dwarfs, very faint stars, or even objects such as planets. However, until now, the studies, using gravitational microlensings, show that such objects do not represent a significant percentage of the required amount of mass. Finally, particles called axions are still hypotheticals and are being searched. \\  

\begin{figure}[!h]
\centering
\includegraphics[width=0.7\textwidth]{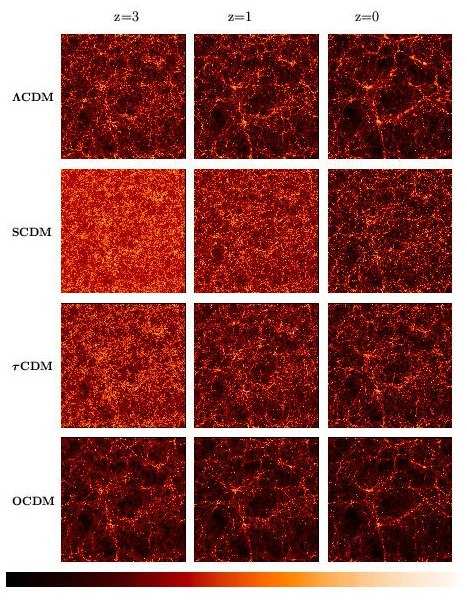}
\caption[CDM model simulations]{CDM model simulations \citep[][for the Virgo Collaboration]{1998ApJ...499...20J,1998MNRAS.296.1061T}.}
\label{CDM_models}
\end{figure}

\begin{figure}[!h]
\centering
\includegraphics[width=0.8\textwidth]{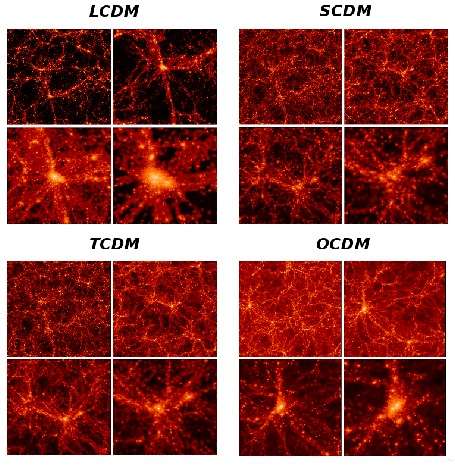}
\caption[Different CDM models simulations at z=0]{LCDM, SCDM, TCDM and OCDM model simulations showing different scales at z=0. \citep[][for the Virgo Collaboration]{1998ApJ...499...20J,1998MNRAS.296.1061T}.}
\label{CDM_models_scales0}
\end{figure}

\begin{figure}[!h]
\centering
\includegraphics[width=0.6\textwidth]{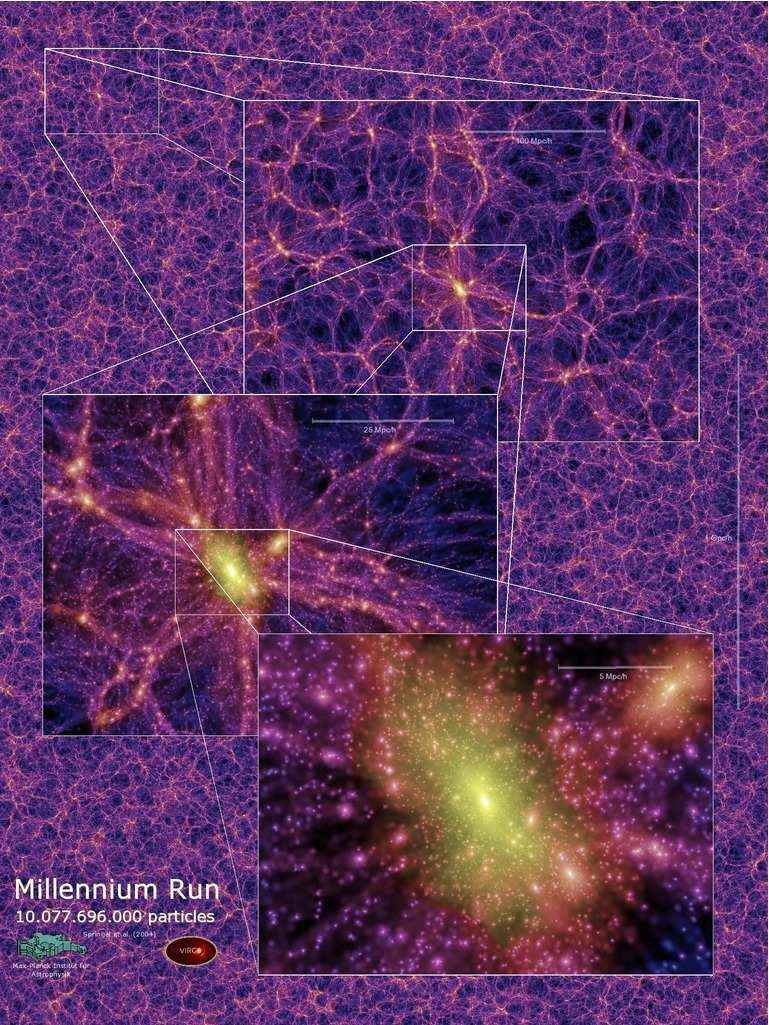}
\caption[Illustration of the Millenium simulation]{Illustration of the Millenium simulation \citep{2005Natur.435..629S}. It shows the dark matter density field on various scales at redshift z=0 with a thickness slice of 15 Mpc/h. The zoom sequence displays consecutive enlargements by factors of four, centred on one of the many galaxy cluster halos present in the simulation. Colors are coded by density and local dark matter velocity dispersion (Credit: Virgo Consortium).}
\label{Millenium_simulation}
\end{figure}

\begin{figure}[!h]
\centering
\includegraphics[width=0.6\textwidth]{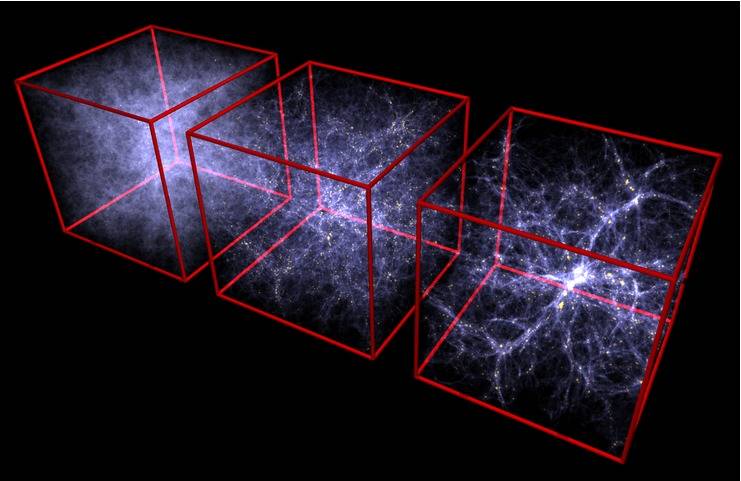}
\caption[3D dark matter simulation showing the structure formation in the gaseous component of the universe]{Structure formation in the gaseous component of the universe (within the dark matter density field), in a simulation box 100 Mpc/h on a side. From left to right: z=6, z=2, and z=0. Formed stellar material is shown in yellow (Credit: Volker Springel).}
\label{Dark_energy_simulation}
\end{figure}

In any case, according to the different (assumed) nature of the dark matter, a family of CDM models has been developed during the past three decades: SCDM (Standard-CDM), OCDM (Open-CDM), $\tau$CDM, $\nu$CDM, HCDM, WCDM, and the $\Lambda$CDM \citep[see][for a quick review about the CDM models]{1996Sci...274...69D}(see figures \ref{CDM_models} and \ref{CDM_models_scales0})\footnote{Concerning the topology of the Universe, there are also a subclass of big bang models that are Friedmann-Lemaitre solutions in which the space topology is multiply-connected instead of being simply-connected. One example is the Wraparound Universe theory \citep{Luminet...JeanPierre...UC,2008wrun.book.....L}, which proposes a positive spatial curvature within a Poincare Dodecahedral Space (PDS) topology \citep{2003Natur.425..593L,2008arXiv0802.2236L}. In this theory, the Universe has a finite spatial extension without edge. The Wraparound Models have only implications on space geometry, and not on time evolution. They share exactly the same dynamics (time evolution starting from a big bang, and in perpetual accelerated expansion if dominated by dark energy) as the standard model.}. Nonetheless, nowadays, looking for the combination of the parameters that lead to the better agreement with the existing data on both large and small length scales, the more accepted, adopted by a large part of the scientific community, is the $\Lambda$CDM model (see figure \ref{Millenium_simulation}, and also see figure \ref{Dark_energy_simulation} for a 3D visualization). It is described below.  \\

{\textbf{\itshape The Lambda-Cold Dark Matter ($\Lambda$CDM) model}}\\

The $\Lambda$CDM could be seen as an extension of the previous CDM models as it has been developed by changing, adding or suppressing some parameter values to be in agreement with the new observations \citep[see][for the more recent parameter values, as their constraints come from seven years data-observations of the CMB anisotropies by the WMAP satellite]{2010arXiv1001.4635L}. Then, the $\Lambda$CDM model supposes an expanding Universe composed of baryonic and  cold dark matter. This cold dark matter, as it has been said before, is dissipationless and collisionless, which means, respectively, that it can not be cooled by photons and that CDM particles can interact between them and with other particles only through gravitation. The real innovation of the $\Lambda$CDM model is the assumption of a non-zero cosmological constant ($\Lambda$) \citep{1984ApJ...284..439P}. This last is associated with a vacuum energy or dark energy, which would explain the current accelerating expansion of space against the attractive effects of gravity. In the $\Lambda$CDM model the Universe is described by a hot big bang cosmology, where the Einstein's general relativity describes how gravity works, while the standard model of particle physics, with the addition of dark matter and dark energy, explains the material content and the way particles interact.\\

In this context, the numerical simulations of the $\Lambda$CDM model show the more recent view of the evolution history of our Universe. It is known as the concordance model. In some cases, this model is also called the standard cosmological model, which should not be confused with the standard CDM model mentioned above. Therefore, the evolution history of our Universe through the $\Lambda$CDM model could be summarized as follows. \\

After the Big-Bang, we have, in principle, four main epochs. First, the inflationary epoch \citep{1981PhRvD..23..347G,1982PhLB..108..389L,1982PhRvL..48.1437A,1986isos.book..287G}, where the seed density perturbations, which are responsible for the formation of the cosmic structures and all the properties of the Universe we see today, take place. Second, the radiation dominated era, during which the formation of the atomic nuclei ("nucleosynthesis", see below) take place. Third, the matter dominated era, where the recombination, the decoupling, and, the formation of larger structures, in that order, take place. Finally, we would be, eventually, in the dark energy dominated epoch which main characteristic is the Universe accelerating expansion. It is not known what happened before the inflation, if we could say so, and it is neither known what will happen after the dark energy dominated era, if there is such an after.\\

\begin{figure}[!h]
\centering
\includegraphics[width=0.6\textwidth]{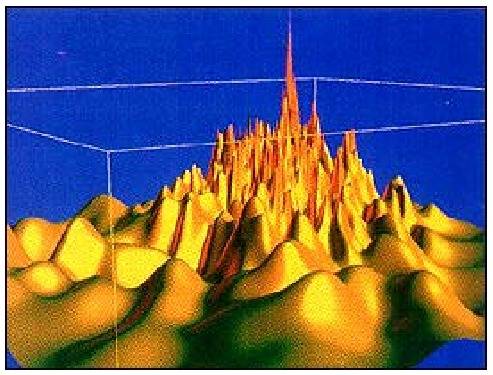}
\caption[Illustration of the primordial scalar field fluctuations]{Illustration of the primordial scalar field fluctuations (Credit: Linde, A.) \citep[see also][]{2008LNP...738....1L}.}
\label{Scalar_FieldFluc}
\end{figure}

The seed density fluctuations during inflation are scale-invariant according to the Harrison-Zeldovich spectrum \citep[see figure \ref{Scalar_FieldFluc};][]{1970PhRvD...1.2726H,1972MNRAS.160P...1Z}. More precisely, if one would have access to a map of the perturbations, one would not be able to distinguish whether the map is one kiloparsec across, one megaparsec across, the size of the whole observable Universe, or even zillion times larger than the whole observable Universe. The Harrison-Zeldovich perturbations are adiabatic, which corresponds to a perturbation in the total density shared amongst the different components (they have the same profile, e.g., where the DM density has a maximum, the densities of all other materials have also a maximum). These adiabatic perturbations are density perturbations (point-to-point variations in density) coming from earlier quantum fluctuations in the original scalar field that dominated the Universe during inflation. This scalar field is known as "the inflaton" and could thus be considered as the original material that later decays to create the material filling the present Universe. In such a case, the Universe, in its origin, presents a thermal equilibrium during which energy is exchanged between the different "materials" until the adiabatic conditions are satisfied. Such a thermal state is important as it forces the perturbations in any kind of material participating in the thermal equilibrium to become adiabatic perturbations. Two properties characterize these perturbations, their length scale (e.g., peak to peak distance) and their amplitude. Then, when the quantum fluctuations begin, they lead to large-scale density energy perturbations. As the inflation continues, smaller density energy perturbations are generated, with the smallest ones near the end. Hence, when the Universe finishes its inflation, the density energy perturbations coexist in a very wide range of length scales, and the quantum nature of the first fluctuations are no longer important as perturbations can be considered to follow classical physics laws. The transition from quantum to classical nature is called "decoherence". \\

\begin{figure}[!h]
\centering
\includegraphics[width=0.5\textwidth]{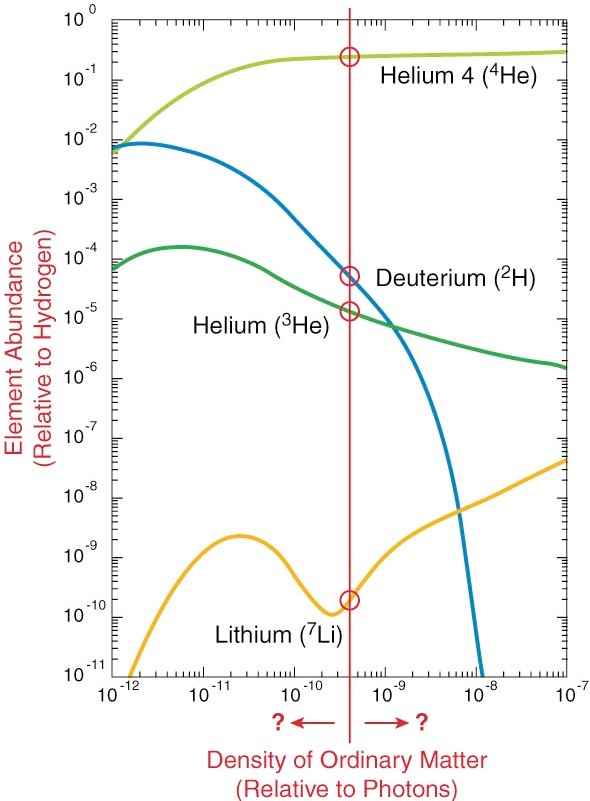}
\caption[The abundance of the main four light elements vs. baryon density in the Universe, as predicted by the theory]{The abundance of the main four light elements vs. baryon density in the Universe, as predicted by the theory. The baryon density is given in $\mu$g/km$^{3}$. The vertical red line indicates a strong interpretation of the observational data \citep[for more details see][]{1998RvMP...70..303S,1995Sci...267..192C}.}
\label{Elem_abundances}
\end{figure}

This is how the radiation dominated era begins. Given the fact that Einstein energy-matter relation (E=mc$^{2}$) allows us to go from matter to energy, and vice versa, the name of "radiation-dominated" only refers to the idea of a relativistic matter domination. During this epoch, the constituent particles (e.g., photons of light or neutrinos) travel at or near to the speed of light. However, it does not mean there is no other constituent. The Friedmann equation shows that the principal quantity that governs the expansion rate of the Universe is the density of the different material which dominate it in a specific epoch. In such a case, the total energy density of relativistic particles dominate, in a large range, over the baryon, dark matter, and dark energy densities. Nevertheless, as the Universe expands, this tendency changes. In any case, even though by the end of the radiation-dominated era only photons and neutrinos were relativistic, in the earlier moments the Universe was hotter and its ambient energy was thus higher, which could allow other particles to be also relativistic. It is during this epoch where the atomic nuclei of hydrogen and helium-4 are formed\footnote{in less amount, atomic nuclei of deuterium, helium-3, and lithium-7 are also formed.} in a process called nucleosynthesis \citep[][see also figure \ref{Elem_abundances}]{PhysRev.70.572.2,PhysRev.73.803}.\\

\begin{figure}[!h]
\centering
\includegraphics[width=1.0\textwidth]{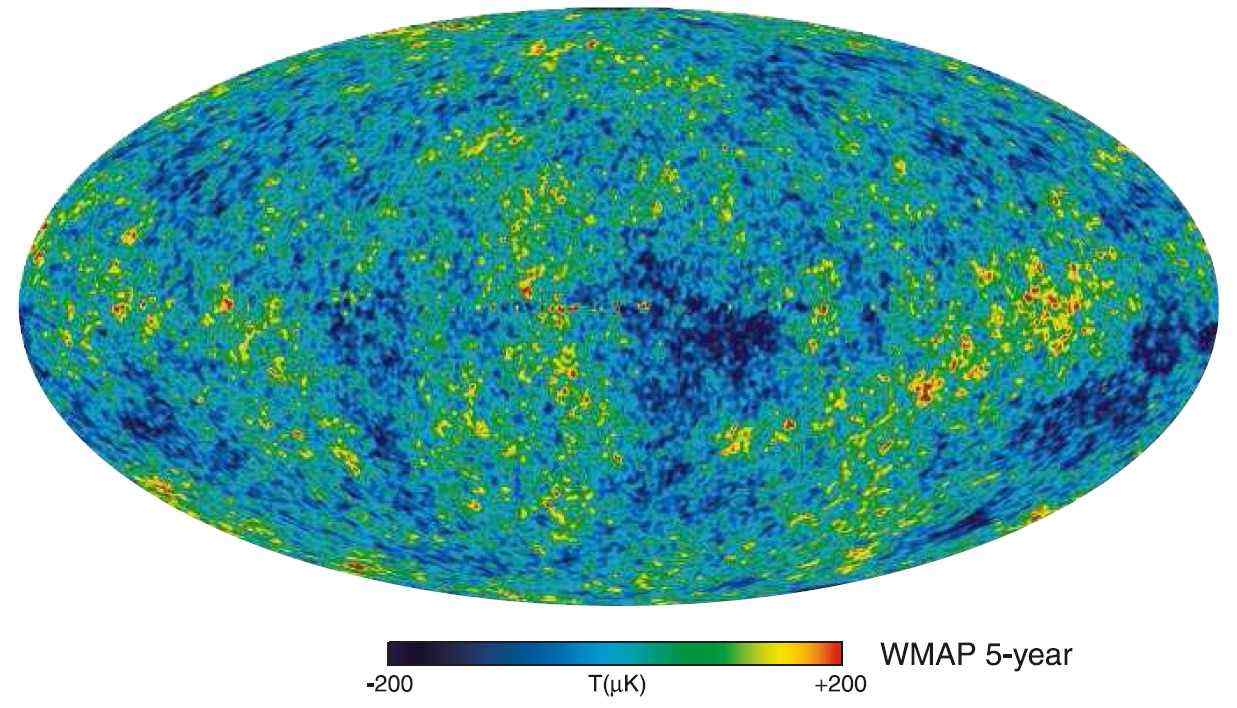}
\caption[CMB observations from WMAP5]{Cosmic Microwave Background (CMB) see by WMAP. The fluctuations in temperature, as determined from five years of WMAP data, are of the order of 10$^{-5}$, and the temperature has an average value of 2.725 K \citep{2009ApJS..180..225H}\citep[see also][for a general explanation]{2009AIPC.1132...86C}.}
\label{CMB_WMAP5}
\end{figure}

As the Universe continues its expansion, it becomes cooler and bigger, with the density of relativistic particles dropping faster than non-relativistic particles density. When the non-relativistic particles (principally, cold dark matter) took over, a long matter-dominated era began, and extending roughly until half of the present age of the Universe. Recombination, decoupling and reionization are key events of this era. The first makes reference to the process when electrons bind to the atomic nuclei formed during nucleosynthesis. Hence, the Universe become gradually populated by atoms, more and more neutral. This phenomena combined to the expansion\footnote{the expansion of the Universe decreases the photon energies and the particle densities. The first reduces the probability of photons to interact, and the second reduces the probability of two particles to meet.} makes the interaction between photons and matter particles negligible, and thus the Universe becomes transparent to radiation. This is the so called decoupling. Before these two process (recombination and decoupling), the competition between gravity, expansion and radiation pressure induces acoustic fluctuations in the primordial plasma that stop matter over-densities to grow. However, after these two processes, matter over-densities are amplified by gravity because the radiation pressure of photons can not stop it any more. As we can imagine, recombination and decoupling take place in a range of redshift ($\sim$200) that is named, abusively, {\it{surface of last diffusion}}. During this period, the radiation is still not totally decoupled from matter, interacting with the first over-densities in formation. What we see on the CMB map (see figure \ref{CMB_WMAP5}) are those fluctuations. Finally, the other key event during the matter-dominated era that has been mentioned above is the reionization. This one refers to the formation of the first massive stars and their supernova explosions, by which considerable amounts of energy are release into the intergalactic medium causing the dissociation of electrons from their atomic nuclei and ionizating the Universe again. Quasars and active galactic nuclei also contribute to this process, which take place at a redshift value of approximatly ten. After this, the formation of structures continues by the influence of gravity, and the Universe arrives to a period where the dark energy dominates, causing the acceleration of the Universe expansion \citep[][see figure \ref{HistoryOfTheUniverse} for an artistic view of the Universe history]{1998AJ....116.1009R,1999ApJ...517..565P,2010A&A...516A..63S}.\footnote{\citep[For more details about the history of the Universe, with a good pedagogical aim, see][]{Liddle...Loveday...TOCC}.}\\

\begin{figure}[!h]
\centering
\includegraphics[width=1.0\textwidth]{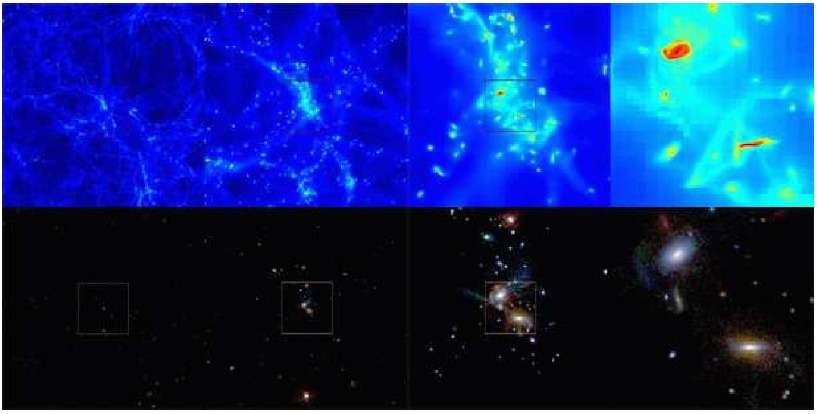}
\caption[Simulated galaxies in the $\Lambda$CDM model]{Simulated galaxies in the $\Lambda$CDM model \citep{2006A&A...445....1R}.}
\label{GalaxiesInLCDMmodel}
\end{figure}

\begin{figure}[!h]
\centering
\includegraphics[width=0.9\textwidth]{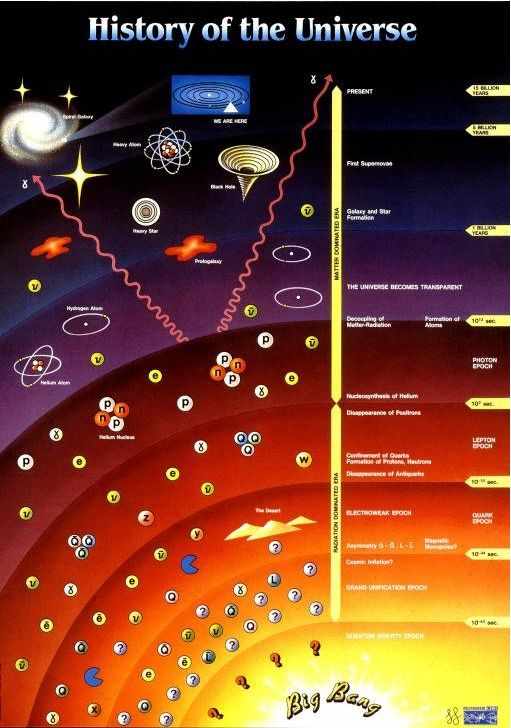}
\caption[Artistic view of the Universe history]{Artistic view of the Universe history (Credit: Microscom-CERN).}
\label{HistoryOfTheUniverse}
\end{figure}

In short, we have that the $\Lambda$CDM model is called hierarchical because structures grow by successive fusion of dark matter halos, which are follow by their baryonic content. When the cold dark matter is decoupled from the primordial plasma, it becomes insensitive to the interactions that bind the other components of that primordial plasma. Hence, it begins to condense, by the action of gravity, to form the dark matter halos. Later, as baryons become free, they are gravitationally caught by these halos, which make the growth of baryonic over-densities easier. At the moment these over-densities exceed the Jeans' mass, their gravitational and kinetic energies can then compensate each other, reaching the equilibrium and fixing their macroscopic properties (e.g., size, temperature), to finally condense by different cooling process \citep{2003Sci...300.1904M}. The first baryonic structures to be formed like this have then a typical size of the present globular clusters (10$^{5}$-10$^{6}$ M$_{\odot}$)\citep{2002Sci...295...93A,2004ARA&A..42...79B}. Inside these halos will form the first stars, which are metal-free and very massive (> 100 M$_{\odot}$), at redshift 20-30. These are named population III stars, and will live for a very short time, exploding in extremely energetic supernovae which may destroy the halos and, at the same time, disperse their metals throughout the universe. The next star generations (populations I and II) are thus metal rich, less massive and have a longer live. Their supernovae being less strong do not destroy the dark matter halos, making the formation of larger gravitational structures possible (galaxies and clusters, for example; see figure \ref{GalaxiesInLCDMmodel}).\\

\chapter{Basic and short introduction to modern imagery and photometry~~~}
\label{IntroImageryPhotometry}
\minitoc

\begin{figure}[!h]
\centering
\includegraphics[width=1.0\textwidth]{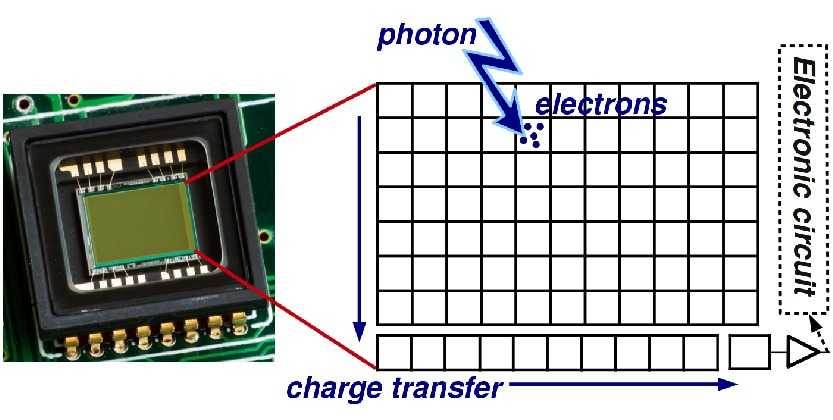}
\caption[CCD diagram to illustrate how it works]{CCD diagram to illustrate how it works.}
\label{CCDchip_illustration}
\end{figure}

\footnote{This appendix \ref{IntroImageryPhotometry} has been written thanks to the more detailed description presented in \citet{PHDthe...Lamareille06}.}In the present, the photometry and the imagery are made thanks to the CCD (Charge Coupled Device). The CCD is composed of a photosensitive cells matrix (see figure \ref{CCDchip_illustration}). Each time a photon hit one of this cells, it produces electrons that are counted later. These cells are called pixels, and they are the basic elements of a digital image. Like this, when the exposition time finishes, the system begins the signal reading by transferring the electric charges collected in each pixel to an electronic circuit where they are counted. The whole reading proceeding can be summarize by two step:

\begin{itemize}   
\item{1.)} The lecture begins by the bottom line, transferring at each time the electric charges from left to right,
\item{2.)} the charges in each line are transferred downwards to the next line, and the cycle return to step 1 (until there is no more line to read). 
\end{itemize}

It is noted that the reading time can represent a substantial constraint for the observers, and may be greater than the exposure time. In any case, some corrections must be applied to the CCD images before use them, in order to rectify some instrument imperfections. The most important are described below:\\
{\bf The bias:} this is the residual background noise, which is a constant noise received by the instrument. An image of the bias can be obtained making an instantaneous exposure with the shutter closed (i.e., in the dark).\\
{\bf The dark:} this corresponds to the bias plus an intrinsic thermal background noise produced by the instrument. It is proportional to the exposure time, and can also be imaged with the shutter closed.\\
{\bf The flatfield:} it allows to normalize the different responses that could have the matrix from one pixel to the other, taking into account that each pixel could react diferently compared to the others to a photon arrival, as well as some of them could be dead (with no response at all). The flatfield image can be obtained making an exposure of a diffuse light source, which is considered to be perfectly uniform. \\

Then, if, for a pixel {\it i}, the bias is noted B$_{i}$, the thermal background noise T$_{i}$ within an exposure time t$_{T}$, the dark D$_{i}$ is defined as,

    \begin{eqnarray}
       D = \frac{B_{i} + T_{i}}{t_{T}}
    \end{eqnarray}

Finally, noting I$_{0}$ the original image taken by the CCD sensor with an exposure time t$_{o}$, F the flatfield image taken with an exposure time t$_{F}$, and m$_{F}$ the mean value of the (F - D$\times$t$_{F}$) image, the final corrected image (I) for each pixel {\it i} would be:\\

    \begin{eqnarray}
       I_{i} = \frac{(I_{0_{i}} - D_{i}\times t_{o}) \cdot m_{F}}{F_{i} - D_{i}\times t_{F}}
    \end{eqnarray}

\section{Filters and photometric calibration}   

\begin{figure}[!h]
\centering
\includegraphics[width=1.0\textwidth]{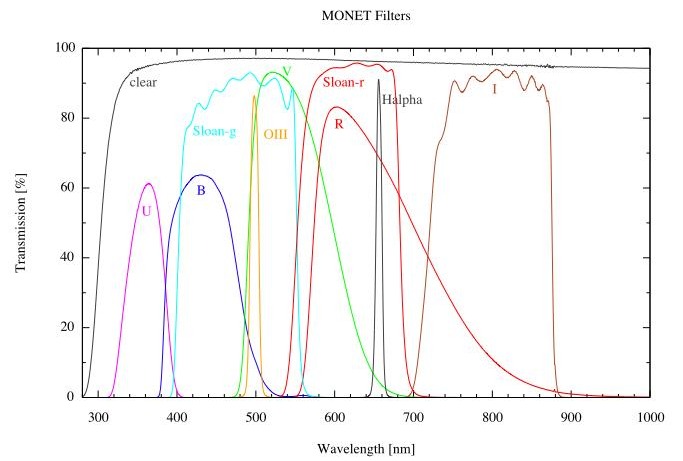}
\caption[Response curve for different filters]{Example of response curves. These correspond to the available filters for the Monet telescope at McDonald Observatory (Credit: Monet telescope webpage).}
\label{CurveResponseFilters}
\end{figure}

The principal role of a filter is to allow a precise vision of the observed object in a specific wavelength. Then, by combining the images obtained using different (wavelength) filters, we could have a better understanding of the object. However, all the filters do not have the same response, and this must be taken into account when comparing different observations. Fortunately, each filter is carefully studied, and finally characterized by a response curve (see figure \ref{CurveResponseFilters}) with a bandwidth $\Delta \lambda$ and a central waveleght $\lambda _{0}$. \\

The photometric calibration lies in finding the relationship between the number of electrons detected at each sensor cell, and the luminosity corresponding to each pixel of the image. In astrophysics, it is common to express the luminosity of an object in magnitude, which is a logarithmic unit that allows to work with a wide range of scales. Moreover, the total luminosity of an astrophysical object can be calculated only if one knows its distance\footnote{Luminosity refers to the mount of electromagnetic energy a body radiates per unit of time, and can be calculated as L = 4$\pi r^{2} F$. Where, r is the object distant and F the flux measured at the distant r.}, which poses some observational problems. Thus, the magnitudes are not calculated from the luminosity, but from the observed flux, which corresponds to the luminosity received per unit area of the detector. \\

Being ADU (Analog-to-Digital Unit) the number of electrons detected per pixel for a given object, and t$_{e}$ the used exposure time, the observed magnitude or "apparent magnitude" $m$ of such object is calculated as follow:

    \begin{eqnarray}
       m = -2,5 \cdot log (\frac{ADU}{t_{e}}) + ZP
    \end{eqnarray}

The constant ZP is called the image "zero-point". It contains various information including the area of each pixel, the atmospheric absorption, the physical units used to represent the observed fluxes, or even the normalization constant of the magnitude system (see below). Rather than calculating theoretically the zero point value, it is more reasonable to take images of reference stars which one already knows the apparent magnitude. It must be stressed that the image of the reference star must be achieved through the same filter as the image we want to calibrate. In addition, we must take into account the atmospheric absorption difference between the image to be calibrated and the reference image. Indeed, the atmospheric absorption is directly proportional to the height (called "airmass") of the observed objects, so to the length of atmosphere that has been crossed (by projection, it is smaller at the zenith than at the horizon). It is unusual that the image to be calibrated and the reference image are taken at the same height. \\

Then, let ADU$_{\star}$ be the number of electrons detected per pixel for the reference star, m$_{\star}$ its apparent magnitude, t$_{\star}$ the exposure time of the reference image (in the same unit as t$_{e}$), a the airmass of the image to be calibrated, a$_{\star}$ the airmass of the reference image, and c$_{a}$ the coefficient of atmospheric absorption which depends on the telescope and the instruments used for the observation. Like this, the zero-point can be calculated by \\

    \begin{eqnarray}
       ZP = m_{\star} + 2,5 \cdot log (\frac{ADU_{\star}}{t_{\star}}) + c_{\star}\times (a_{\star}-a)
    \end{eqnarray}

\section{The magnitude systems}

Now, as mentioned before, it is important to ensure that all magnitudes across the different filters are expressed in the same system. There are several systems of magnitude, each one characterized by a different normalization method. The magnitudes normalization, which is nothing more than the division of the observed flux by a reference flux, is performed for two main reasons: to overcome the physical units used to express the observed flux, and to overcome the response curve of the used filter (see above). This last point is absolutely necessary if one wants to really compare the luminous properties of the object in different range of wavelengths, rather than comparing the response curves of the filters. \\

As a result, an object magnitude through a given filter is defined as

    \begin{eqnarray}
       m = -2,5 \cdot log \Big(\frac{\int_0^\infty T(\lambda)\cdot f_\lambda(\lambda)d\lambda}{\int_0^\infty T(\lambda)\cdot f_\lambda^n d\lambda}\Big) = -2,5 \cdot log \Big(\frac{\int_0^\infty T(\nu)\cdot f_\nu(\nu)d\nu}{\int_0^\infty T(\nu)\cdot f_\nu^n d\nu}\Big) 
    \label{MonochromatiFluxMagSyst}
    \end{eqnarray}

where $\lambda$ is the wavelength of the received light, $\nu$ its frequency, f$_{\lambda}(\lambda)$ or f$_{\nu}(\nu)$ the monochromatic flux\footnote{this is the flux in a precise wavelength or frequency, per wavelength or frequency unit respectively.}, T($\lambda$) or T($\nu$) the response curve of the used filter, and $f_{\lambda}^{n}$ or $f_{\nu}^{n}$ the monochromatic flux of reference. \\

In here, the observed monochromatic flux must be expressed in the same units than the reference one, and it is this last which defines the magnitude system. Even if equation \ref{MonochromatiFluxMagSyst} comes from a different technique (i.e., the spectroscopy) to measure the monochromatic flux, rather than the photometry, in both cases there are two different systems to express the magnitude of an object. These are the Vega system and the AB system. \\

The Vega system is the more ancient, and it is based on the measure of the magnitude of the star Vega ($\alpha$ Lyrae)\footnote{It is one of the brightest stars in the sky, situated in the Lyra constellation.} which is consider as equal to 0. The Vega system thus uses this definition for any employed filter, taking the respective Vega star magnitude (or the monochromatic flux in equation \ref{MonochromatiFluxMagSyst}) as the reference.\\

Differently to the Vega system, the AB system \citep{1974ApJS...27...21O} does not depend on the observation of a particular object. The AB system is defined by having a constant monochromatic flux as reference expressed per frequency unit as follow\footnote{1 jansky (Jy) = 10$^{-23} erg \cdot s^{-1} \cdot cm^{-2} \cdot Hz^{-1}$ = 10$^{-28} J \cdot m^{-2}$}:
    \begin{eqnarray}
       f_\nu(AB) (\nu) = 3631 Jy 
    \end{eqnarray}
It is the most popular system nowadays. Then, equation \ref{MonochromatiFluxMagSyst} becomes
    \begin{eqnarray}
       m_{AB} =  -2,5 \cdot log \Big(\frac{\int_0^\infty T(\nu)\cdot f_\nu(\nu)d\nu}{\int_0^\infty T(\nu) d\nu}\Big) - 48,6
    \end{eqnarray}
with $f_{\nu}(\nu)$ in $erg \cdot s^{-1} \cdot cm^{-2} \cdot Hz^{-1}$. Curiously, the value of the monochromatic flux has been chosen in order to have a zero magnitude value for the star Vega through the V filter.\\

Finally, we can hence go from one system to the other by 
    \begin{eqnarray}
       m_{AB} = m_{Vega} +  m_{AB}(Vega)
    \end{eqnarray}
for which the values of $m_{AB}(Vega)$ for different filters are shown in table \ref{MagnitudeABofthestarVega}.

\begin{table}[!h]
\begin{center}
\begin{tabular}{|c|c|c|c|c|c|c|c|c|c|}
\hline
Filter & U & B & V & R & I & J & H & K  \\
\hline
$m_{AB}(Vega)$ & 1,0 & -0,09 & 0,0 & 0,18 & 0,46 & 0,9 & 1,37 & 1,88 \\
\hline \hline
Filter & u & g & r & i & z &  &  &    \\
\hline
$m_{AB}(Vega)$ & 0,96 & -0,09 & 0,16 & 0,39 & 0,54 & & & \\
\hline
\end{tabular}
\caption[Magnitude AB of the star Vega]{Magnitude AB of the star Vega through different common filters from \citet{PHDthe...Lamareille06}.}
\label{MagnitudeABofthestarVega}
\end{center}
\end{table}


\chapter{Kinematic studies of distant galaxies using integral field spectroscopy}
\label{KinematicStudies}
\minitoc

The first 3D kinematic studies of a representative sample of intermediate redshift galaxies, using FLAMES/GIRAFFE (see figure \ref{FlameGirrafe_Ifus} and section \ref{Introduction_chapterV}), were carried out by \citet{2006A&A...455..107F} and \citet{2008A&A...477..789Y}. Here I present a short description of the kinematic analysis, as well as some preliminary results of a small sample I studied during my PhD. \\

    \section{Toward a kinematic classification of distant galaxies}

\citet{2006A&A...455..107F} developed a simple kinematic classification scheme for distant galaxies based on their 3D kinematics and their HST optical images. It relies on the fact that at low spatial resolution, a rotating disk should show a well defined peak in the center of the $\sigma$-map, which corresponds to the convolution of the large scale motion (i.e., the rotation) with the (relatively small) dispersion of the gas in the disk or in the bulge. Indeed, the central parts of the galaxy, where the rotation curve rises most quickly, are not spatially resolved with ground-based optical spectroscopy and our classification fully accounts for its convolution with the actual PSF (point spread function). To summarize, we distinguish between the following classes:

\begin{itemize}
\item Rotating disks (RD): the VF shows an ordered gradient, and the dynamical major axis is aligned with the morphological major axis. The $\sigma$-map indicates a single peak close to the dynamical center;
\item Perturbed rotations (PR): the kinematics shows all the features of a rotating disk (see above), but the peak in the $\sigma$-map is either absent or clearly shifted away from the dynamical center;
\item Complex kinematics (CK): neither the VF nor the $\sigma$-map are compatible with regular disk rotation, including VFs that are misaligned with the morphological major axis.
\end{itemize}

\begin{figure}[!h]
\centering
\includegraphics[width=1.0\textwidth]{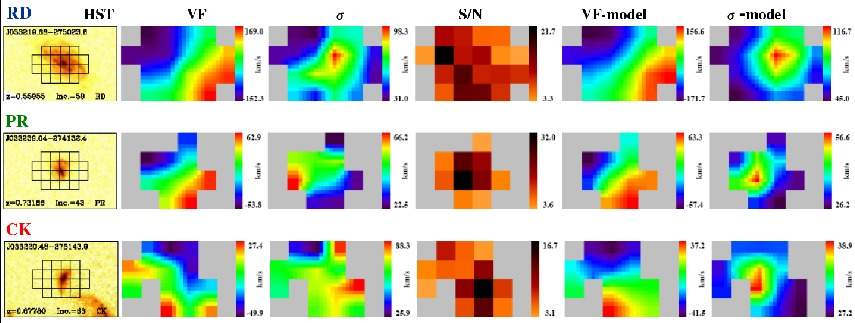}
\caption[Examples of different kinematic classes]{From {\it top} to {\it bottom}: Rotating disk, perturbed rotation, and complex kinematics. From {\it left} to {\it right}: HST/ACS image, measured velocity field, measured FWHM map, S/N map, simulated velocity field, simulated FWHM map.}
\label{DifferentClasses}
\end{figure}

Examples of each class are shown in figure \ref{DifferentClasses}. Interestingly, \citet{2006A&A...455..107F} find that the large scatter of distant TFR shown in previous studies is due to non-relaxed systems while the pure rotational disks exhibit a TFR that is similarly tight as that of local spirals.

\section{The kinematic analysis using GIRAFFE}
\label{KinematicStudiesGIRAFFE}

This is a topic well developed by \citet{PhDThe..PuechM06}, and \citet{PhDThe..Neichel08}. However, during my thesis I also had the opportunity to make the kinematic analysis of 15 galaxies using all the observed lines in each galaxy spectrum, and not just the O[II]$\lambda$3727 line. The procedure is the following. \\

First, we must accentuate that the special size of the pixel in the GIRAFFE IFU makes the width of the detected lines to be the result not only of the random movements of gas particles, but the product of the convolution between them and the tidy movements at larger scales \citep[for example, the rotation in the case of spiral galaxies;][]{2008A&A...484..173P}. Thanks to the resolution of FLAME/GIRAFFE (R$\sim$9 000), we can deblend the O[II] doublet (see figure \ref{Line_interface2}) for galaxies having a uniform movement \citep[RD; see][]{2008A&A...484..173P,2008A&A...477..789Y,2006A&A...455..107F}. If the convolution of large and small scale motions become more important, even with the high resolution of FLAMES/GIRAFFE, the lines are blended. Such cases are related to the signature of perturbed movements inside the galaxy \citep{PhDThe..PuechM06}, and makes impossible to calculate a barycenter for each line of the doublet. In kinematic studies of local galaxies, only those spatial pixels with S/N (signal to noise ratio) $>$ 3.0 are taken into account \citep{2006A&A...455..107F}. In those studies the S/N is defined as the ratio between the barycenter flux and the pseudo-continuum noise. As a result of the facts cited in the above phrases of this paragraph, a such S/N definition would severely under-estimate the signal to noise ratio in our case because of the trough separating the two lines of the [OII] doublet. We have hence chosen to define the signal to noise ratio as the total flux in the doublet normalized by the root mean square deviation of the pseudo-continuum (noise) surrounding the doublet. We then normalized again this S/N by the number of spectral resolution elements (N), over which the doublet [OII] emission line is spread, to obtain the mean S/N ratio: $<S/N>$ = $\frac{\sum_{i}^{N} S_{i}}{\sqrt{N} \times \sigma}$, where S$_{i}$ is the signal in the pixel {\itshape i} and $\sigma$ is the dispersion of the noise in the continuum of the spectra. For simplicity, we call this "$<S/N>$" simply "spectral S/N" or even "S/N".\\

\begin{figure}[!h]
\centering
\includegraphics[width=0.8\textwidth]{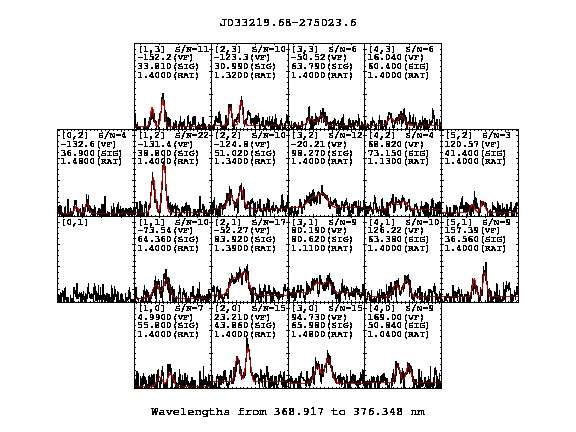}
\caption[OII GIRAFFE detection data-cube]{[OII] GIRAFFE data-cube detections.}
\label{OII_GIRAFFE_detectionDataCube}
\end{figure}

\begin{figure}[!h]
\centering
\includegraphics[width=1.0\textwidth]{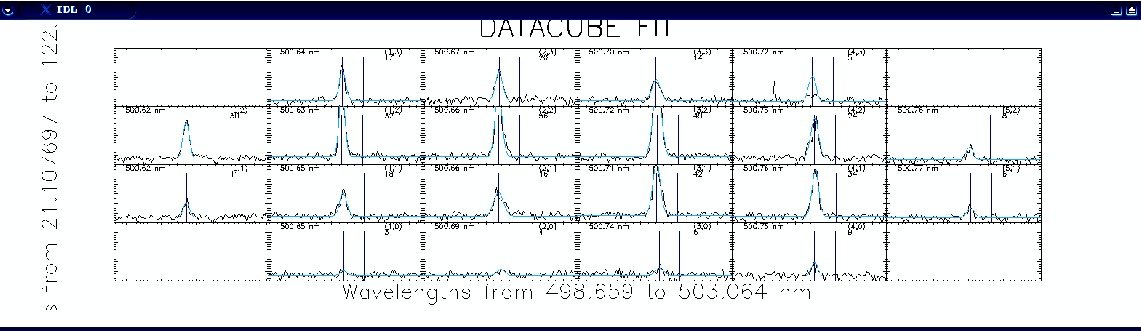}
\caption[Spatial data-cube for a low redshift galaxy (z=0.14)]{Spatial data-cube for a low redshift galaxy (z=0.14).}
\label{Cube_fit}
\end{figure}

\begin{figure}[!h]
\centering
\includegraphics[width=0.7\textwidth]{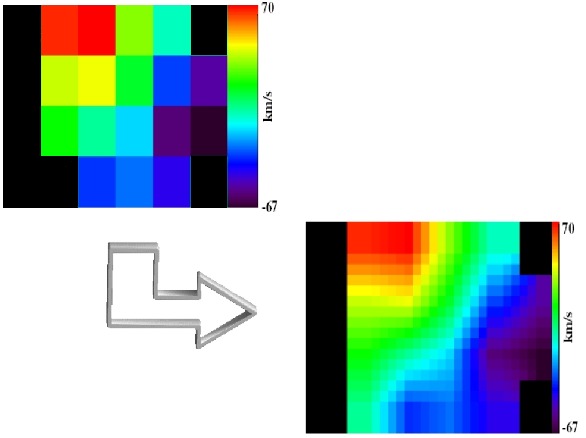}
\caption[Example of a velocity field from FLAMES/GIRAFFE]{{\it Left}: Example of a velocity field we can get with FLAMES/GIRAFFE. {\it Right}: Linear interpolation, for a better view of the velocity field. In the same way, we can also get the FWHM map (sigma map) and S/N map.}
\label{Velocity_field}
\end{figure}

We fit the O[II] doublet with a double Gaussian function satisfying the following constraints: distance between the lines = 2.783 $\AA$ ($\lambda _{1} - \lambda _{2} = 2.783 \AA$ \ at rest), same FWHM for both lines ($\sigma _{1} = \sigma _{2}$), and ratio between the two lines of the doublet free (see figures \ref{Line_interface2} and \ref{OII_GIRAFFE_detectionDataCube}). Therefore, the result from the fitting gives us the doublet position, the common FWHM, and the line ratio. In the case the fitting fails, we set the line ratio equal to 1.4, which is the median value observed in the integrated spectra \citep{2006A&A...455..119P,2006ApJ...653.1027W}(R(3729/3727) = 1.4 is also a good approximation for the maximum line ratio in gas regions, as estimated from \citet{2006agna.book.....O}; see more details in \citet{2006A&A...455..131P} and references therein)\footnote{Other works concerning the [OII] line ratio are presented by \citet{1992MNRAS.257..317K} and  \citet{1987MNRAS.227..161B}.}. This fitting failure is caused either by the low S/N, or by the difficulty to de-blend the [OII] doublet in some complex systems. It affects only $\sim$10-15$\%$ of the measured pixels and thus cannot significantly alter our conclusions \citep{2006A&A...455..107F,2008A&A...477..789Y}. In each case, we have checked by eye if the derived fit was acceptable. In some cases, we found blended [OII] spectra, in spite of the relatively high spectral resolution (R $\sim$ 10 000). This effect is independant of the S/N and is thus not an instrumental/observational artefact \citep{PhDThe..PuechM06}. This could be due to high local velocity gradients, multiple structures in the velocity field and/or high local extinction. Blended doublets are present in most of the complex kinematic systems, rather than in the more regular ones (rotating disks and perturbed rotation)\citep{2006A&A...455..107F,2008A&A...477..789Y}. Then, after interactively fitting the spectrum coming from each IFU pixel (see figure \ref{Line_interface2}), we get a 3 dimension data-cube as shown in figure \ref{Cube_fit} (I show here an example of low redshift galaxy, using the $O[III]\lambda 5007$ emission line). Finally, we can thus construct the respective velocity field (see figure \ref{Velocity_field}), and, in the same way, the FWHM map and the S/N map. A 5x5 linear interpolation is also applied to the velocity field and the FWHM map. However, this linear interpolation only has a visual impact, as its only utilization is limited to the visual interpretation of the internal kinematics of the galaxy. \citep{2008A&A...484..173P} shows that the rotation velocities retrieved through the IFU pixels of GIRAFFE have a systematic uncertainty within $\sim$0.02 dex \citep[see also the independet study of][]{2010MNRAS.401.2113E}. \\

\begin{figure}[!h]
\centering
\includegraphics[width=1.0\textwidth]{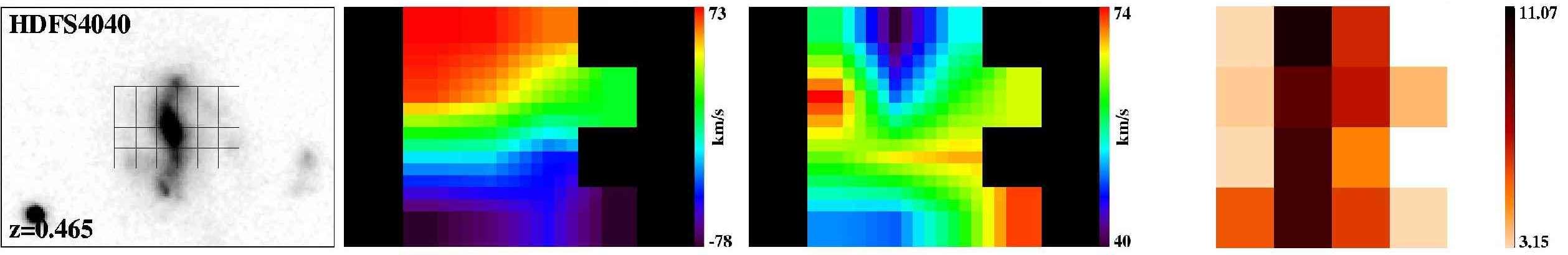}
\caption[HST/ACS image, velocity field, FWHM map (sigma-map), and S/N map for one galaxy]{From {\it left} to {\it right}: HST/ACS image of the galaxy, velocity field, FWHM map (sigma-map), and S/N map.}
\label{DifferentMaps}
\end{figure}

Concerning other emission lines (e.g., H$\gamma$, H$\beta$, [OIII]$\lambda$4959, [OIII]$\lambda$5007), the fitting is done by a simple Gaussian function. In this case, we keep the same S/N definition because it allows us to better exploit the detection of emission lines in the distant galaxies. For each line presents in each galaxy we keep only those IFU pixels with $<S/N>$ $\geq$ 3 during the confection of different maps (velocity field, sigma-map, S/N map, see figure \ref{DifferentMaps}). Nevertheless, for a map to be considered as significantly representative of the internal kinematic of a given galaxy, it must contain at least four GIRAFFE spatial pixels with S/N $\geq$ 4. \\

In short, following the above procedure, I constructed 3 dimension data-cubes (see figure \ref{Cube_fit}) using all the observed emission lines in each galaxy spectrum with and without background subtraction. For this, an IDL procedure has been implemented. Subsequently, from these data-cubes, the {\itshape velocity fields}, {\itshape sigma-maps} and {\itshape S/N maps} have been build also using IDL procedures that follow the definitions and explanations detailed before.\\

\section{Some preliminary results}

From the sample of 15 galaxies, 12 of them are kinematically exploitable. They are in the redshift bin 0.11 $<$ z $<$ 0.40, and they have absolute B magnitude between -19.65 < M$_{B_{Vega magnitude}}$ < -16.40. Their velocity fields, sigma maps, and S/N maps are shown in figure \ref{ChampsCartesSimula1} at the end of this appendix.  These results interestingly show that galaxies with complex kinematics (called DC in the figure) present the same complex signature through the same ionised emission lines (e.g., [OIII]$\lambda$4959 and [OIII]$\lambda$5007), while different emission lines (e.g., H$\beta$ and [OIII]) show different complex kinematics. In contrast, rotating disks (called DR in the figure) have the same kinematic signature using any emission line.\\

\begin{figure}[!t]
\centering
\includegraphics[width=0.5\textwidth]{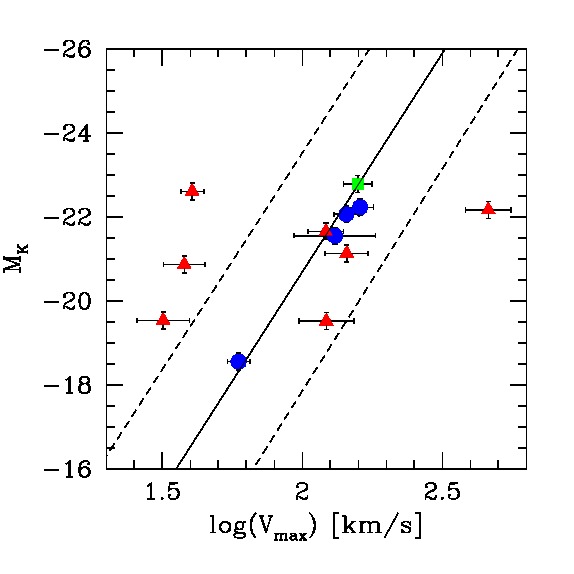}
\includegraphics[width=0.5\textwidth]{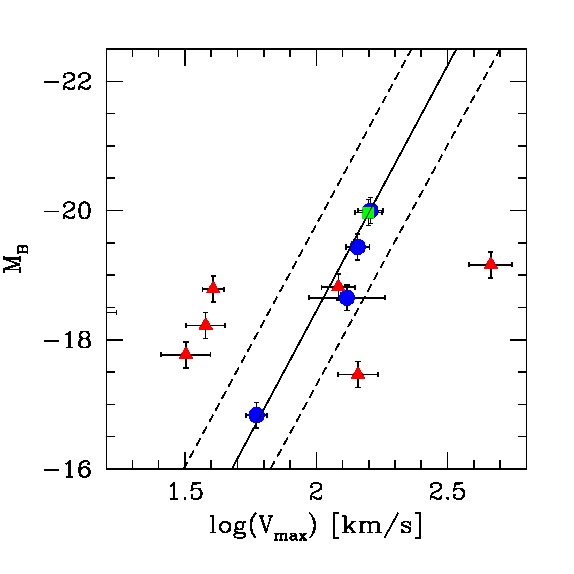}
\caption[Preliminary results: Tully-Fisher relation]{Tully-Fisher relation, for the sample of 12 galaxies, in K band (Vega magnitudes) and in B band (Vega magnitude, corrected for extinction), as in \citet{2006A&A...455..107F}. Red triangles, green squares and blue dots represent complex kinematics, perturbed rotations and rotating disks, respectively. Full and dotted lines represent the local TF and its 3 sigma scatter amplitude \citep[following][]{2005ApJ...628..160C}\citep[see also][]{2001ApJ...563..694V}.}
\label{TFrelationfromMySample}
\end{figure}

Another preliminary result is shown in figure \ref{TFrelationfromMySample}. Similar to precedent results these plots show that all the scatter of the TF relation is caused by interloper galaxies with kinematics classified as complex. Considering only rotating disks, the TF relationship at 0.11 $<$ z $<$ 0.40 is similar to that in the local universe.\\

\newpage

\begin{figure}[!ht]
\centering
\caption[3D kinematic analysis of M$_{B}$ < -16.40 galaxies]{For each galaxy in the sample I present here their HST/ACS image, measured velocity field, measured FWHM map, S/N map, simulated velocity field, simulated FWHM map (in that order), corresponding to each emission line that could be analyze.}
\label{ChampsCartesSimula1}
\end{figure}

\newpage
\begin{changemargin}
\thispagestyle{empty}
\includegraphics[angle=180]{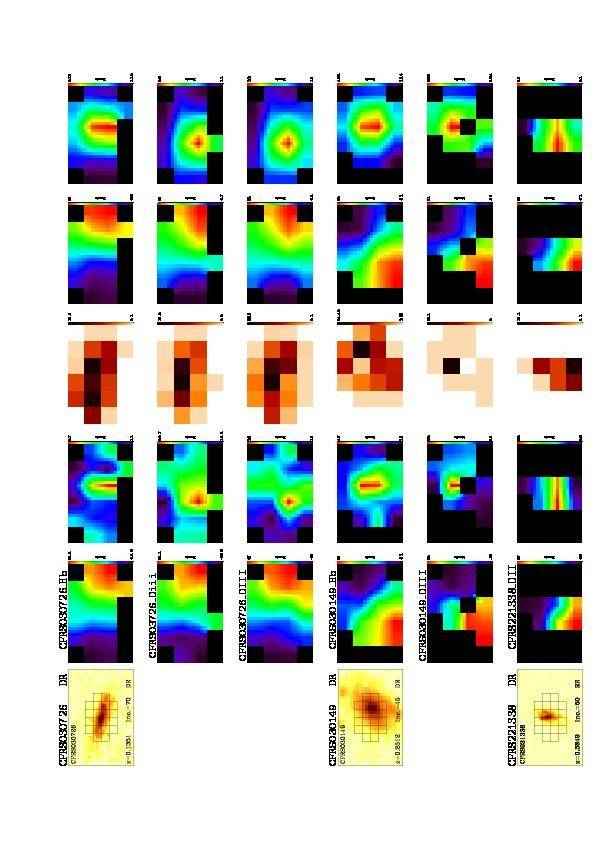}
\newpage
\thispagestyle{empty}
\includegraphics{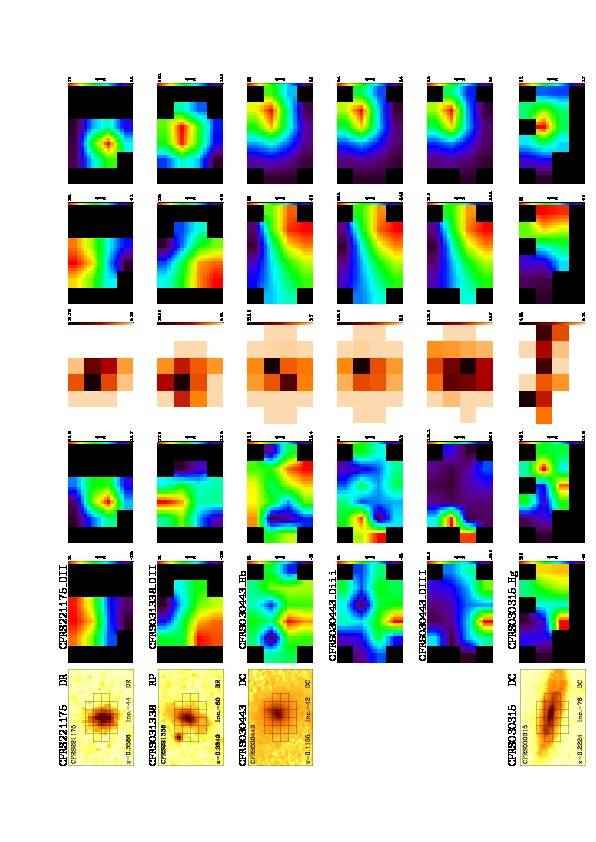}
\newpage
\thispagestyle{empty}
\includegraphics[angle=180]{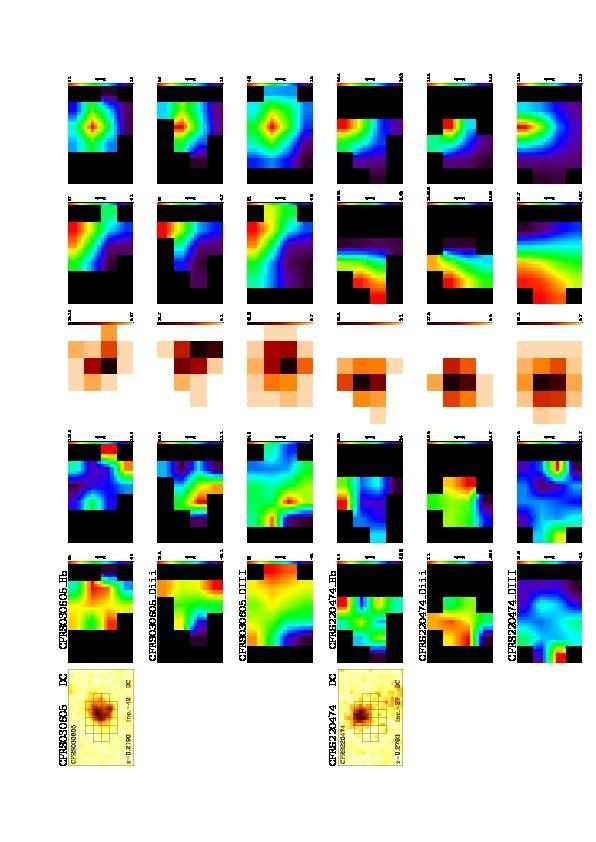}
\newpage
\thispagestyle{empty}
\includegraphics{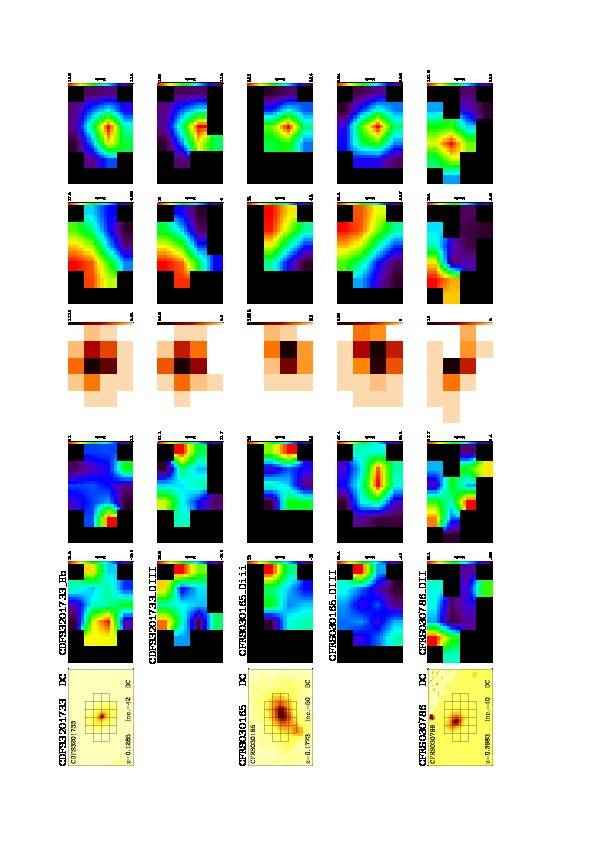}
\end{changemargin}


\chapter{Building an Observatory in Panama}
\label{PanamaObservatory}
\minitoc

\section{A very short history}

If I must give a starting date of the project, I would say "solar eclipse of February 1998". Thanks to the opportunity we were going to have in Panama to observe such total solar eclipse, a delegation of the "Uranoscope de France" traveled to Panama. Once they were in place, and after having a nice time watching the eclipse, they made contact with some government authorities and with the APAA ("Asociaci{\'o}n Paname{\~n}a de Aficionados a la Astronom{\'i}a"\footnote{Panamanian Association of Amateur Astronomy.}). The Uranoscope then let them know about the Uranoscope idea to help the interested people in Panama to construct an astronomical observatory there. With this in mind, the Uranoscope proposed to offer a telescope to Panama. However, Panama should take the responsibility to build the appropriate structure.\\

It was until 2000 when I arrived at the School of Physics at the University of Panama. Since then I was interested in studying Astrophysics, and with other classmates I formed a group with the aim of organizing talks and other activities.  At the end of that year, professor Bernardo Fernandez told me about the posibility to make an agreement of mutual collaboration between the University of Panama and the Uranoscope de France. Immediately, professor Fernandez and I began to work on that, and the following year (2001), it was done and signed.\\

Between 2001 and 2005, a lot of activities were developed. Amongst them, the biggest one took place once a year, and was called "The French scientific fortnight in Panama". In its organization participated the "Alliance Francaise au Panama", the French Embassy, the University of Panama, and the Technological University of Panama.\\

Finally, on April 2005 a hybrid solar eclipse was visible in Panama\footnote{Yes, another one, so I have seen 3 solar eclipses in my life without the need to leave my country (11 July 1991, 26 February 1998, and 08 April 2005).}. The Uranoscope visited again Panama, and made, through the French Embassy, the official donation of a telescope to the Technological University of Panama. It was now our turn to continue with the project. As a result, since then, we have been working on it, and the observatory is almost finished (see figures \ref{Panama_ObservatoryMarch} and \ref{Panama_ObservatoryJuly}). Moreovere, a new agreement of mutual collaboration between the Technological University of Panama and the Uranoscope de France was signed in 2008.

\section{Location}

\begin{figure}[!h]
\centering
\includegraphics[width=0.9\textwidth]{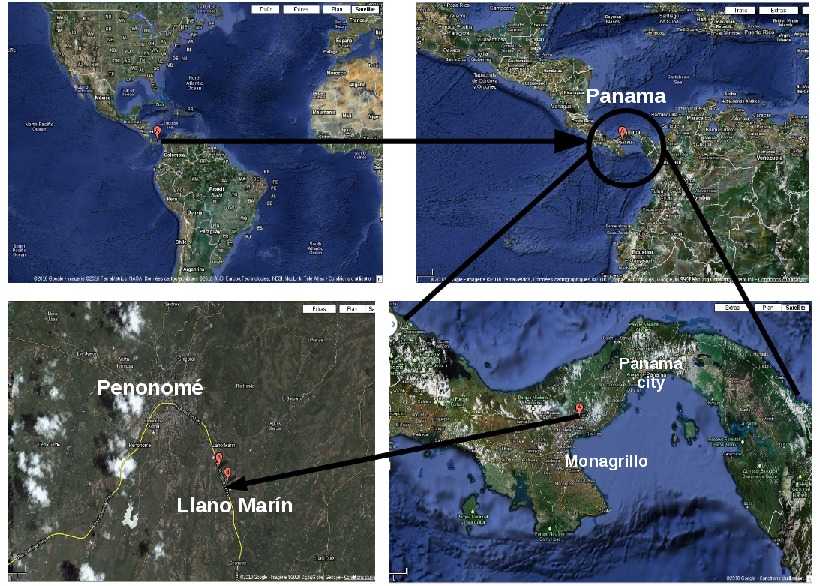}
\caption[Panama observatory - Location]{How to arrive to the Panama observatory.}
\label{Panama_ObservatoryLocation}
\end{figure}

The new astronomical observatory of Panama has been built inside a regional center of the Technological University of Panama, located at Llano Marin at $\sim$5 km from Penonome city (latitude: 08$^{\circ}$30'00"N, longitude:80$^{\circ}$19'48"W; see figure \ref{Panama_ObservatoryLocation}).\\

\section{General description}

\begin{figure}[!h]
\centering
\includegraphics[width=0.9\textwidth]{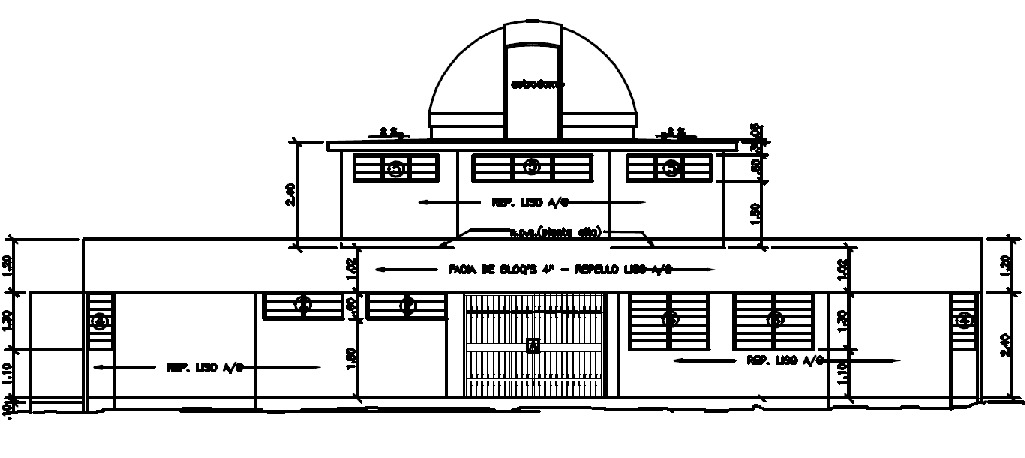}
\caption[Panama observatory - Architectural scheme]{Architectural scheme of the Panama observatory building.}
\label{Panama_ObservatoryArchi}
\end{figure}

The building has three floors including the dome, and measures 5.52 m from the ground to the base of the dome (see figure \ref{Panama_ObservatoryArchi}). The dome dimensions are the following: 3.69 m high, and a diameter of 4.57 m at the base. \\

\section{The Instruments}

Nowadays, the observatory owns a Meade LX200GPS telescope. It has an aperture diameter of 35.56 cm. The telescope will be mounted on a Astro-Physics 1200 mount, which is already in Panama. \\

The rest of the instruments are still being purchased. Indeed, we plan to have them in Panama before the end of the year. Amongst such instruments we have a CCD SXVF-H16 Starlight Xpress, a LHIRES III spectrograph with three gratin modules (1200, 600, and 500 line/mm), a Lodestar Stand-Alone autoguider, and an active optics system. We will also have other essential accessories as a set of Sloan photometric filters, field corrector, focal reducer, polarizing filters, bandpass CCD imaging filters, light pollution suppression filters, barlows, a last generation eyepiece set, as well as a Zenithstar APO telescope for a better follow-up control, and a Coronado telescope. A series of other amateur telescope are also included to be used in educational activities.\\

\section{The Projects}

The first projects to be developed with our first generation of instruments are the following:\\

\begin{figure}[!h]
\centering
\includegraphics[width=0.8\textwidth]{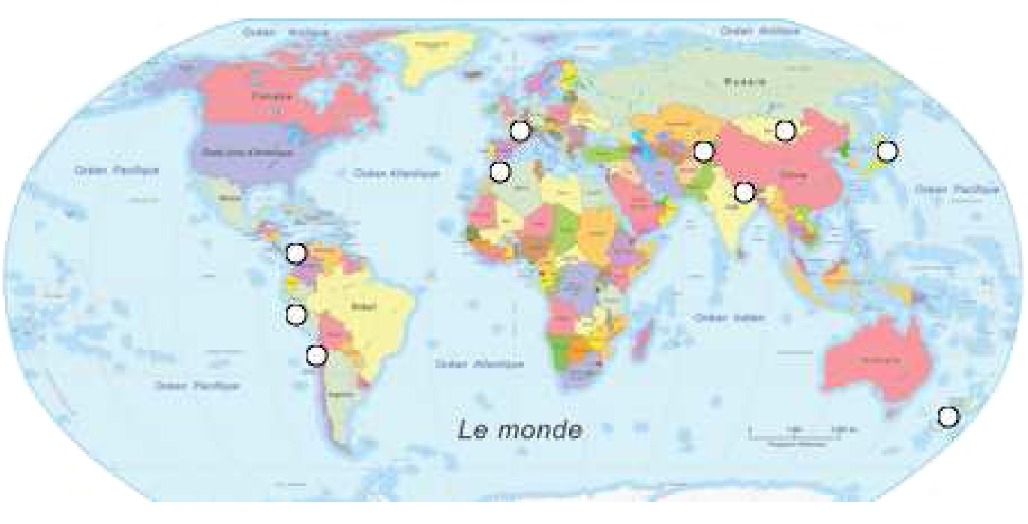}
\caption[Observatories international network]{Observatories international network to study the lunar flash.}
\label{Observatories_netwokFlash}
\end{figure}

\begin{figure}[!h]
\centering
\includegraphics[width=0.45\textwidth]{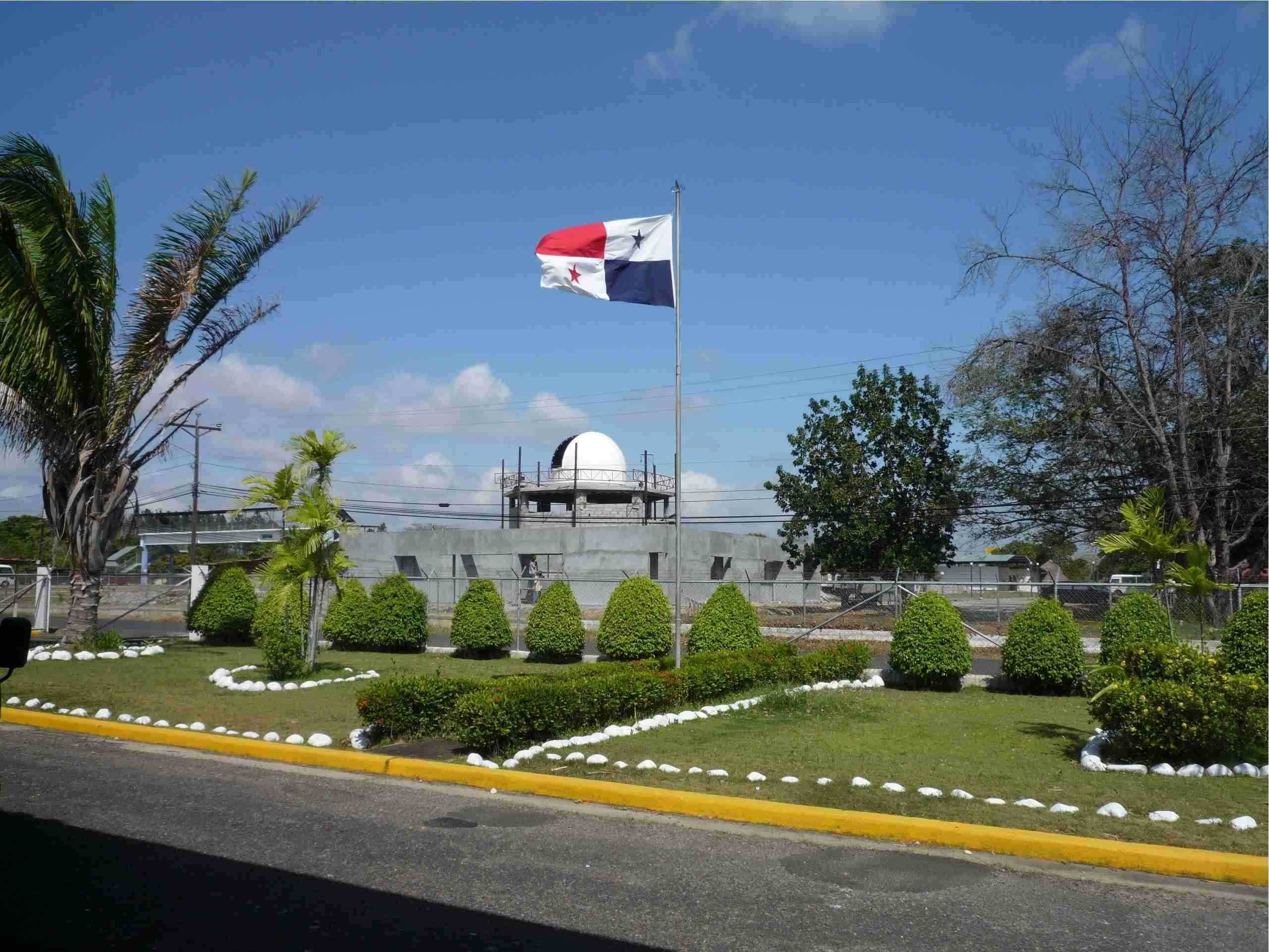}
\includegraphics[width=0.45\textwidth]{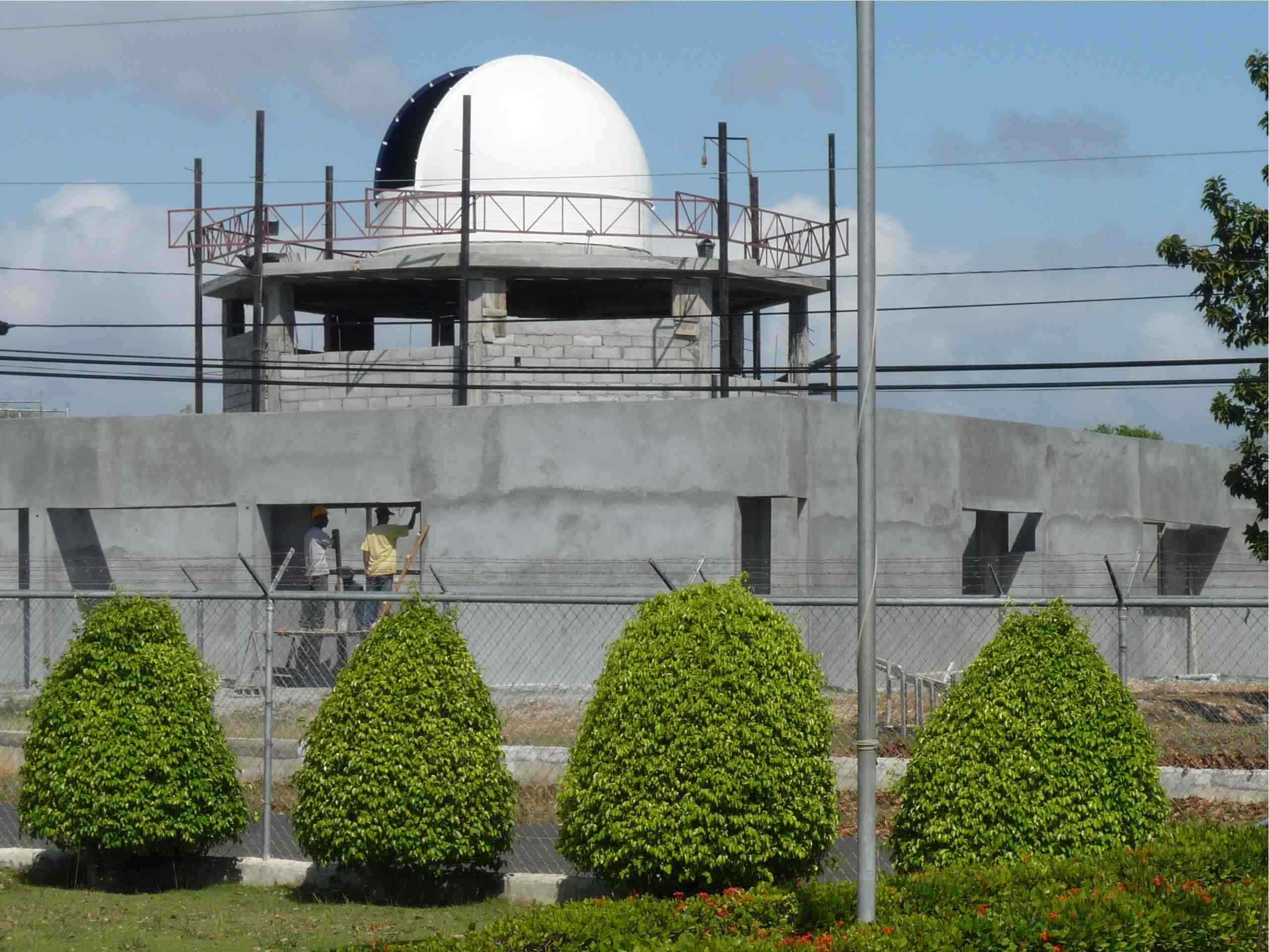}\\
\vspace*{0.2cm}
\includegraphics[width=0.45\textwidth]{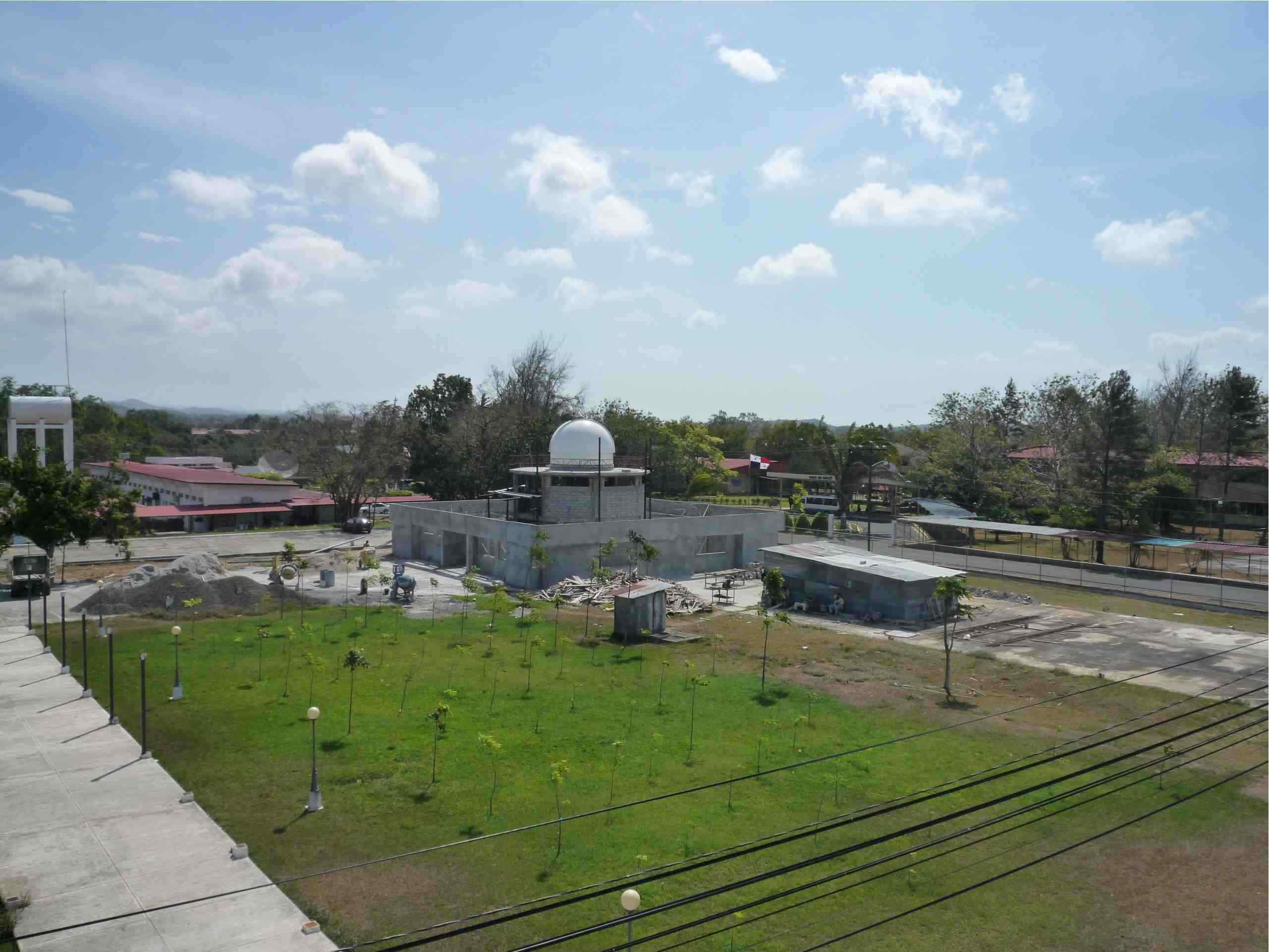}
\includegraphics[width=0.45\textwidth]{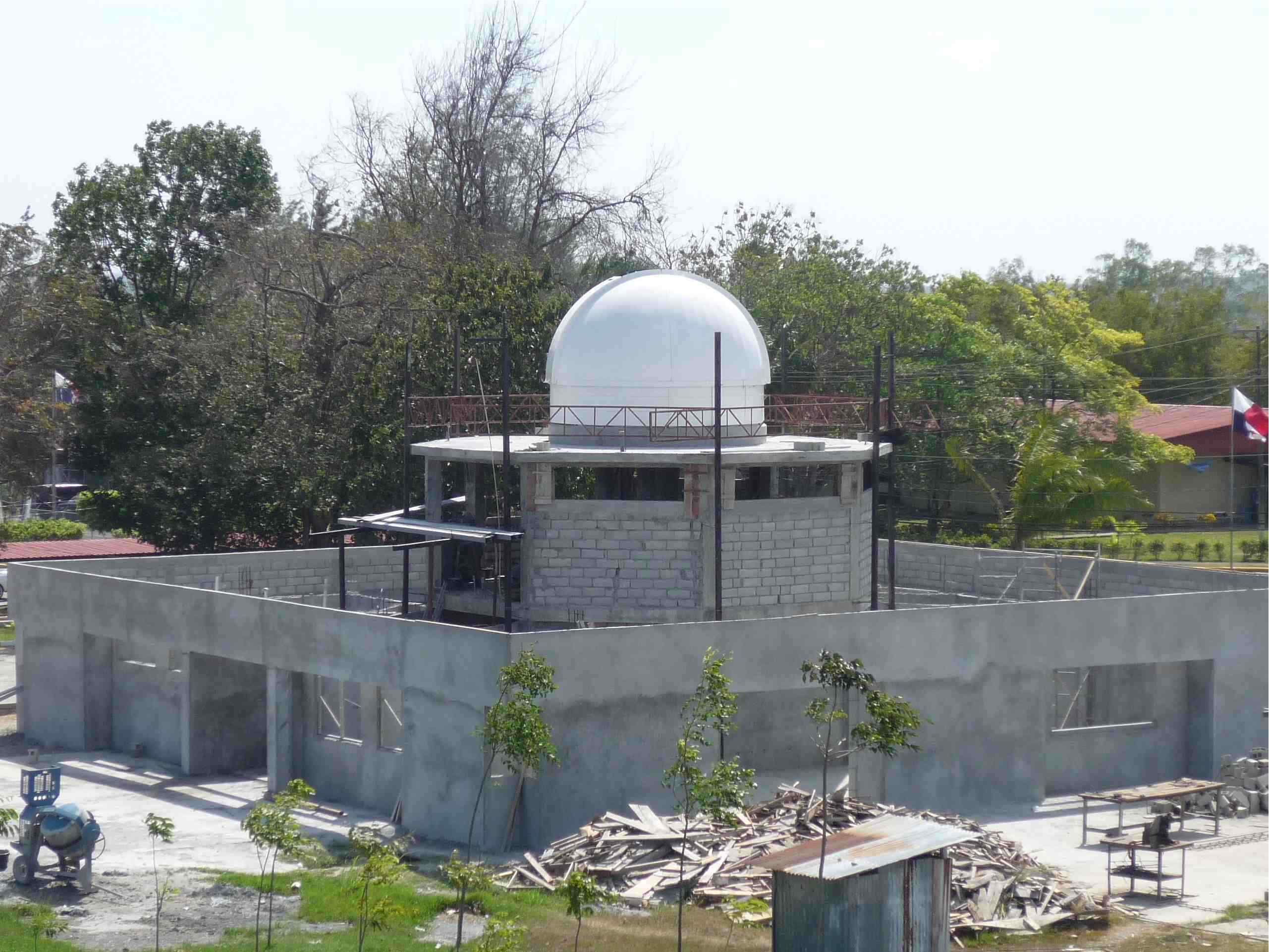}
\caption[Panama observatory - on March 2010]{Images of the observatory during March 2010.}
\label{Panama_ObservatoryMarch}
\end{figure}

\begin{figure}[!h]
\centering
\includegraphics[width=0.8\textwidth]{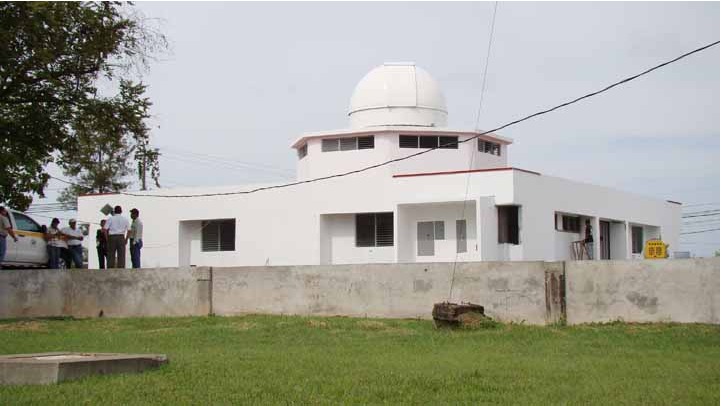}\\
\vspace*{0.2cm}
\includegraphics[width=0.8\textwidth]{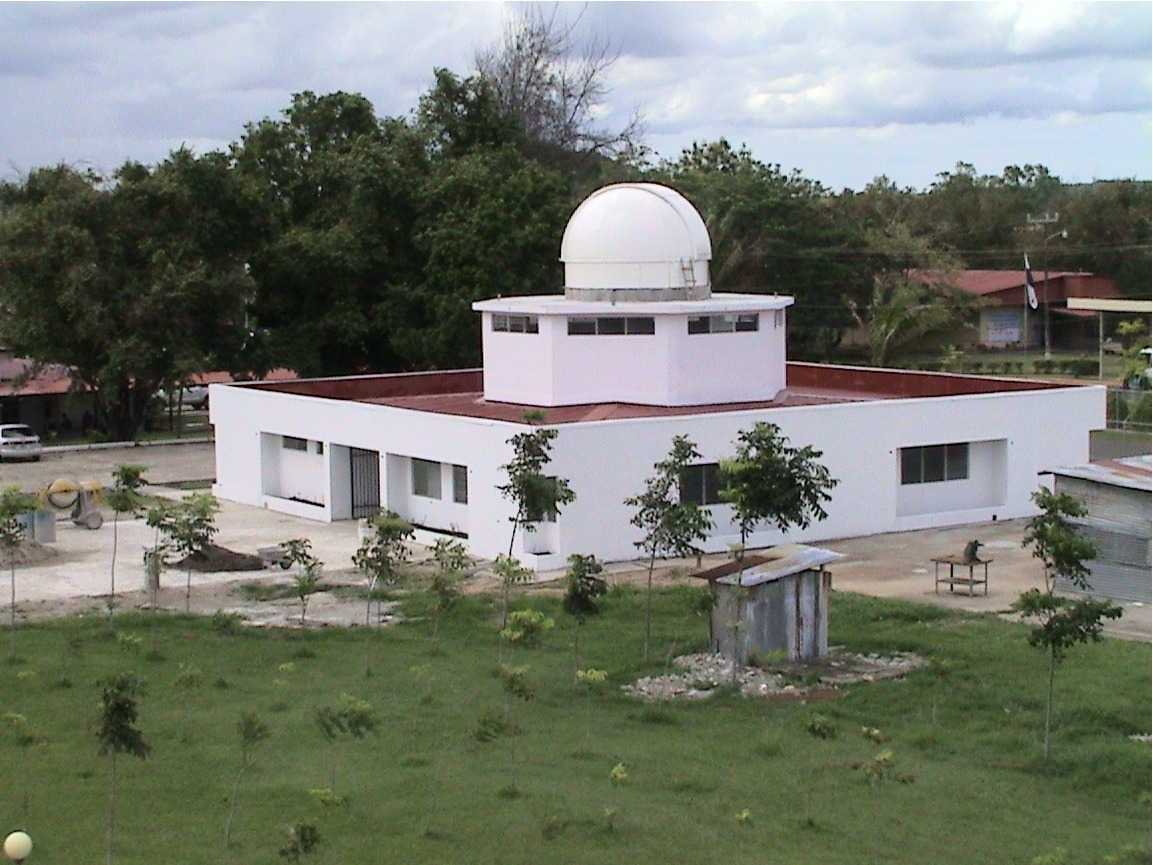}\\
\vspace*{0.2cm}
\includegraphics[width=1.0\textwidth]{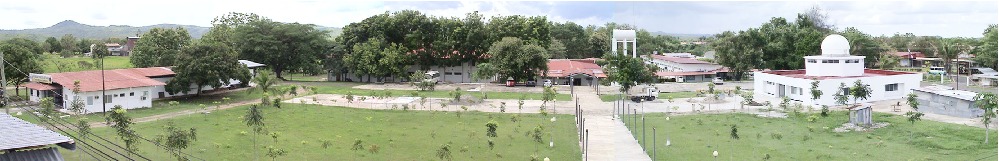}
\caption[Panama observatory - on July 2010]{Images of the observatory during July 2010.}
\label{Panama_ObservatoryJuly}
\end{figure}

\begin{itemize}
\item 1. Imagery and photometry of tidal stream in nearby galaxies (see prospective 2 in chapter 7).\\
\item 2. Follow-up of lunar events: to the hunt of lunar flashes. It is an international project integrated by different observatories (see figure \ref{Observatories_netwokFlash}) to track meteoroid impacts on the moon. Its principal scientific issues are, for example, to derive the amount of kinetic energy being radiated and the fraction being transmitted through seismic energy at each impact, as well as measuring the crater dimensions in function of the produced flash. This data will be very important in order to be compared with the information that has been and will be obtained by different missions on the moon (measurements "in situ" with spectrometers, magnetometers, seismic detectors, gravimeters, etc.).  \\
\item 3. NEOs (Near-Earth Objects) detection and tracking (in collaboration with the IMCCE of Paris Observatory).\\
\end{itemize}

Other projects will be focused on the education (in collaboration with Paris Observatory, Plan{\`e}te Science Institute, Uranoscope de France, and with the Palais de la Decouverte-Minister de la Culture de France).\\

Another idea is also to test automatisation systems, such as Tarot (T{\'e}lescope {\`a} Action Rapide pour les Objets Transitoires) system, and Hatnet, for example. Undoubtedly, any other project and/or collaboration is happily welcome. \\

A long (but not so long) term project is the installation of an observatory, hosting a bigger (at least 1 m diameter) telescope, at an altitude higher than 1300 m in Panama, and having an automatic control system.\\

\bibliographystyle{ThesisStyleWithEtAl}
\bibliography{Biblio}

\begin{thebibliography}{xxxx}
\bibitem[{Braun} \textit{et al.}, 2009]{2009arXiv0906.1615B}
{Braun} J., {Hubert} D. and {for the IceCube Collaboration} .
\newblock {\em {Searches for WIMP Dark Matter from the Sun with AMANDA}}.
\newblock ArXiv e-prints, June 2009.
\end{thebibliography}

\onecolumn
\newpage
\cleardoublepage
\begin{vcenterpage}
\noindent\rule[2pt]{\textwidth}{0.5pt}
\begin{center} 
{\large\textbf\small{The Evolution of the Hubble Sequence: morpho-kinematics of distant galaxies\\}} 
\end{center} 
{\large\textbf\small{Abstract:}}
{\small The main objective of my thesis was to provide us, for the first time, with a reliable view of the distant Hubble sequence, and its evolution over the past 6 Gyr. To achieve this goal, we have created a new morphological classification method which (1) includes all the available observational data, (2) can be easily reproduced, and (3) presents a negligible subjectivity. This method allows us to study homogeneously the morphology of local and distant galaxies, and has the main advantage of presenting a good correlation between the morphological type and dynamical state of each galaxy. 
\\
The first step has been to study the evolution of galaxies using the IMAGES survey. This survey allowed us to establish the kinematic state of distant galaxies, to study the chemical evolution of galaxies over the past 8 Gyr, and to test important dynamical relations such as the Tully-Fisher relation. Kinematics is, indeed, a crucial information needed to guarantee a robust understanding of the different physical processes leading to the present day Hubble sequence. Using Integral Field Spectroscopy, which provides a complete kinematic diagnosis, we have been able to test our new morphological classification against the kinematic state of each galaxy. We found that the morpho-kinematic correlation is much better using our classification than other morphological classifications. Applying our classification to a representative sample of galaxies at $z \sim 0.6$, we found that 4/5 of spiral galaxies are rotating disks, while more than 4/5 peculiar galaxies are not in a dynamical equilibrium.  
\\
Applying our morphological classification to a representative sample of both local and distant galaxies, having equivalent observational data, we obtained a Hubble sequence both in the local and distant Universe. We found that spiral galaxies were 5/2 times less abundant in the past, which is compensated exactly by the strong decrease by a factor 5 of peculiar galaxies, while the fraction number of elliptical and lenticular galaxies remains constant. It strongly suggests that more than half of the present-day spirals had peculiar morphologies, 6 Gyr ago. 
\\
Finally, I present further studies concerning the history of individual galaxies at $z < 1$, combining kinematic and morphological observations. I also present the first ever-estimated distant baryonic Tully-Fisher relation, which does not appear to evolve over the past 6 Gyr. In the coming years, our morphological classification and these studies will be extended to galaxies at $z >> 1$, thanks to the future ELTs.
\\
{\large\textbf{Keywords:}}
Galaxy Formation and Evolution - Galaxy Morphology and kinematics.
}
\\
\noindent\rule[2pt]{\textwidth}{0.5pt}
\end{vcenterpage}

\end{document}